\documentclass[12pt]{article}
\usepackage{hyperref}
\usepackage{cite}
\usepackage{color}
\usepackage{graphicx}
\usepackage{amsmath}
\usepackage{amssymb}
\usepackage{multirow}

\makeatletter
\@addtoreset{equation}{section}

\makeatletter
\renewcommand\section{\@startsection {section}{1}{\z@}%
                                   {-3.5ex \@plus -1ex \@minus -.2ex}
                                   {2.3ex \@plus.2ex}%
                                   {\normalfont\large\bfseries}}
\renewcommand\subsection{\@startsection{subsection}{2}{\z@}%
                                     {-3.25ex\@plus -1ex \@minus -.2ex}%
                                     {1.5ex \@plus .2ex}%
                                     {\normalfont\bfseries}}

\def\baselinestretch{1.2}
\numberwithin{equation}{section}

\newcommand{\be}{\begin{equation}}
\newcommand{\ee}{\end{equation}}

\newcommand{\tr}{{\rm Tr}}
\newcommand{\gone}[1]{{}}
\newcommand{\bl}{\noindent $\bullet\ $}

\newcommand{\A}{\mathcal{A}}
\newcommand{\X}{\mathcal{X}}
\newcommand{\Y}{\mathcal{Y}}

\newcommand{\stup}{|\! \! \uparrow \rangle}
\newcommand{\stdown}{|\! \!\downarrow \rangle}

\newcommand{\bR}{\ensuremath{\mathbb{R}}}

\newcommand{\scN}{\ensuremath{\mathcal{N}}}

\newcommand{\beq}{\begin{eqnarray}}
\newcommand{\eeq}{\end{eqnarray}}

\newcommand{\rot}[3]{\left[{#1}\atop{#2}\right]_{#3}}

\begin{document}
\begin{titlepage}
\begin{flushright}
MAD-TH-14-03\\
IPMU-14-0094\\
\end{flushright}

\vfil

\begin{center}

{\large{\bf Boundaries and Defects of ${\cal N}=4$ SYM with 4 Supercharges} \\
\bigskip
Part I: Boundary/Junction Conditions}

\vfil

Akikazu Hashimoto$^a$, Peter Ouyang$^b$, and Masahito Yamazaki$^{c,d}$

\vfil

$^a$ Department of Physics, University of Wisconsin, Madison, WI 53706, USA

$^b$ Department of Physics, Purdue University, West Lafayette, IN 47907, USA

$^c$ Institute for Advanced Study, Princeton, NJ 08540, USA

$^d$ Kavli IPMU (WPI), University of Tokyo, Kashiwa, Chiba 277-8583, Japan

\vfil

\end{center}

\begin{abstract}

\noindent We consider ${\cal N}=4$ supersymmetric Yang Mills theory on
a space with supersymmetry preserving boundary conditions. The
boundaries preserving half of the 16 supercharges were analyzed and
classified in an earlier work by Gaiotto and Witten. We extend that
analysis to the case with fewer supersymmetries, concentrating mainly
on the case preserving one quarter. We develop tools necessary to
explicitly construct boundary conditions which can be viewed as taking
the zero slope limit of a system of D3 branes intersecting and ending
on a collection of NS5 and D5 branes oriented to preserve the
appropriate number of supersymmetries.  We analyze how these boundary
conditions constrain the bulk degrees of freedom and enumerate the
unconstrained degrees of freedom from the boundary/defect field theory
point of view.  The key ingredients used in the analysis are a
generalized version of Nahm's equations and the explicit
boundary/interface conditions for the NS5-like and D5-like impurities
and boundaries, which we construct and describe in detail.  Some bulk
degrees of freedom suggested by the naive brane diagram considerations
are lifted.

\end{abstract}
\vspace{0.5in}

\end{titlepage}
\renewcommand{\baselinestretch}{1.05}  

\tableofcontents 

\section{Introduction}

Impurities, defects, and boundaries are important objects in the study
of field theories.  The dynamics of the field theory itself can
generate defects nonperturbatively, and the existence or nonexistence
of certain defect solutions can serve as a probe for the phase
structure of the theory.  It can also be interesting to study boundary
conditions abstractly, in terms of conditions imposed on the fields.
In the presence of boundaries and impurities, one often encounters
edge effects and localized degrees of freedom which can
give rise to interesting physics.

The generic study of defects and boundaries is an enormous subject,
which touches on the physics of essentially every field theory.  To
sharpen this study, it is useful to restrict to supersymmetric field
theories and boundaries which preserve some fraction of the bulk
supersymmetry.  For example, by studying the supersymmetric boundaries
of two dimensional superconformal field theory, one uncovers the
existence of D-branes and other worldsheet boundary states.

Another natural class of systems one can consider is the set of
boundaries preserving half of the supersymmetries of the maximally
supersymmetric Yang-Mills (SYM) theory with gauge group $G$ in 3+1
dimensions.  These boundaries were studied extensively by Gaiotto and
Witten (GW) \cite{Gaiotto:2008sa,Gaiotto:2008ak}.  Their treatment
begins with a study of the BPS field configurations on a half-space
with translation invariance broken in one direction; the relevant
Bogomolny-type equations are the Nahm equations \cite{Nahm:1979yw}.
Using intuition from string theory realizations \cite{Hanany:1996ie},
GW formulated a low-energy classification of the boundary conditions
in terms of a triple
$(\rho,H,\mathfrak{B})$\cite{Gaiotto:2008sa,Gaiotto:2008ak}.  Here
$\rho$ is an embedding of $\mathfrak{sl}(2)$ into $\mathfrak{g}$ (the
Lie algebra of $G$) representing the Nahm pole, gauge group $G$ is
broken to a subgroup $H$ of the commutant of $\rho$ at the boundary,
and $\mathfrak{B}$ is a 3$d$ $\mathcal{N}=4$ boundary theory with
global symmetry $H$.  Moreover, they gave a recipe for relating the
action of S-duality on a given 1/2 BPS boundary condition to the
action of mirror symmetry on a related 3$d$ theory, which they
constructed by coupling a given boundary condition to a 3$d$ theory
called $T[SU(N)]$ \cite{Gaiotto:2008ak}.

Our work initially stemmed out of a rather innocent inquiry: how does
the structure of boundaries and their classification generalize if 
they preserve less than half of the original supersymmetry?
Generally, we expect richer physics when the amount of supersymmetry
is reduced. Would we discover an object, generalizing the NS5-branes
or the D5-branes, on which a D3 can end and preserve less than half of
the supersymmetries? Is there a simple generalization to the triple
$(\rho,H,\mathfrak{B})$ that one can formulate to generalize the
GW classification?

We have encountered a number of subtleties in answering these
questions.  Below, we will review the formalism of GW, and show that
attempts to construct an elementary 1/4 BPS object analogous to an NS5
brane or a D5 brane do not lead to anything new. We can construct
boundaries and defects preserving less than half of the
supersymmetries by including 5-branes oriented in such a way that each
of the 5-branes break different components of the supersymmetries. We
can attempt to classify different configurations of stacks of NS5 and
D5 branes arranged to preserve some fraction of supersymmetries. Such
a classification, however, will be more complicated than the results
in the 1/2 BPS case. The main reason is the fact that in the 1/2 BPS
analysis, changing the positions of the 5-branes does not change the
low energy physics. This allowed GW to define a canonical ordering of
the 5-branes, and this feature was used to dramatically reduce the set
of possible boundaries. In the cases with less than 8 supersymmetries, 
however, the changes in the ordering of the 5-branes
can, in general, change the low energy physics.  This does not
necessarily imply that a classification for these boundaries is impossible, but it
does imply that such a classification will be far more intricate than
when 8 supersymmetries are preserved.

Even if the classification scheme for 1/2 BPS boundaries does not
generalize easily to the case of 1/4 BPS, it is interesting to explore
the rich dynamics of boundaries with less supersymmetries.  In this paper,
we will construct several examples of boundaries preserving 1/4 of the
supersymmetries of ${\cal N}=4$ supersymmetric Yang-Mills theory by
combining a collection of NS5 and D5-branes. These structures possess
localized as well as delocalized degrees of freedom, which in a loose
sense could be considered the moduli space of the boundary.  Through
explicit analysis of the Lagrangian of this system in the classical
limit, we map out these deformations, which have a natural K\"{a}hler
structure as a result of having 4 supercharges.

These boundary conditions can then be used as building blocks for
engineering ${\cal N}=2$ field theories in 2+1 dimensions, for
instance, by 
considering the 4$d$ $\mathcal N = 4$ theory on an interval with boundary conditions
at both ends.  On the
interval, there are no issues with the delocalized modes and the notion
of a moduli space is well defined. Strictly speaking, our analysis is
limited to classical dynamics. However, for ${\cal N}=4$ theory in 3+1
dimensions on an interval, we can also analyze the moduli space of the
S-dual configuration in the classical limit.  Depending on the pattern
of gauge symmetry breaking, which can be mapped nontrivially by
duality, we can gain access to some {\it quantum corrected} features of the
model by doing {\it classical} computations in the S-dual.

In addition, there are non-renormalization theorems for systems with
${\cal N}=2$ supersymmetries in 2+1 dimensions
\cite{Aharony:1997bx}. We expect superpotentials to be protected from
corrections at the perturbative level.  
However, superpotential terms can be generated dynamically
at the non-perturbative level through
instantons as was shown in \cite{Affleck:1982as}. On the other hand,
we expect the instanton corrections to be absent in branches where all of the
gauge symmetries are spontaneously broken. Generally, the information
contained in the $D$-terms (which encodes the metric data of the
moduli space through the K\"{a}hler potential) are subjected to
corrections, but there are situations where even the K\"{a}hler
potential is protected from quantum corrections
\cite{Aganagic:2001uw,Intriligator:2013lca}.

One interesting application of our program is to map out (as much as possible) the fully
quantum corrected moduli space of systems using the data available
from the combination of S-duality and the assortment of
non-renormalization theorems.\footnote{As was emphasized in
  \cite{Hanany:1996ie}, S-duality in this context is closely related
  to mirror symmetry. The advantage of embedding the 2+1 theories into
  boundary/defect systems in 3+1 dimensions is the fact that the
  Lagrangian of the system and the S-dual can be read off
  systematically. In other words, we have direct access to a
  microscopic description of the mirror pair.}  In this paper, we will
begin this program by performing a detailed analysis of boundaries
which preserve 1/4 of the supersymmetry.  The application to ${\cal
  N}=2$ theories in 2+1 dimensions by placing the $\mathcal N =4$
theory on a finite interval with boundary conditions on both ends of
the interval 
will appear in a separate paper \cite{ToAppear}. We
will assess the power of this approach and attempt to extract some
general lessons by working out numerous concrete examples with varying
degrees of complexity.  Although we mostly study the case where $1/4$
of the supersymmetries are preserved, some of our methods extend
easily to the cases with $3/8$ or $1/8$ of the supersymmetries being
preserved.

\section{Basic Construction of Boundaries in 4$d$ ${\cal N}=4$ SYM}\label{sec2}

In this section, we describe the construction of supersymmetric
boundary conditions in ${\cal N}=4$ SYM in 3+1 dimensions.  We will
begin by reviewing the construction of boundaries preserving half of
the supersymmetries of the bulk theory (section \ref{sec21}),
following the treatment of \cite{Gaiotto:2008sa}, whose conventions we
adopt. We then describe the generalization to the case preserving a
quarter of the supersymmetries (section \ref{sec22}).  We also comment
on more general composite boundary conditions (section \ref{sec23})
and on the classification of 1/4 BPS boundary conditions (section
\ref{sec24}).

Before going into the details, we will make a few general remarks about
our analysis:

\begin{enumerate}

\item Throughout this paper, we distinguish between ``bulk''
  properties which refer to the 3+1-dimensional theory and
  ``boundary'' or ``interface'' properties which refer to
  2+1-dimensional defects embedded in the 3+1-dimensional bulk.

\item In this section, we assume that the boundaries themselves do not
  carry any dynamical degrees of freedom of their own. As we will see,
  this assumption often does not hold in practical applications, and
  we will relax this assumption later in section \ref{sec:localized}.

\item 
We will not make use of the conformal symmetry (or its fermionic
  counterpart) of the the $\mathcal{N}=4$ SYM theory 
  in the discussion of the boundary conditions.  Our
  boundary conditions in general break conformal symmetry, which is
  recovered only in the IR limit.  This is in contrast with the
  discussion of boundary conformal field theories, which relies
  crucially on conformal symmetry.

\item While our discussion in this paper deals with 4$d$
  $\mathcal{N}=4$ SYM, the approach we follow is rather
  general.  It would be interesting to apply the same methods to
  supersymmetric boundary conditions for other theories in various
  dimensions; for example, 1/2 BPS boundary conditions of 4$d$
  $\mathcal{N}=2$ theories \cite{Gaiotto:2009we} should also give rise
  to 3$d$ $\mathcal{N}=2$ theories associated with 3-manifolds
  \cite{Terashima:2011qi,Dimofte:2011jd}.
\end{enumerate}

Let us first recall the basics of 4$d$ $\mathcal{N}=4$ theory, and set
up some notation.  The ${\cal N}=4$ SYM in 3+1 dimensions can be
efficiently obtained from ${\cal N}=1$ SYM in 9+1 dimensions.  Our
convention is that the 10$d$ metric $g_{IJ}$ ($I, J=0, \ldots, 9$) has
signature $(-,+,\cdots +)$ and the Clifford-Dirac algebra is
$\{\Gamma_I,\Gamma_J\} = 2 g_{IJ}$. The 10$d$ chirality operator is
$\bar \Gamma \equiv \Gamma_0 \Gamma_1 \cdots \Gamma_9$.  The 10$d$ SYM
action is
\be I = {1 \over g_{\rm YM4}^2} \int d^{10} x \, \tr \left({1 \over 2} F_{IJ} F^{IJ} - i \bar \Psi \Gamma^I D_I \Psi \right) ,
\ee 
where $A_I$ is a gauge field whose field strength we defined to be
$F_{IJ} \equiv \partial_I A_J - \partial_J A_I {-} [A_I,A_J]$, and
$\Psi$ is a Majorana-Weyl fermion satisfying $\Psi = \bar \Gamma
\Psi$.  We use the convention that the bosonic fields are
anti-hermitian, so the field strength $F_{IJ}$ is defined without a
factor of $i$ in front of the commutator.  The fields transform under
supersymmetry as
\beq \delta A_I &=& i \bar \varepsilon \, \Gamma_I \Psi  \ , \\ 
\delta \Psi & = & {1 \over 2} \Gamma^{IJ} F_{IJ} \varepsilon  \ .
\eeq
The supersymmetry generator $\varepsilon$ is also Majorana-Weyl and
satisfies $\varepsilon = \bar{\Gamma} \varepsilon$.  The supercurrent
associated with these transformations is
\be J^I = {1 \over 2} \tr\, \Gamma^{JK} F_{JK} \Gamma^I \Psi. \label{supercurrent10d}\ee

The 3+1-dimensional ${\cal N} =4$ SYM theory can now be obtained by
dimensional reduction of the 10$d$ theory, with the ansatz that the
fields depend only on $x^0, \cdots , x^3$.  It is conventional to
re-label the components of the gauge fields as $A_{3+i} = \Phi_i$ for
$i=1\ldots 6$; these transform as scalars under the $SO(3,1)$ Lorentz
symmetry.

We will take the boundary to be flat, extended in $x_0$, $x_1$, and
$x_2$ coordinates, and localized at some fixed $y\equiv x_3$
coordinate.  This breaks translation invariance in the $y$
direction.

The condition that the boundary preserves supersymmetry is that the
flux of the supercurrent through the boundary vanishes, or in other
words
\be 
J^3\big|= \tr \, \bar \varepsilon\, \Gamma^{IJ} F_{IJ} \Gamma_3 \Psi \big|=0 \ , \label{boundary_half} \ee
where the symbol $\big|$ means that the equation holds at the
boundary.  Since the 3+1 dimensional Lorentz is broken by the presence
of the boundary, we cannot impose \eqref{boundary_half} for all the
sixteen supercharges; to preserve half (a quarter) of the
supersymmetry, we impose \eqref{boundary_half} for 8 (4) of the 16
components of $\varepsilon$.

The boundary condition eliminates half of the degrees of the freedom
at the boundary.  For bosons, we could choose Dirichlet or Neumann, or
more generally mixed boundary conditions.  For the fermions, the
boundary condition sets half of the components of $\Psi$ to some fixed
values (using the Dirac equation, one sees that Dirichlet boundary
conditions for half the fermionic degrees of freedom imposes Neumann
boundary conditions for the other half.)  We then need to see which of
these boundary conditions are consistent with \eqref{boundary_half}.

\begin{figure}
\centerline{\includegraphics[scale=0.8]{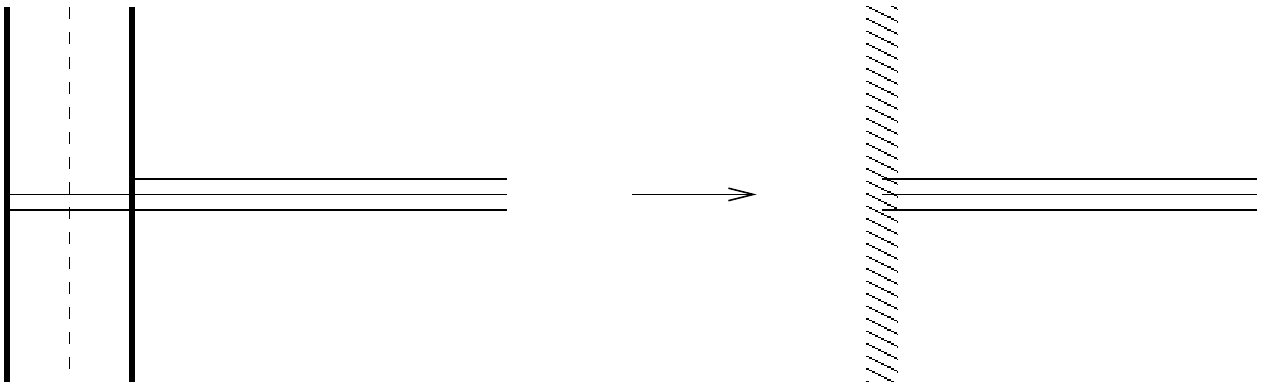}}
\caption{By collapsing the boundary conditions we can construct more
  complicated boundary conditions.  For this analysis it is necessary
  to keep track of the bulk degrees of freedom between the two
  defects, which is constrained by \eqref{bulk_half}.}
\label{defect_collide}
\end{figure}

While we are mainly interested in boundary conditions in this paper,
it is also useful to consider the BPS condition in the bulk.  
The bulk equations are especially important if we wish to construct complicated boundary
conditions by starting with several boundary/junction conditions
separated by the bulk theory on the interval, and then take the limit
where the defects collide, as illustrated in figure
\ref{defect_collide}. This limit is equivalent to the taking the IR
limit of the composite boundary.  These composite boundary conditions
contain the bulk degrees of freedom on the interval, which are crucial
for the analysis of the moduli space of the boundary condition.  The
same remark applies to the analysis of the 3$d$ field theory discussed
in our second paper \cite{ToAppear}.

The bulk BPS equation is given by
\be \, \delta \bar \Psi= \bar \varepsilon\, \Gamma^{IJ} F_{IJ} = 0  \ ,\label{bulk_half}\ee
where the right hand side refers to $\varepsilon$ but not to $\Psi$. 

Clearly, the bulk BPS condition \eqref{bulk_half} is stronger than the
boundary BPS condition for fermions \eqref{boundary_half}, in that the
former restricted to the boundary implies the latter.  This is not
necessarily the case for scalars.  In some cases (for example for the
Neumann boundary condition for the scalar field) the actual boundary
condition is contained in the supersymmetry condition
\eqref{boundary_half}; in other cases (for example for the Dirichlet
boundary condition) we have to impose the boundary condition
separately.  We will see concrete examples momentarily.

\subsection{Boundaries Preserving Half of the Supersymmetries}\label{sec21}

Let us now consider the construction of boundaries preserving half of
the 16 supersymmetries.

To proceed further in analyzing these equations, it is useful to
parametrize $\varepsilon$ and $\Psi$ to maximally reflect the
symmetries of the problem. Here, the key feature is that the symmetry
of the bulk boundary system is equivalent to the symmetry of ${\cal
  N}=4$ supersymmetry in 2+1 dimensions, whose $R$ symmetry is $SO(4)
\simeq SO(3)_X \times SO(3)_Y$. As the notation suggests, we can
identify these $SO(3)$'s as acting on two sets of three transverse
scalars which we label
\be (X_1,X_2,X_3) \equiv (\Phi_1, \Phi_2, \Phi_3)\ , \qquad (Y_1, Y_2, Y_3) \equiv (\Phi_4, \Phi_5, \Phi_6) \ . \ee
Let us define
\begin{align}
\begin{split} B_0 &\equiv \Gamma_{456789} \ ,  \cr
B_1 & \equiv  \Gamma_{3456}  \ , \cr
B_2 & \equiv  \Gamma_{3789}  \ .
\end{split}
\end{align}
These form an $SL(2, \mathbb{R})$ algebra.
In the representation of appendix \ref{appendixA}, these
can be represented by 
\beq 
B_0 &=& \left(\begin{array}{cc} 0 & 1 \\ -1 & 0 \end{array}\right) \ , \cr
B_1 &=& \left(\begin{array}{cc} 0 & 1 \\ 1 & 0 \end{array}\right) \ , \label{SL2Rrep} \\
B_2 &=& \left(\begin{array}{cc} 1 & 0 \\ 0 & -1 \end{array}\right) \ . \nonumber
\eeq
Moreover these matrices commute with the $SO(3)_X\times SO(3)_Y$
symmetry, as well as the (2+1)-dimensional Lorentz symmetry $SO(2,1)$.
The supercharges, which are in {\bf 16} of $SO(1,9)$, can then be
represented as $V_8 \otimes V_2$.  Here $V_2$ is the two dimensional
space on which the representation (\ref{SL2Rrep}) of $SL(2,
\mathbb{R})$ acts, and $V_8$ is the $({\bf 2},{\bf 2},{\bf 2})$ of
$SO(1,2) \times SO(3)_X \times SO(3)_Y$. Some of these details will be
reviewed in the appendix \ref{appendixA}.

Armed with this amount of structure, we parametrize
\be \varepsilon = v \otimes \varepsilon_0 \in V_8\otimes V_2 \ ,
\label{varepAnsatz} \ee
and similarly
\be \Gamma_3 \Psi = \Psi' \otimes \vartheta \in V_8\otimes V_2  \ .
\label{PsiAnsatz}
\ee
Here $\varepsilon_0$ and $\vartheta$ are {\it specific}, {\it fixed} 2
component vectors, and $v$ and $\Psi'$ are {\it arbitrary} eight
component vectors.  The choice of $\varepsilon_0$ specifies which 8 out
of 16 components of the supersymmetry generator $\varepsilon$ are preserved. The choice of $\vartheta$, on
the other hand, specifies the components of $\Psi$ which are allowed
to take arbitrary values at the boundary. Components orthogonal to
$\Psi' \otimes \vartheta$, on the other hand, must vanish at the
boundary.

We can now substitute \eqref{varepAnsatz}, \eqref{PsiAnsatz} back into
the boundary BPS condition (\ref{boundary_half}) and the bulk BPS
condition (\ref{bulk_half}) For the boundary condition
(\ref{boundary_half}), we arrive at the following set of conditions
(see appendix \ref{appendixA} for more details)
\beq
 \bar{\varepsilon}_0 \left( F_{\mu \nu} -\epsilon_{\mu \nu \lambda} F^{3\lambda} B_0\right)\, \vartheta \, \big| \label{halfbps1} &=& 0 \ ,\\
 D_{\mu} X_a \, \left(\bar{\varepsilon}_0 B_2 \vartheta\right)\label{halfbps2}\big| &=& 0 \ , \\
D_{\mu} Y_m \, \left(\bar{\varepsilon}_0 B_1 \vartheta\right)\label{halfbps3}\big| &=& 0 \ , \\
 \lbrack X_a,Y_m \rbrack \, \left(\bar{\varepsilon}_0 B_0 \vartheta\right)\big| &=& 0 \label{halfbps4} \ , \\
 \bar{\varepsilon}_0\left( \lbrack X_b, X_c \rbrack -\epsilon_{abc} D_3 X_a B_1 \right) \vartheta \,\big| &=& 0 \label{halfbps5} \ ,\\
 \bar{\varepsilon}_0\left( \lbrack Y_m, Y_n \rbrack -\epsilon_{mnp} D_3 Y_p B_2 \right) \vartheta \,\big| &=& 0\label{halfbps6} \ .
\eeq

Since we have the choice of the $\varepsilon_0$ and $\vartheta$ up to
their overall normalization, we could in principle have a 2-parameter
family of boundary conditions.  It turns that only one parameter
survives.  Two special points in the 1-parameter space corresponds to
D5-like\footnote{The term ``D5-like'' first appeared in
  \cite{Gaiotto:2008sa} and refers to the fact that these boundary
  conditions arises naturally in the field theory ($\alpha'
  \rightarrow 0$) limit of D3-branes ending on D5-branes. Since these
  structures exist in the zero slope limit, however, the concept does
  not rely on string theory. Nonetheless, it is convenient to
  associate the string theory origin of these constructions as they
  are more familiar to many.}  and NS5-like boundary conditions, which
we discuss in turn.

\subsubsection{D5-like Boundary}

Consider setting
\beq
\bar{\varepsilon}_0 &=& \frac{1}{\sqrt{2}} (1 \;\;\;  1) \ , \\
\vartheta &=& \frac{1}{\sqrt{2}}\left( \begin{array}{c}
 1\\
1
\end{array}\right) \ .
\eeq
%
Then
\beq
\bar{\varepsilon}_0 B_0\vartheta = \bar{\varepsilon}_0 B_2\vartheta = 0  \ .
\eeq
Then, the bulk BPS equation (\ref{bulk_half}) reads
\beq
F_{\mu \nu} &=& 0 \ , \label{bulk1}\\
F_{3\lambda} &=& 0 \ , \label{bulk2}\\
D_{\mu} X_a &=& 0 \ ,\label{bulk3}\\
D_{\mu} Y_m &=& 0 \ ,\label{bulk4}\\
\lbrack X_a,Y_m \rbrack &=& 0 \ ,\label{bulk5}\\
 \lbrack X_b, X_c \rbrack - \epsilon_{abc} D_3 X_a &=& 0 \ ,\label{bulk6}\\
\lbrack Y_m, Y_n\rbrack  &=& 0 \ ,\label{bulk7}\\
D_{3} Y_m &=& 0 \ . \label{bulk8}
\eeq
The most important part of this equation is the Nahm equation for the $X_a$:
\beq
D_3 X_a  - \frac{1}{2} \epsilon_{abc}\lbrack X_b, X_c \rbrack  &=& 0\label{Nahmeq}  \ .
\eeq
The appearance of Nahm's equations in D-brane physics was originally
noted in \cite{Diaconescu:1996rk}.  This equation will play crucial
roles in our subsequent analysis.

The boundary BPS equations (\ref{boundary_half}) are a slightly weaker
subset of these equations:
\beq
F_{\mu \nu} \big|&=& 0  \ ,\label{d51}\\
D_{\mu} Y_m \big| &=& 0 \ ,\label{d52}\\
\left( \lbrack  X_b, X_c \rbrack - \epsilon_{abc} D_3 X_a\right) \big| &=& 0 \ ,  \label{d53} \\
\lbrack Y_m, Y_n\rbrack \big| &=& 0  \ .\label{d54}
\eeq
The boundary conditions consistent with these are
\beq
\textrm{D5-like}: \quad F_{\mu \nu}\big|
=\left( D_3 X_{a} - \frac{1}{2}\epsilon_{abc} \lbrack X_b, X_c \rbrack  \right) \Big|
= Y_{m}\big|=0 \ .
\label{D5like}
\eeq

When $[X_a, X_a]\big|=0$, this means that we have Dirichlet boundary
condition for $(A_{\mu}, X_a)$ and Neumann boundary condition for
$(A_3, Y_m)$; in 3$d$ $\mathcal{N}=4$ language each of these makes up a
3$d$ $\mathcal{N}=4$ hypermultiplet.  Note that the boundary condition
for gauge fields should be imposed in a gauge invariant manner, and
hence we have for example $F_{\mu\nu}\big|=0$, and not
$A_{\mu}\big|=0$.

Since $X_a$ ($Y_m$) obey Neumann (Dirichlet) conditions, the boundary
condition \eqref{D5like} can be interpreted as a boundary condition
for the D5-brane extended along the $012456$ directions.  This is the
reason why the boundary condition \eqref{D5like} was called
``D5-like.''

If the D3 brane extends on both sides of the D5, 
there
will be some additional localized degrees of freedom. The boundary
condition \eqref{D5like} can be recovered as a limit of this more
  general junction condition which we will review in section
  \ref{sec:localized}.

When the commutator term $[X_b, X_c]$ in \eqref{D5like} is nonzero, a
new structure emerges.  After setting $A_3=0$ by a choice of gauge,
the equation \eqref{D5like} becomes
\beq
 \frac{dX_a}{dy}  - \frac{1}{2} \epsilon_{abc} \lbrack X_b, X_c \rbrack  &=& 0\label{Nahmeq-A0}  \ ,
\eeq
which has a singular solution of the form
\beq
X_a \sim \frac{\rho(t_a)}{y}+\mathcal{O}(y^0) \ ,
\label{Xsingular}
\eeq
where we have chosen the boundary to be at $y=0$, and $t^a$ are three
matrices satisfying the $\mathfrak{sl}(2)$ commutation relation $[t_a,
  t_b]=-\epsilon_{abc} t_c$, and $\rho$ is an embedding of
$\mathfrak{sl}(2)$ into the gauge group $\mathfrak{g}$ (this is the
same $\rho$ as appears in introduction).  This means we can impose
\eqref{Xsingular} instead of the standard Neumann boundary condition.
This $1/y$ singularity is often called a ``Nahm pole'' in the
literature.

While the singular boundary condition with a pole might unfamiliar to
some of the readers, the singularity \eqref{Xsingular} naturally
describes the funnel of D3-branes ending on D5-branes
\cite{Diaconescu:1998ua,Constable:1999ac}.  It is also the case the
the singular boundary conditions of this kind are required in order
for the S-duality to NS5-like boundaries to work in detail. Let us now
turn to that example.

\subsubsection{NS5-like Boundary}

Another choice for specializing (\ref{halfbps1})--(\ref{halfbps6}) is to  set\footnote{
In appendix \ref{appendixA} NS5-like boundary condition 
corresponds to 
\begin{eqnarray*}
\bar{\varepsilon}_0 &=& (1 \;\;\;  0)  \ , \\
\vartheta &=& \left( \begin{array}{c}
0\\
1
\end{array}\right) \ .
\end{eqnarray*}
This represents the NS5-brane along the 012456-directions. When we
rotate the NS5 to 012789-directions, we obtain
\eqref{varepNS}--\eqref{varthetaNS}.}
\beq
\bar{\varepsilon}_0 &=& \frac{1}{\sqrt{2}} (1 \;\;\;  1) \label{varepNS}  \ , \\
\vartheta &=& \frac{1}{\sqrt{2}}\left( \begin{array}{c}
- 1\\
1
\end{array}\right) \label{varthetaNS} \ ,
\eeq
%
so that 
\beq
\bar{\varepsilon}_0 \vartheta = \bar{\varepsilon}_0 B_1\vartheta = 0  \ .
\eeq
Because the choice of $\varepsilon_0$ is the same as before, this
system should preserve the same set of supersymmetries. The bulk
equation (\ref{bulk_half}) is therefore the same as before. However,
because $\vartheta$ is different, the boundary condition changes. In
fact, the boundary BPS condition (\ref{boundary_half}) will now read
\beq
F^{3 \lambda} \big| & = & 0  \ ,\\
D_\mu X_a \big|& = & 0  \ ,\\
\lbrack X_a,Y_m \rbrack \big| & = & 0  \ ,\\
D_3 Y_m \big| &=& 0 
\ .
\eeq
The boundary condition consistent with these BPS conditions are 
\beq
\textrm{NS5-like}: F_{3 \mu}\big|=X_{a}\big|=D_3 Y_{m}\big|=0 \ ,
\label{NS5}
\eeq
which can be represented as an NS5-brane extended along the
$012789$-directions.

The NS5-like boundary condition, when we exchange $456$-directions
with $789$-directions, might appear to be somewhat similar to the
D5-like boundary conditions, in that one of $X, Y$ obeys Dirichlet
boundary conditions, and the other Neumann.  However, note that we do
not have the commutator $[Y_m, Y_n]$ this time for the Neumann
boundary condition, and hence the singular solution \eqref{Xsingular}
is not allowed for $Y_m$. The generalization to the case where some
number of D3 branes ends on both sides of the NS5 brane will be
discussed in section \ref{sec:localized}.

\subsubsection{$(p,q)$ 5-brane-like Boundary}

One other possibility considered by GW in \cite{Gaiotto:2008sa} can be
interpreted as ``$(p,q)$ 5-brane like.''  Somewhat interestingly,
$p/q$ does not necessarily need to be a rational number, in which case
$p/q$ can still be interpreted as the $\theta$-angle of the
$\mathcal{N}=4$ SYM.  One can also consider various rotations of
5-branes in the 456789 coordinates.  These appear to correspond, in
some sense, to the exhaustive set of elementary boundaries with no
explicit boundary degrees of freedom. The D5, the NS5, and the $(p,q)$
also impose conditions on the set of allowed gauge transformations at
the interface, as we will illustrate in more detail below.  More
sophisticated boundaries are then constructed by introducing multiple
sets of NS5 and D5 branes, as was described in
\cite{Gaiotto:2008sa,Gaiotto:2008ak}.

\subsection{Boundaries Preserving 1/4 of the Supersymmetries}\label{sec22}

We have now accumulated enough tools to study the properties of
boundaries breaking all but one quarter of supersymmetries of ${\cal
  N}=4$ SYM in 3+1 dimensions.  The problem we want to solve is to
repeat the analysis of the boundary supersymmetry condition
(\ref{boundary_half}) but with the requirement that only four of the
components of $\varepsilon$ need to be nonzero.

An efficient way to do the computation is to insert a projection
operator $\hat{P}$ which annihilates half the components of
$\varepsilon$, and study the resulting supersymmetry condition:
\be \tr \, \bar \varepsilon \hat P \Gamma^{IJ} F_{IJ} \Gamma_3 \Psi=0 \label{boundary_q} \ . \ee
The advantage of working with the projection operators is that we can
avoid explicitly writing the supersymmetry generators and work instead
with the algebra of Dirac matrices.  Under the decomposition
$\varepsilon = v \otimes \epsilon_0$, we see that $\hat P$ must act on
$v$ since $\epsilon_0$ has already been fixed to pick out 8
independent components. This projection operator should further break
the $SO(3)_X \times SO(3)_Y$ R-symmetry. This is to be expected since
the amount of supersymmetry left unbroken in the 2+1 dimensional sense
is that of ${\cal N}=2$ supersymmetry, whose R-symmetry group,
$SO(2)$, is much smaller. We will however consider mostly the case
where there is an accidental global symmetry $SO(2) \times SO(2)$
which corresponds to orienting the NS5 and D5 branes at right angles.

Let us pause to make a comment about the notation.  When discussing
constructions with one quarter of supersymmetries, we use the notation
\beq (X_4, X_5, \ldots, X_9)\equiv (\Phi_1, \Phi_2,
\ldots, \Phi_6) \ .  \eeq
When discussing constructions preserving one half of the
supersymmetries, we will continue to use $X_a$ and $Y_a$ with the
index $a$ ranging from 1 to 3.  At later stages, we will also combine
some of these components into complex combinations. Care has been made
to make sure that this issue of notation is clear from context.

Just as we had some freedom in choosing the $\varepsilon_0$ and
$\vartheta$, we have the freedom to chose $\hat P$ which in essence
is the choice of which  components of supersymmetry to preserve. One
natural candidate we will consider is to take
\be \hat P = 1 - \Gamma_{4578} \label{proj} \ee
which projects out half of the components of $v$.\footnote{It is also
  possible to consider other projection operators which would project
  $v$ to other components, with possibly 2, 4, or 6 independent
  components (c.f.\ appendix \ref{appendixA}).}  Note that this
projection is compatible with an $SO(2) \times SO(2)$ global symmetry
corresponding to rotations in the 45 and 78 planes.

For this choice of $\hat P$, the boundary condition (\ref{boundary_q}) is given by
\beq
D_3 X_4 \left(\bar{\varepsilon}_0 B_1 \vartheta\right)-\left[ X_5, X_6  \left(\bar{\varepsilon}_0 \vartheta\right)-X_9 \left(\bar{\varepsilon}_0 B_0\vartheta\right)\right] &=& 0\label{d3x4} \ , \\
D_3 X_5 \left(\bar{\varepsilon}_0 B_1 \vartheta\right)+\left[ X_4, X_6  \left(\bar{\varepsilon}_0 \vartheta\right)-X_9 \left(\bar{\varepsilon}_0 B_0\vartheta\right)\right] &=& 0\label{d3x5}\ , \\
D_3 X_7 \left(\bar{\varepsilon}_0 B_2 \vartheta\right)-\left[ X_8, X_9  \left(\bar{\varepsilon}_0 \vartheta\right)-X_6 \left(\bar{\varepsilon}_0 B_0\vartheta\right)\right] &=& 0\label{d3x7}\ , \\
D_3 X_8 \left(\bar{\varepsilon}_0 B_2 \vartheta\right)+\left[ X_7, X_9  \left(\bar{\varepsilon}_0 \vartheta\right)-X_6 \left(\bar{\varepsilon}_0 B_0\vartheta\right)\right] &=& 0\label{d3x8}\\
\bar{\varepsilon}_0 B_0 \vartheta\left(\left[X_4,X_7\right]-\left[X_5,X_8\right]\right) &=& 0\ ,  \\
\bar{\varepsilon}_0 B_0 \vartheta\left(\left[X_4,X_8\right]+\left[X_5,X_7\right] \right)&=& 0\ , \\
\bar{\varepsilon}_0 B_0 \vartheta\left(\left[X_6,X_9\right]\right) &=& 0\ ,  \\
D_3 X_6 \left(\bar{\varepsilon}_0 B_1 \vartheta\right)-\left[ X_4, X_5\right] \left(\bar{\varepsilon}_0  \vartheta\right)+D_3 X_9 \left(\bar{\varepsilon}_0 B_2 \vartheta\right)-\left[ X_7, X_8\right] \left(\bar{\varepsilon}_0  \vartheta\right)&=&0\ ,
\eeq
along with equations with Lorentz indices which do not play an
important role for the Lorentz-invariant solutions we are interested
in.

Now, let us choose
\be \bar{\varepsilon}_0 = \frac{1}{\sqrt{2}} (1 \;\;\;  1) 
\ee
as we did in the previous section. We can then read off the bulk equations
\beq
D_3 X_4 -\lbrack X_5, X_6 \rbrack &=& 0 \ , \label{bulkf}\\
D_3 X_5 -\lbrack X_6, X_4 \rbrack &=& 0 \ ,\\
D_3 X_6 -\lbrack X_4, X_5 \rbrack -\lbrack X_7, X_8 \rbrack&=& 0 \ ,\\
D_3 X_7 -\lbrack X_8, X_6 \rbrack &=& 0 \ ,\\
D_3 X_8 -\lbrack X_6, X_7 \rbrack &=& 0 \ ,\\
\left[X_4,X_7\right]-\left[X_5,X_8\right] &=& 0 \ , \\
\left[X_4,X_8\right]+\left[X_5,X_7\right] &=& 0 \ ,\\
D_3 X_9 &=& 0 \ ,\\
\lbrack X_9, X_i \rbrack &=& 0  \ .
\label{bulkl}
\eeq
The first five equations generalize the Nahm equations.
Indeed, if we set $X_{7,8,9}=0$ we have the Nahm equation for
$X_{4,5,6}$:
\beq
D_3 X_4 -\lbrack X_5, X_6 \rbrack &=& 0 \ , \\
D_3 X_5 -\lbrack X_6, X_4 \rbrack &=& 0 \ ,\\
D_3 X_6 -\lbrack X_4, X_5 \rbrack &=& 0 \ . 
 \label{Nahm}
\eeq
Similarly, we have the Nahm equation for $X_{6,7,8}$ if we set
$X_{4,5,9}=0$.  In this sense the first five equations of the bulk
equation \eqref{bulkl} can be thought of as a composite of two Nahm
equations in $X_{4,5,6}$ and $X_{6,7,8}$, the two being coupled
through the common scalar $X_6$.
While this work was in progress, a closely related system of equations appeared in the context of the
Hitchin equation in \cite{Xie:2013gma} (see also
\cite{Bonelli:2013pva}).  

Just as Nahm's equations can be viewed as a dimensional reduction of
the self-dual Yang-Mills equations in 4$d$, the generalized Nahm's
equations can be understood as a particular dimensional reduction of
the Donaldson-Uhlenbeck-Yau equations
\cite{Donaldson:1985zz,UhlenbeckYau}.  We will elaborate more on these
equations and their solutions when we explore the moduli space of
non-Abelian systems in a follow up paper \cite{ToAppear}.

Let us now examine the consequence of choosing
\beq
\vartheta &=& \frac{1}{\sqrt{2}}\left( \begin{array}{c}
 1\\
1
\end{array}\right) \ .
\eeq
This gives rise to the boundary constraint
\beq
D_3 X_4 -\lbrack X_5, X_6 \rbrack &=& 0 \ , \label{bc1f}\\
D_3 X_5 -\lbrack X_6, X_4 \rbrack &=& 0 \ ,\\
D_3 X_6 -\lbrack X_4, X_5 \rbrack -\lbrack X_7, X_8 \rbrack&=& 0 \ ,\\
\lbrack X_9, X_7 \rbrack =\lbrack X_9, X_8 \rbrack &=& 0 \ . \label{bc1l}
\eeq
Alternatively, setting
\beq
\vartheta &=& \frac{1}{\sqrt{2}}\left( \begin{array}{c}
- 1\\
1
\end{array}\right)
\eeq
gives rise to the boundary constraints
\beq
D_3 X_7 -\lbrack X_8, X_6 \rbrack &=& 0 \ , \label{bc2f}\\
D_3 X_8 -\lbrack X_6, X_7 \rbrack &=& 0 \ ,\\
\left[X_4,X_7\right]-\left[X_5,X_8\right] &=& 0 \ , \\
\left[X_4,X_8\right]+\left[X_5,X_7\right] &=& 0 \ ,\\
D_3 X_9 & = & 0 \ , \\
\lbrack X_9, X_4 \rbrack =\lbrack X_9, X_5 \rbrack =\left[X_9,X_6\right]&=& 0 \ .  \label{bc2l}
\eeq

These boundary constraints, as in earlier cases, are implied by the
bulk equations. We see that boundary conditions
(\ref{bc1f})--(\ref{bc1l}), combined with the bulk equations
(\ref{bulkf})--(\ref{bulkl}), support either D5-branes oriented along
012456 or NS5-branes oriented along 012459, which we will refer to as
NS5$^\ensuremath{\prime}$-branes. Similarly,
(\ref{bc2f})--(\ref{bc2l}) are compatible with an NS5-brane oriented
along 012789 and a D5$^\ensuremath{\prime}$-brane oriented along
012678. These are the 5-branes we expect to find when supersymmetry is
projected from 8 supercharges to 4 using the projection operator
(\ref{proj}). Perhaps not too surprisingly, we do not find any
candidate configuration corresponding say to a D7-brane oriented along
01245678. Such a configuration, in the presence of a D3 brane along
0123, will not preserve any supersymmetry.


\subsection{Composite Boundary Conditions}\label{sec23}

\begin{figure}
\centerline{\includegraphics[scale=0.8]{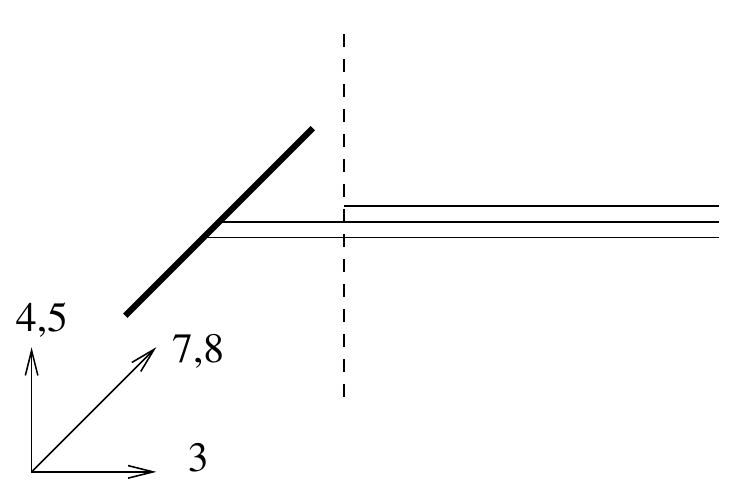}}
\caption{1/2 BPS boundary for a $U(3)$ theory preserving $U(2)$ gauge
  symmetry.
\label{fig.composite}}
\end{figure}

So far we have mostly considered boundary conditions which one might
consider as arising from the field theory limit of D3-branes ending on
a single object, whether it be a D5, an NS5, or a $(p,q)$ five brane.
As discussed in figure \ref{defect_collide}, a broader class of
boundary condition arises from considering a system with more than one
component, as we will discuss throughout the rest of this paper.  In
this subsection we  describe part of this compositeness in the
formulation of \cite{Gaiotto:2008sa}.  A simple example might be to
consider a D5 and an NS5 in combination, as is illustrated in figure
\ref{fig.composite}, where the bulk gauge group $G=U(3)$ is broken to
a subgroup $H=U(2)$.

In such a situation, the boundary conditions are imposed somewhat
differently between the unbroken subgroup $H$ and the rest of the
gauge group $G$. To illustrate this idea more precisely, let us
decompose the Lie algebra of $G$ as $\mathfrak{g}=\mathfrak{h}\oplus
\mathfrak{h}^{\perp}$, where $\mathfrak{h}$ is the Lie algebra of $H$
and $\mathfrak{h}^{\perp}$ is the complement.  For an adjoint-valued
field $X$, we could decompose the field as $X=X^{+}+X^{-}$, with
$X^{+}\in \mathfrak{h}$ and $X^{-}\in \mathfrak{h}^{\perp}$.

For the example illustrated in figure \ref{fig.composite}, $G$ is
given by $U(3)$, we can take $H= U(2)$, as residing in the $2\times 2$
upper left block of $G$.  Then $X^+$ is the $2\times 2$ block part of
the $3\times 3$ matrix $X$, while $X^-$ is the all the other
coefficients of the matrix.

We then impose two different types of boundary conditions for $+$ and
$-$.  Note that this is a generalization of the boundary conditions we
have considered previously.  In particular when $H$ is a trivial
subgroup, i.e., just an identity, and then $\mathfrak{h}=0,
\mathfrak{h}^{\perp}=\mathfrak{g}$.  and we have $\Phi^{-}=0, \Phi^{+}=\Phi$.

For example, let us choose the NS5-like boundary condition for $X^{+}$
and D5-like boundary condition for $X^{-}$.  Then the 1/2 BPS boundary
condition, with gauge symmetry breaking allowed, is
\begin{align}
\begin{split}
X_{7,8,9}^{-}\big| &=0\, \\
\left(D_3 X_{a}^- -\epsilon_{abc}[X_{b}^-, X_{c}^-]^-\right) \big|&=0 \quad (a,b,c=4,5,6)\ ,  \\
X_{4,5,6}^{+} \big| &= 0\, \\
D_3 X_{7,8,9}^{+} \big| &= 0 \ .
\label{halfbreaking}
\end{split}
\end{align}
Note that we in general have a commutator term $[X_{b}^-,
  X_{c}^-]^-$ in the  boundary condition.\footnote{This term
  is not emphasized in \cite{Gaiotto:2008sa}.} This term vanishes if
$G/H$ is a symmetric coset, i.e., $[\mathfrak{h}^{\perp},
  \mathfrak{h}^{\perp}]\subset \mathfrak{h}$. 
Note also that there is no commutator term $[X_{4,5,6}^+,
  X_{4,5,6}^+]^-$, since $\mathfrak{h}$ is a subalgebra and hence
satisfies $[\mathfrak{h}, \mathfrak{h}] \subset
\mathfrak{h}$.\footnote{As this example shows, $\mathfrak{h}$ and
  $\mathfrak{h}^{\perp}$ are not on an equal footing in this gauge
  symmetry breaking.} 

We can apply the same idea to junction conditions---when $N$ D3-branes
and $M$ D3-branes meet at a D5 (or NS5)-branes, say with $N<M$, 
we can think of the junction condition
as a boundary condition for the $U(N)\times U(M)$ theory with $H=U(N)$.
As we shall see in the next section, however, we need to take into account
localized degrees of freedom at the junction when $N=M$.

Along similar lines, one can contemplate more complicated composite
boundaries, possibly involving more branes and further reducing the
number of supersymmetries.

\subsection{Comments on Classification of Boundary Conditions}\label{sec24}

Let us conclude this section by commenting briefly on the status of
classifying boundaries preserving supersymmetries specified by
$\varepsilon_0$ and $\hat P$. What we have learned from the analysis
leading up to this point is that for $\hat P$ given by (\ref{proj}),
boundaries can be constructed by stacking NS5,
NS5$^\ensuremath{\prime}$, D5, and D5$^\ensuremath{\prime}$ branes
with orientations specified in the last subsection.  The problem of
interest is to classify the possible supersymmetric boundary
conditions by their effect on the low energy modes of the bulk theory.

\begin{figure}
\centerline{\includegraphics[scale=0.8]{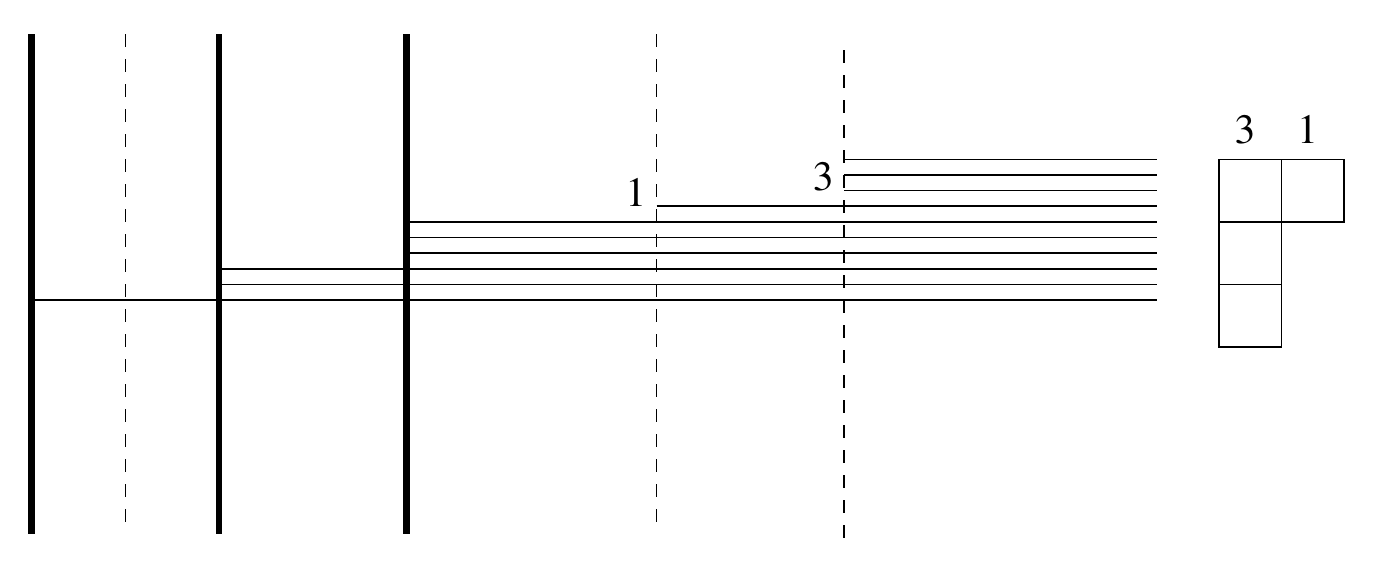}}
\caption{An example of $1/2$ BPS boundary for the ${\cal N}=4$ $U(10)$
  SYM with $H=U(6)$, $\mathfrak{B}$ is $U(1) \times U(3)$ gauge theory
  with 1 and 6 flavors of quarks, respectively. The data $\rho$
  characterizes the sequence non-decreasing linking numbers which for
  this example is $\{1,3\}$ and encodes the pattern of symmetry
  breaking $10 \rightarrow 7 \rightarrow 6$. This data can also be
  represented by the Young diagram as shown.
\label{figa}}
\end{figure}

It is useful to first recall the case considered by GW
\cite{Gaiotto:2008sa,Gaiotto:2008ak} where the boundary preserves one
half of the supersymmetries, i.e.\ $\hat P$ is the identity.  In that
case, in general, one considers a system consisting of NS5(012789)
branes and D5(012456) branes arranged in some arbitrary order,
eventually terminating with all D3 branes ending on either a D5 or an
NS5 on one end, and some number of D3 extending indefinitely on the
other.  The key ingredient in the classification of
\cite{Gaiotto:2008sa,Gaiotto:2008ak} is the well-established
observation that the positions of 5-branes along the $y=x_3$ direction
decouple in the infrared. This means that one can move the 5-branes in
the $y$-direction freely, as long as one accounts for the brane
creation effect when the NS5 and the D5-branes cross
\cite{Hanany:1996ie}. With all this in mind, GW prescribed the
following canonical ordering. Let us suppose that we are considering a
boundary on the left in the $y=x_3$ axis so that the system can be
represented by D3 branes extending semi-infinitely on the right side
and terminating on a collection of D5 and NS5 branes on the
right. Then, the procedure is to:
\begin{enumerate}
\item take all the D5 branes on which there are some net number of D3 branes terminating on them, and move them all to the right of all of the NS5 branes, and
\item arrange the NS5 and D5 branes so that their linking number is non-decreasing from left to right. 
\end{enumerate}
Once the 5-branes are arranged in this canonical order, the meaning of
the triple $(\rho,H,\mathfrak{B})$ becomes apparent. First, collect
the 5-branes into two groups: the D5's with some number of D3's ending
on them, and the rest. Because of the canonical ordering, all of the
5-branes in the first group will be to the right of the second
group. The first group of D5's encode a pattern of breaking of gauge
group $G$ to $H$. The data $\rho$ characterize the embeddings of
solutions to Nahm's equations and encode the sequence of breaking of
gauge symmetry as D5 branes are crossed from right to left among the
first group of D5 branes. Finally, the second group of 5-branes,
consisting of NS5 branes, some D5 branes with no D3's ending on them,
and a set of semi-infinite D3's extending to the right is represented
by a boundary theory $\mathfrak{B}$. The global symmetry of this
boundary theory $\mathfrak{B}$ contains $H$ as a subgroup which is
gauged when coupling to the rest of the boundary characterized by
$\rho$.  An example of the canonically ordered brane configuration is
shown in figure \ref{figa}.  Note that although the classification is
motivated by considering branes, there is no particular requirement
that $\mathfrak{B}$ is constructed from 5-branes; it could be any
$\mathcal{N}=4$ 3$d$ gauge theory with global symmetry $H$.

An interesting question is whether a similar statement classifying
boundaries preserving one quarter of supersymmetries of the ${\cal
  N}=4$ theory can be formulated. One way to frame this question is as
the problem of classifying low energy effective dynamics of a system
of D3 branes ending on collection of NS5(012789),
NS5$^\ensuremath{\prime}$(012459), D5(012456), and
D5$^\ensuremath{\prime}$(012678)
branes\cite{deBoer:1997ka,Elitzur:1997hc}:\footnote{ See appendix
  \ref{appendixA3} for more details on rotate brane configurations.
  In in table \ref{KOOtable2}, we can more generally consider D5 and
  NS5-branes rotated in the same angle in the $47$ and $58$
  directions. The rotation angle in practice corresponds to the mass
  deformation for the 3$d$ $\mathcal{N}=2$ theory, and for the
  consideration of the IR physics we can specialize to the case of the
  D5, NS, D5$^\ensuremath{\prime}$ and NS$^\ensuremath{\prime}$
  branes.  }
\begin{align}
\begin{tabular}{c||ccc|c|ccc|ccc}
       & 0& 1  & 2& 3& 4& 5& 6& 7& 8& 9 \\
       \hline
D3 & $\circ$ &  $\circ$ & $\circ$ & $\circ$ &    &   &    &   &   &     \\
D5 & $\circ$  &  $\circ$ & $\circ$ &  &  $\circ$  & $\circ$  & $\circ$   &   &   &     \\
D5$^\ensuremath{\prime}$  & $\circ$  &  $\circ$ & $\circ$ &  &    &   &  $\circ$  &  $\circ$  & $\circ$  &    \\
NS5 & $\circ$ &  $\circ$ & $\circ$ &  &   &  &  & $\circ$  & $\circ$  & $\circ$    \\
NS5$^\ensuremath{\prime}$ & $\circ$  &  $\circ$ & $\circ$ & & $\circ$    &$\circ$    & &    &   & $\circ$       \\
\end{tabular}
\label{branetable}
\end{align}

The experience from the case of preserving half of the supersymmetries
suggests that one should start by sorting these 5 branes in some
canonical order, but there is a problem with this approach. In the
case where only the D5 and the NS5 are present, it is the case that
exchanging their order do not affect the low energy physics
\cite{Hanany:1996ie}. In the presence of NS5$^\ensuremath{\prime}$ and
D5$^\ensuremath{\prime}$ branes, however, the story is
different. Exchanging the order of D5 and NS5$^\ensuremath{\prime}$ or
D5 and D5$^{\ensuremath{\prime}}$ gives rise to phase transitions
where the number of infrared degrees of freedom can change. Some
examples illustrating this effect were described in
\cite{Aharony:1997ju}.  The classification of boundaries constructed
starting form general configuration of NS5, NS5$^\ensuremath{\prime}$,
D5, an D5$^\ensuremath{\prime}$ branes will necessarily be more
complicated, forcing one to map the equivalences and dualities among
various configurations.

\begin{figure}
\centerline{\includegraphics[scale=0.8]{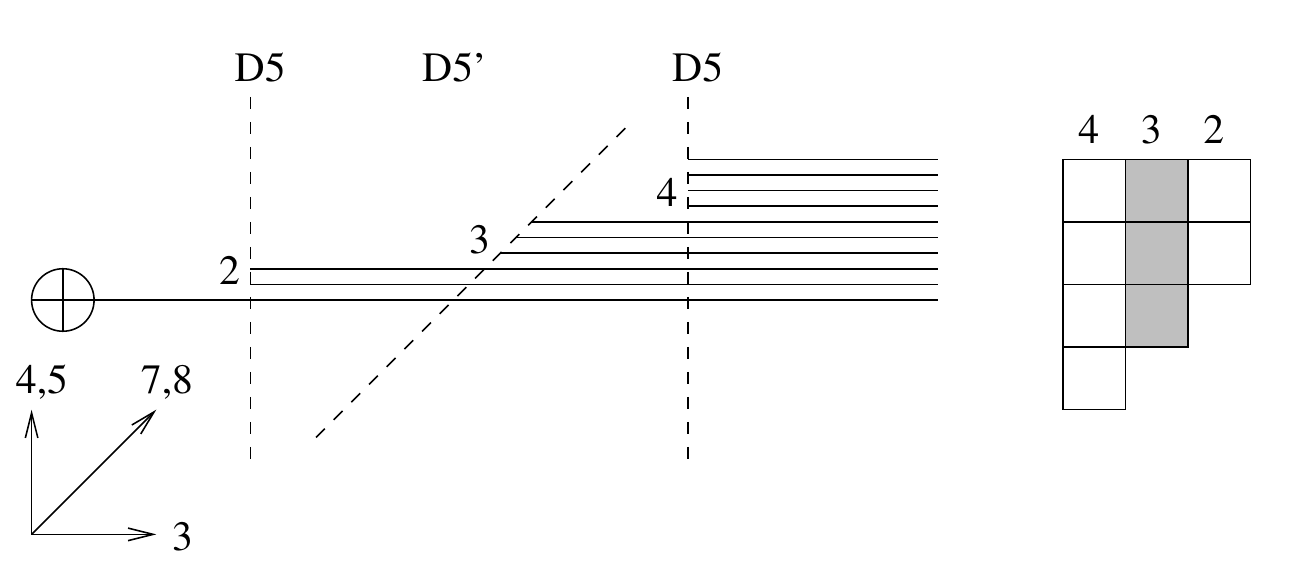}}
\caption{The generalization of $\rho$ when D5 and D5$^{\ensuremath{\prime}}$ branes are
  present. We have restricted our attention to the case where the
  linking number is non-decreasing so that a generalized Young diagram
  can be drawn as illustrated, although there are no a priori reason
  to only consider this case.
\label{fig.rhotilde}}
\end{figure}

A slightly less ambitious problem, perhaps, is to classify the
boundary conditions where some of the brane ordering prescription is
implemented a priori. For example, we can take all the D5 and
D5$^\ensuremath{\prime}$ with D3 ending on them to be to the right of
rest of the 5-branes, and order the 5-branes to have non-decreasing
linking numbers going from left to right. Then, one can characterize
the boundary condition in terms of a triple $(\tilde
\rho,H,\mathfrak{B})$ where $\mathfrak{B}$ is now an ${\cal N}=2$
superconformal field theory, and $\tilde \rho$ must encode the fact
that there are two (or more) distinct types of D5's.  An example for a
configuration of D5 and D5$^{\ensuremath{\prime}}$ giving rise to such
a setup is illustrated in figure \ref{fig.rhotilde}. The Young diagram
associated with the $\tilde \rho$ will now need to also encode the
fact that they might arise either from a D5 or a
D5$^{\ensuremath{\prime}}$.\footnote{Also, a priori, there is no
  reason that for the Young diagram to be monotonic in height.} A
structure very similar to this was also discussed in
\cite{Xie:2013gma}.  The classification of $\mathfrak{B}$ with ${\cal
  N}=2$ supersymmetries is itself a non-trivial problem.  Even the
classification of ``good,'' ``bad,'' and ``ugly'' quivers is
complicated by the non-trivial dynamics of field theory in 2+1
dimensions with ${\cal N}=2$ supersymmetries
\cite{Kapustin:2010xq,Safdi:2012re}.  It would therefore appear that
the problem of classifying boundary conditions preserving one quarter
of ${\cal N}=4$ supersymmetric Yang Mills theory in 3+1 dimensions is
too ambitious at the present time. It would be interesting to revisit
this problem when a more thorough understanding of theories in 2+1
dimensions with ${\cal N}=2$ supersymmetries is available.

\section{Boundaries with Localized Degrees of Freedom}
\label{sec:localized}

In the previous section, we saw that a natural class of supersymmetric
boundary conditions arose from demanding that the supercurrent have
vanishing flux through the boundary.  However, this analysis depended
on a restrictive assumption, that the only degrees of freedom in the
problem live in the bulk.  In this section, we will explore some
interesting classes of boundary degrees of freedom and the boundary
conditions they imply for the bulk fields.

An astute reader might have wondered whether we are required
to choose the particular form of the supercurrent in
eq. (\ref{supercurrent10d}).  Just as is the case for the
energy-momentum tensor, we could have considered adding conserved
``improvement'' terms to the supercurrent. They can generically be
constructed by adding total derivative terms to the Lagrangian before
implementing the Noether procedure; thus it is clear that they do not
affect the local analysis of supersymmetry, but they do affect the
global analysis. In particular, the improvement terms are important in
the presence of boundaries, where the total derivatives give rise to
Lagrangians defined at the boundary, potentially with their own
degrees of freedom.

\begin{figure}
\centerline{\includegraphics[scale=0.8]{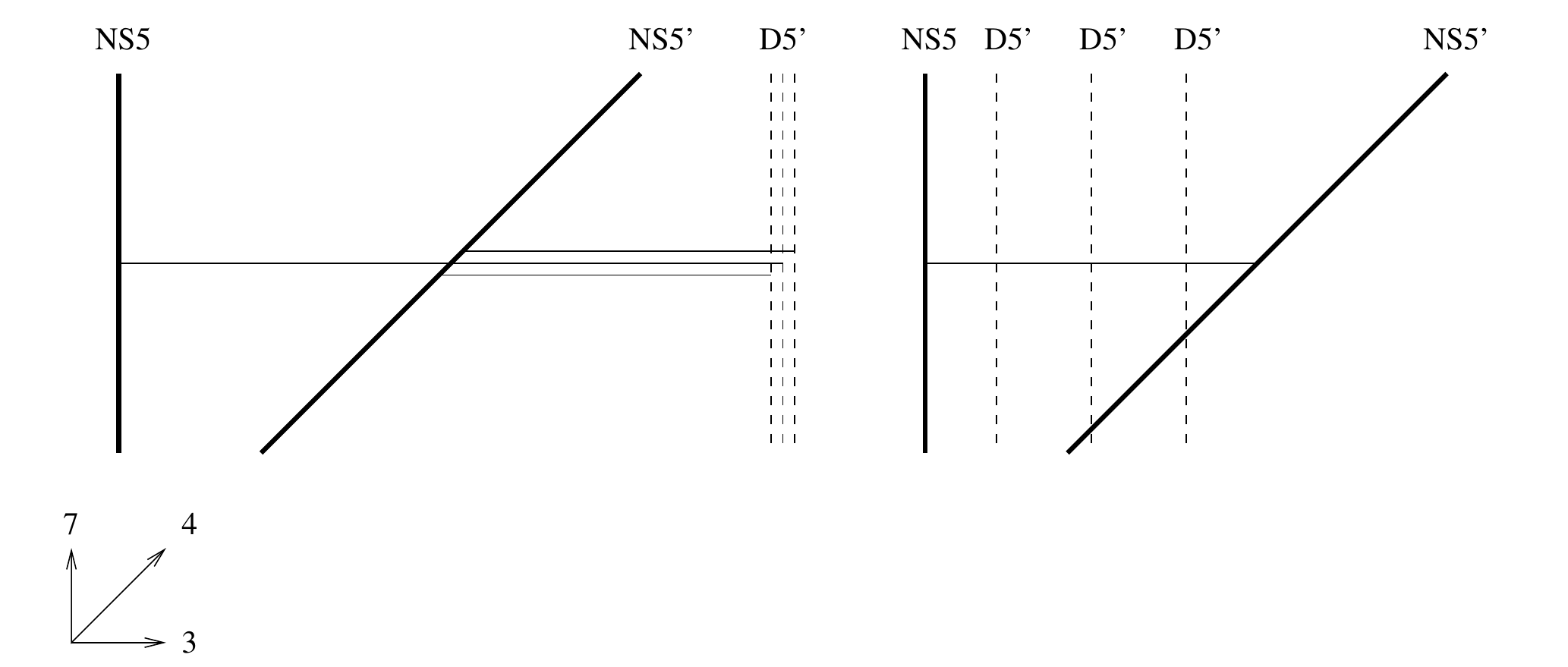}}
\caption{Brane construction of ${\cal N}=2$ $U(1)$ $N_f=3$ theory in
  two different brane orderings.
\label{figb}}
\end{figure}

To visualize the kind of construction we want to study, let us start
by recalling the construction of ${\cal N}=2$ Abelian theory in 2+1
dimensions using branes. As a concrete example, we show two ways of
constructing a theory with $U(1)$ gauge symmetry and $N_f=3$ flavors
in figure \ref{figb}.

Configurations like these were used to engineer ${\cal N}=2$ theories
in \cite{deBoer:1997ka,deBoer:1997kr}. The 5-branes and the 3-branes
can then be allowed to move, with their positions being interpretable
as masses, FI parameters, and components of the moduli. These features
are explained in detail in \cite{deBoer:1997ka,deBoer:1997kr}.

The two configurations in figure \ref{figb} describe equivalent
physics in the IR, in that they are related by the exchange of the
relative positions of an NS5 brane and the D5 brane.  Although what is
illustrated in figure \ref{figb} is a string theory construct, one can
also consider a decoupling limit and think of the system as consisting
of defects and boundaries of a gauge field theory in 3+1 dimensions.
The distance $L$ between the NS5 and the NS5$^\ensuremath{\prime}$
brane encodes the 2+1 gauge coupling via the relation
\be g_{\rm YM3}^2 = {g_{\rm YM4}^2 \over L} \ . 
\ee
It is also common to engineer the theory in 2+1 dimensions by pushing
the D5 branes infinitely to the right so that the configuration
resembles a 3+1 theory bounded by an NS5$^\ensuremath{\prime}$ and an
NS. This picture, however, obscures the decoupling between the 2+1
dimensional physics and 3+1 dimensional physics due to light modes
living on the segment between the NS5 and the
D5$^\ensuremath{\prime}$s.  This issue is related to the subtlety in
formally defining the notion of the ``moduli space of the boundary''
as a local notion, independent of the bulk physics and the ``boundary
condition on the other side.''

Our goal in this section is to specify boundary conditions for ${\cal
  N}=4$ SYM in a sufficiently precise way that we can compute the
moduli space of configurations like the one in figure \ref{figb} in
terms of an explicit computation for a defect version of the ${\cal
  N}=4$ theory.

\subsection{Bulk Equations}

In this section, we will describe the conditions imposed on the bulk
fields by supersymmetry.

Before proceeding, we will rewrite the bulk supersymmetry equations
(\ref{bulk1})--(\ref{bulk8}) in a form where holomorphy is manifest:
we have 3$d$ $\mathcal{N}=2$ theory and the moduli space is
K\"{a}hler.  A consistent truncation that is sometimes useful to
consider is to set $\vec{Y}=0$, and to suppress all terms with Lorentz
indices, leaving precisely the Nahm equation \eqref{Nahm}.
It will turn out to be convenient to combine $X_4$ and $X_5$ into
complex combination
\be {\cal X} \equiv X_4 + i X_5 \label{calXdef}  \ ,
\ee
as well as $A_3$ and $X_6$ into
\be {\cal A} \equiv  A_3 + i X_6  \ , 
\label{calAdef}
 \ee
so that the two of the three Nahm equations can be written as a
complex equation
\beq
\frac{\mathcal{D}\mathcal{X}}{\mathcal{D}y}\equiv \frac{d\mathcal{X}}{dy} + \lbrack \mathcal{A}, \mathcal{X}\rbrack = 0 \ ,
\label{cplxnahm}
\eeq
and the third Nahm equation is
\beq
\frac{d}{dy}\left(\mathcal{A}-\bar{\mathcal{A}}\right) - \left[\mathcal{A},\bar{\mathcal{A}}\right]+ \left[\mathcal{X},\bar{\mathcal{X}}\right] =0 \ ,
\label{nahm3}
\eeq
where the barred fields are $\bar {\cal X} \equiv X_4 - i X_5 $ and
$\bar {\cal A} \equiv A_3 - i X_6$.  This is a standard form of
presenting the Nahm equation, and can be an effective form for
analyzing moduli spaces as quotient space as we will review later in
this section.

In the case where only one quarter of supersymmetry (\ref{proj}) is
preserved, one should generally expect the the $Y$ fields to mix as
seen in (\ref{bulkf})--(\ref{bulkl}).  In this case, the useful
truncation is to consider is $X_9=0$.  Then the bulk equations in
terms of complex fields are given by three naturally complex
equations,
\beq
\frac{\mathcal{D}\mathcal{X}}{\mathcal{D}y} &=& 0 \ ,\label{qnahm1}\\
\frac{\mathcal{D}\mathcal{Y}}{\mathcal{D}y} &=& 0 \ ,\label{qnahm2}\\
\left[\mathcal{X},\mathcal{Y}\right] &=& 0 \ , \label{qnahm3}
\eeq
and one real equation,
\beq
\frac{d}{dy}\left(\mathcal{A}-\bar{\mathcal{A}}\right)-\left[\mathcal{A},\bar{\mathcal{A}}\right] +\left[\mathcal{X},\bar{\mathcal{X}}\right] + \left[\mathcal{Y},\bar{\mathcal{Y}}\right] &=& 0\ ,\label{qnahm4}
\eeq
where
\beq
\mathcal{Y} &\equiv & X_7+i X_8  \ .\label{calYdef}
\eeq

In addition to these equations, we also have BPS conditions for the
sixth scalar $X_9$:
\beq
D_3 X_9 &=& 0 \ ,\\
\left[X_6 , X_9\right] &=& 0 \ ,\\
\left[ \mathcal{X},X_9 \right] &=& 0 \ ,\\
\left[ \mathcal{Y},X_9 \right] &=& 0 \ .
\eeq
The equations involving $X_9$ do not obey the same division into
``naturally real" and ``naturally complex" parts as with the other BPS
equations;\footnote{$X_9$ combines with the field strength of the 3$d$
  gauge field into a 3$d$ linear multiplet.} because of this, when
$X_9$ is nonzero, the correct K\"{a}hler structure of the moduli space
can be obscured.

One simple class of solutions can be obtained by simply setting
$X_9=0$, so that the BPS equations are indeed those of a K\"{a}hler
quotient.  In fact this turns out to be the most interesting case, but
it is not the most general.  For example, we might consider a single
D3-brane suspended between two NS5-branes (extended in the 789
directions), for which we would expect that there is indeed a modulus
associated with motions of the D3-brane in the $X_9$-direction, and
also a modulus associated with the ``dual scalar'' from the reduction
of the 4$d$ theory to a 3$d$ theory.  This example with NS5-branes is
prototypical; the NS5-brane boundary conditions allow $X_9$ to
contribute to the moduli space, but they also preserve an unbroken
abelian gauge symmetry which in this case allows the dual scalar to
also be part of the moduli space.

It will turn out that the boundary conditions which preserve some
amount of unbroken gauge symmetry are those for which the scalar $X_9$
is active.  This is a subtle issue when quantum effects are taken into
account, but sometimes it can be useful to keep track of the
classical moduli associated with $X_9$ and the dual scalar, even if
they receive quantum corrections.

\subsection{Junction Conditions}

In order to discuss junction conditions, we need to incorporate the
localized degrees of freedom at the D5 and NS5 defects.
In the case where the NS5 and D5 break the same half of the
supersymmetry, this issue has been worked out in the treatment of
GW \cite{Gaiotto:2008sa}.

Specifically, for the D5 brane at $x^3=x^7=x^8=x^9=0$ intersecting $N$
D3-branes, equation (3.27) of \cite{Gaiotto:2008sa} reads
\be {D \vec X \over D y} + \vec X \times \vec X + \delta(y) \vec \mu_{\omega} = 0 \ , \ee
and for the NS5-brane at $x_3=x_4=x_5=x_6=0$, equation (3.79) of
\cite{Gaiotto:2008sa} reads
\beq - \vec X_L(0) + \vec \mu_L &=& 0 \label{muL}  \ ,\\
\vec X_R(0) + \vec \mu_R & = & 0 \label{muR} \ . 
\eeq
Here, $\mu_{\omega}$, $\mu_L$, and $\mu_R$ refers to the contribution
of the matter fields localized on the defect to the moment maps. These
relations would be all that we need in principle, except for one
critical ingredient: in the half BPS case, we usually restrict to the
case that $\vec Y=0$.  When we consider 1/4 BPS configurations,
however, it is no longer appropriate to ignore the $\vec Y$ fields.

In this subsection, we will review the arguments deriving the junction
condition for the $\vec X$ and the $\vec Y$ fields across the D5 and
NS5 domain walls. Along the way we will reproduce the explicit form
for the moment functions $\mu_{\omega}$, $\mu_L$, and $\mu_R$. Once
these are worked out, it is straightforward to generalize the junction
conditions to the NS5$^\ensuremath{\prime}$ and the
D5$^\ensuremath{\prime}$ defects.

\subsubsection{D5-like Junction}

We begin by reviewing the D5 boundary and defects. We start by
considering $N$ D3 branes ending on both sides of a D5-brane
interface. The junction condition for $N$ D3 on one side and $M<N$ D3
on the other side of the D5 can be inferred by pushing some of the
D3's to infinity along the world volume of D5.  As we have seen
already, even for 1/4 BPS cases the junction condition at the D5-brane
is 1/2 BPS, which was already discussed in \cite{Gaiotto:2008sa}.
However for the application to the 1/4 BPS cases it is crucial to work
out the conditions of all the scalars involved, in particular
$\mathcal{Y}$.

It is intuitively clear that there are some modes localized at the
D3-D5 intersection, when $M=N$: these are the the strings connecting
D3 and D5-branes.  From the viewpoint of the D3-brane the strings give rise to
 a chiral multiplet $Q$ in the fundamental representation or
$\tilde{Q}$ in the anti-fundamental representation, depending on the
orientation of the strings.  Since the D3-D5 intersection locally is
1/2 BPS, $Q$ and $\tilde{Q}$ combine to form a 3$d$ $\mathcal{N}=4$
hypermultiplet ${\omega}=(Q, \tilde{Q})$.

The localized fields $Q, \tilde{Q}$ should couple to the bulk degrees
of freedom.  For this purpose we can write the fields of 4$d$
$\mathcal{N}=4$ SYM in terms of 3$d$ $\mathcal{N}=2$ superspace,
following the formalism of \cite{DeWolfe:2001pq,Erdmenger:2002ex}.

The 4$d$ $\mathcal{N}=4$ vector multiplet can be decomposed into four 3$d$
$\mathcal{N}=2$ multiplets.  First, we have one $\mathcal{N}=2$ vector
multiplet $V$, or rather the associated linear multiplet
\be \Sigma=\frac{1}{2i } \epsilon^{\alpha\beta} \bar{D}_{\alpha}(e^{-V} D_{\beta }e^V) \ ,
\ee
which contains the topological current $J^{\mu}\equiv \frac{1}{2}\epsilon^{\mu\nu\rho} F_{\nu\rho}$ in one
of its components.  Here the role of the vector multiplet adjoint
scalar is played by the adjoint scalar $X_9$.  Second, we have three
$\mathcal{N}=2$ chiral multiplets $\mathcal{A}, \mathcal{X},
\mathcal{Y}$, which are the complex combinations of $A_3$ and
$X_{4,5,6,7,8}$.  The latter three coincide with our previous
definition given in \eqref{calXdef},\eqref{calAdef}, and
\eqref{calYdef}.  Each of these four $\mathcal{N}=2$ multiplets
depends on the $y$-coordinate.

We can write the bulk $\mathcal{N}=4$ Lagrangian in terms of 3$d$
$\mathcal{N}=2$ superfields.  The answer is given by
\cite{Erdmenger:2002ex}
\begin{align}
\label{Lbulk}
\begin{split}
\mathcal{L}_{\rm bulk}& =
\int d^4\theta\,  \textrm{Tr}
\left[
\Sigma^2+\frac{1}{2}(e^{-V} \bar{\mathcal{D}} e^V + \mathcal{A})^2
-e^{-V} \bar{\mathcal{X}} e^V \mathcal{X}
-e^{-V} \bar{\mathcal{Y}} e^V \mathcal{Y}
\right] \\
& \qquad +\int d^2 \theta\,  
 \textrm{Tr}\left( \mathcal{X} \mathcal{D}  \mathcal{Y}
- \mathcal{Y} \mathcal{D} \mathcal{X} 
\right)
+\int d^2 \bar{\theta}\,  
 \textrm{Tr}\left(
- \bar{\mathcal{X}} \bar{\mathcal{D}}  \bar{\mathcal{Y}}
+ \bar{\mathcal{Y}} \bar{\mathcal{D}} \bar{\mathcal{X}} 
\right) \ ,
\end{split}
\end{align}
where $\mathcal{D}\equiv \partial_y+\mathcal{A}, 
\bar{\mathcal{D}}\equiv \partial_y+\bar{\mathcal{A}}$ as in \eqref{cplxnahm}.
Recall that in our convention $\mathcal{A}, \X, \Y$ are anti-Hermitian and 
hence for example $\X^{\dagger}=-\bar{\X}$.

Note that the $F$-term contains the commutator term
$\textrm{Tr}([\mathcal{X}, \mathcal{Y}]\mathcal{A})$, which 
is related to 
the superpotential term of the 4$d$ $\mathcal{N}=4$ theory.  Note
also that the $F$-term equations for this bulk Lagrangian give
\begin{align}
\mathcal{D}\mathcal{X}=\mathcal{D}\mathcal{Y}=0
\end{align}
which are nothing but the complex part of the Nahm equations.

We can also vary the vector multiplet $V$ in \eqref{Lbulk}, and if we
neglect the covariant derivatives along the 3$d$ directions we have the
real part of the Nahm equations.

Let us now come back to the coupling with the $Q, \tilde{Q}$.  The
coupling between the bulk and the boundary should preserve 3$d$
$\mathcal{N}=4$ supersymmetry.  This determines the bulk-boundary
interaction to be
\begin{align}
\mathcal{L}_{\rm defect}= 
\left[ \int d^4 \theta  \left(Q^{\dagger} e^V Q+  \tilde{Q}e^{-V} \tilde{Q}^{\dagger}   \right)
 + \int \! d^2 \theta \,  \left[  \tilde{Q}  \mathcal{Y} Q\right]
 - \int \! d^2 \bar{\theta} \,  \left[ Q^{\dagger} \bar{\mathcal{Y}} \tilde{Q}^{\dagger}  \right]
\right] \delta(y-y_0) \ ,
\label{Ldefect1}
\end{align}
where $y_0$ represents the position of the defect along the
$y$-direction.  This takes the form of the Lagrangian of 3$d$
$\mathcal{N}=4$ theory, where the role of the adjoint 3$d$
$\mathcal{N}=2$ chiral (vector) multiplet is played by the bulk chiral
multiplet $\mathcal{Y}$.

Note that the superpotential term $\tilde{Q} \mathcal{X} Q$ has the
correct dimension $2$; this is because the bulk field $\mathcal{Y}$,
being a field in four dimensions, has mass dimension $1$ ($Q$ and
$\tilde{Q}$ have canonical mass dimension in 3$d$, namely $1/2$).

Let us now see how this affects our analysis of the moduli space.
When we vary the bulk fields, the equations of motion for the bulk
fields read
\begin{align}
&V:  \quad \frac{d} {dy}(\mathcal{A}-\bar{\mathcal{A}}) -[\mathcal{A}, \bar{\mathcal{A}}]
+[\mathcal{X}, \bar{\mathcal{X}}]+[\mathcal{Y}, \bar{\mathcal{Y}}]+
 \mu_{\omega, \mathbb{R}}\, \delta(y-y_0)=\zeta \ , \\
&\mathcal{X}: \quad \frac{\mathcal{D}\mathcal{X}}{\mathcal{D} y} +  \, \mu_{\omega, \mathbb{C}} \, \delta(y-y_0) =0  \ ,
\end{align}
and $\mu_{\mathbb{R}}, \mu_{\mathbb{C}}$ are real and complex moment
maps defined by
\begin{align}
\mu_{\omega, \mathbb{R}}&\equiv Q Q^{\dagger} - \tilde{Q}^{\dagger} \tilde{Q} \ , \\
\mu_{\omega, \mathbb{C}}&\equiv Q\tilde{Q}  \ .
\end{align}

We can combine these equations into
\beq
 D_y X_a - \frac{1}{2} \epsilon_{abc} [X_b,X_c] - \mu_{\omega, a} \delta(y-y_0) =0\ ,  \label{jump} 
\eeq
where 
\beq
\omega\equiv  \frac{1}{\sqrt{2}}\left( \begin{array}{cc}  Q & \tilde{Q}^{\dag} \end{array} \right) 
 \ ,
\qquad \omega^{\dag} \equiv \frac{1}{\sqrt{2}}\left( \begin{array}{c} Q^{\dag}   \\ \tilde{Q}\end{array} \right)
 \ ,
\eeq
and
\begin{align}
\vec{\mu}_{\omega}\equiv \omega^{\dag} \vec{\sigma} \omega
\end{align}
is a triplet of moment maps, making manifest the hyperK\"{a}hler
structure (which comes from 3$d$ $\mathcal{N}=4$ supersymmetry locally
present at the junction.) Note that the triplet structure is not
manifest in the 3$d$ $\mathcal{N}=2$ superspace formulation, see however
appendix \ref{sec.boundaryN1} for a 3$d$ $\mathcal{N}=1$ superspace
formulation which makes the triplet structure manifest.

The moment maps $\vec \mu_{\omega}$ cause $\vec X_a$ to jump at the D5 and
are sometimes referred to as the ``jumping conditions.'' Two of the
three components of $\vec{\mu}_{\omega}$ can also be found in (3.33) of
\cite{Gaiotto:2008sa}. The equations are known in the context of the Nahm
equations, see \cite{Hurtubise:1989qy} and also
\cite{Kapustin:1998pb,Tsimpis:1998zh}.

In general the supersymmetry is broken to 1/4 BPS with other
boundary/junction conditions, and we have only 3$d$ $\mathcal{N}=2$
superspace.  Even in these cases, the $F$-term part is protected from
perturbative quantum corrections thanks to the non-renormalization
theorem for 3$d$ $\mathcal{N}=2$ theory.

There is an additional constraint imposed on the fields
$\mathcal{Y}$ due to the superpotential (\ref{Ldefect1}).  The
$F$-term equation for the fields $Q, \tilde{Q}$ gives
\begin{align}
{\cal Y}  Q &= 0 \ , \label{YQtilde1st} \\
\tilde{Q} {\cal Y} &  = 0  \ . \label{QY1st}
\end{align}
This is part of the triplet of the equations (see appendix
\ref{sec.boundaryN1})
\beq
Y_a \sigma^a \omega = \omega^{\dag}  Y_a \sigma^a= 0 \ ,
\label{D5Ycondition1}
\eeq
which includes extra conditions
\be Y_3   Q = \tilde Q Y_3  = 0  \ .\ee
For the most part, we can consistently set
\be Y_3 =X_9= 0 
\ee
for both half and quarter supersymmetric cases. Exceptions will be
discussed briefly below, but as long as $Y_3$ is set to zero, the
constraint on $X$ and $Y$ can be written more concisely as
\beq \Delta \mathcal{X}(y_0) &=& Q \tilde{Q} \ , \label{dX}  \\
i\Delta X_6 &=& Q Q^{\dag}-\tilde{Q}^{\dag} \tilde{Q} \ , \label{dX6} \\
{\cal Y}Q   &  =& 0  \ ,\label{YQtilde} \\
\tilde Q {\cal Y} &  =& 0  \ .  \label{QY}
\eeq
We see that the new conditions \eqref{YQtilde}, \eqref{QY} imply that
$\X$ is only allowed to jump if $\Y=0$, or in other words, when the
D3-branes intersect the D5-brane.

With these ingredients, one can easily understand the case of $N$
D3-branes on one side and $N-1$ on the other by considering limiting
forms of the fundamental quarks $Q, \tilde{Q}$ (this essentially
follows the analysis of \cite{Chen:2002vb} in the context of monopole
solutions.) One possibility is to parametrize
\be {\cal X}_L = \left(\begin{array}{ccc|c}
&&& {}\\
& A && B \\
&&& {} \\ \hline
& C && D 
\end{array}\right) , \qquad
{\cal X}_R = \left(\begin{array}{ccc|c}
&&& {}\\
& A' && 0 \\
&&& {} \\ \hline
& 0 && \alpha 
\end{array}\right) \ ,  \label{structure}
\ee
with
\be Q^\dag = (u_1, \ldots u_{N-1}, M) \ , \qquad  \tilde Q = (v_1, \ldots v_{N-1}, M) \ .
\label{QLimitForm} \ee

We can then solve (\ref{dX}) by setting
\be |M^2| =  \alpha - D \ , \qquad   u_i = {\lambda \over M} B^*_i   \ , \qquad  v_i = {1 \over \lambda^* M^*}  C_i \ .
\ee
We can consider the case
\be i \Delta X_6 = Q Q^\dag - \tilde Q^\dag \tilde Q = 0  \ ,\ee
which amounts to not moving the D3's in the $X_6$ direction to fix
$\lambda$. These two conditions determine $Q$ and $\tilde Q$
completely up to the phase of $\lambda$ and $M$. These phases,
however, are irrelevant when we take the limit $|M|
\rightarrow \infty$. In that limit, $Q$ and $\tilde Q$ are determined
uniquely. Also, in that limit, we find that the $(N-1) \times (N-1)$
block
\be A = A' \ee
is continuous. Finally, we observe that in the limit, $B$, $C$, and
$D$ are arbitrary.
This is the same feature found in (3.42) and (3.43)
of \cite{Gaiotto:2008sa}.

The D5-brane interface conditions imply restrictions on the allowed
gauge transformations when the numbers of D3-branes on each side of
the interface are unequal.  Specifically, the form (\ref{QLimitForm})
only preserves a $U(N-1)$ gauge symmetry of the bulk $U(N)$; note also
that because some of the fields are continuous across the interface,
the gauge transformations must also be continuous.  We see that the
allowed gauge transformations take the block form
\beq
g = \left(\begin{array}{ccc|c}
&&& {}\\
& * && 0 \\
&&& {} \\ \hline
& 0 && 1
\end{array}\right).
\eeq

The other structure that we need is the condition imposed on the $Y$
fields. This was covered only implicitly in \cite{Gaiotto:2008sa} but
we can read off the relevant detail from (\ref{YQtilde}) and
(\ref{QY}). Since one component of $Q$ and $\tilde Q$ is blowing up
while the others are going to zero, we infer that the row and the
column of ${\cal Y}$ associated with the divergent component of $Q$
and $\tilde Q$ must vanish. Physically, this is simply the statement
that the D3 brane ending on a D5 brane must have its transverse
coordinates coincident with the transverse coordinate of the D5-brane.

Now that we have worked out the case of D5 junction with $N$ D3 on one
side and $N-1$ on the other, we can extend to the case with $N$ on one
side and $N-M$ on the other with $M > 1$. For the ${\cal X}$ fields,
the structure is similar to (\ref{structure}) except that now, $B$ and
$C$ are $M \times (N-M)$ and $D$ is $M \times M$. $B$ and $C$ are
required to be finite, whereas $D$ will have the structure of Nahm
poles close to the D5 brane. In the limit $M=N$, $A$, $B$, and $C$
goes away and we have the standard Nahm pole boundary condition. These
are basically the findings reported in \cite{Gaiotto:2008sa}. We also
impose the condition that the $N-M$ rows and column of ${\cal Y}$
field vanish.

We close this subsection by pointing out that this analysis can easily
be extended to the junction condition for the D5$^\ensuremath{\prime}$
brane oriented along the 012678 direction. They are related to the
case of D5-brane oriented along 012456 simply by exchanging ${\cal X}$
and ${\cal Y}$ while leaving $X_6$ alone:
\beq \Delta \mathcal{Y}(y_0) &=& Q \tilde{Q} \ , \label{dY} \\
i\Delta X_6 &=& Q Q^{\dag}-\tilde{Q}^{\dag} \tilde{Q} \ , \\
{\cal X} Q &=& 0 \ , \label{XQtilde} \\
\tilde Q   {\cal X}  &  =& 0 \ .  \label{QX}
\eeq

\subsubsection{NS5-like Junction}

Next, let us turn our attention to the case of NS5-branes intersecting
$N$ D3-brane from the left and $M$ D3-branes from the right.  There
should again be localized degrees of freedom, this time representing
the strings between D3-branes on the left and on the right.  They are
bifundamentals under the $U(N)\times U(M)$ gauge symmetry; we have
$\mathcal{N}=2$ bifundamental chiral multiplets $A$ and $B$
transforming as (${\bf N,  \bar{M}}$) and (${\bf \bar{N}, M}$),
which together make up an $\mathcal{N}=4$ hypermultiplet.

We can again write down the Lagrangian representing the coupling of
the localized field $A, B$ to the bulk:
\begin{align}
\label{Ldefect2}
\mathcal{L}_{\rm defect}&=
\delta(y-y_0)\int d^4 \theta \left(\textrm{Tr}_R  A^{\dagger} e^{V_L} A e^{V_R}+ \textrm{Tr}_L B^{\dagger} e^{V_R} B e^{V_L} \right) \nonumber\\
&\quad +\delta(y-y_0)
\left( 
\int d^2 \theta \left[ \textrm{Tr}_L AB \mathcal{Y}_L+\textrm{Tr}_R BA \mathcal{Y}_R \right]
+\textrm{(h.c.)} 
\right) \ ,
\end{align}
where as before $y_0$ represents the position of the defect along the
$y$-direction.  Note that again this interaction is completely fixed
by the requirement of 3$d$ $\mathcal{N}=4$ supersymmetry We can check
that the superpotential terms have correct dimension $2$, since $A$,
$B$ have 3$d$ canonical dimension $1/2$, whereas the bulk fields
$\mathcal{X}, \mathcal{Y}$ dimension $1$.

We can derive the junction conditions again from
\eqref{Ldefect2}. However, some  care is needed since there is also a
boundary contribution from the bulk Lagrangian \eqref{Lbulk}, which
contains the term, after integrating by parts,
\begin{align}
\int d^3 x \int_{y<y_0} \! dy\, \textrm{Tr}(\mathcal{X}_L \mathcal{D} \mathcal{Y}_L- \mathcal{Y}_L \mathcal{D} \mathcal{X}_L)
=\int d^3 x \int_{y<y_0} \! dy \, \textrm{Tr}(2 \mathcal{X}_L \mathcal{D} \mathcal{Y}_L)
-\int d^3 x \,  \textrm{Tr}\,  \mathcal{X}_L  \mathcal{Y}_L \Big|_{y=0}  \ .
\label{parts}
\end{align}
The integration by parts are done in such a way that the bulk
contribution vanishes at the boundary in light of ${\cal D Y}$ being
zero at the interface.  When we have no localized degrees of freedom
at the boundary, we should impose $\X_L=0$ on the NS5-like boundary
and hence \eqref{parts} vanishes. However for our purposes $\X_L=0$
turns to be too strong, and we should keep the boundary contribution
$\textrm{Tr}\, \mathcal{X}_L \mathcal{Y}_L$.  There is also a similar
contribution from the region $y>y_0$ (this time with the opposite sign
due to orientation reserval), and collecting there we have the
boundary contribution from the bulk:
\begin{align}
\mathcal{L}_{\partial \textrm{(bulk)}}=\int d^2\theta\, \textrm{Tr}\, \left[ -\mathcal{X}_L  \mathcal{Y}_L + \mathcal{X}_R  \mathcal{Y}_R \right] \delta(y-y_0)  \ .
\end{align}

We can now derive the $F$-term constraints, which gives the same equations as in (3.79) and (3.80) of \cite{Gaiotto:2008sa}: 
\begin{align} {\cal X}_L &= \mu_{L, \mathbb{C}} \equiv AB \ ,\label{XAB1} \\
{\cal X}_R& = -\mu_{R,  \mathbb{C}} \equiv  -BA \ . \label{XAB2} 
\end{align}

We can again supplement \eqref{XAB1}--\eqref{XAB2} by the equation involving real
part of the triplet of the moment maps, making the hyperK\"ahler
structure at the junction manifest (c.f.\ appendix
\ref{sec.boundaryN1}):
\beq
i X_{6,L} &=& \mu_{L, \mathbb{R}} \equiv A A^\dag -  B^\dag B  \ ,\\
i X_{6,R} & = & \mu_{R, \mathbb{R}}\equiv  A^\dag A - B B^\dag \ ,
\eeq
where the we have introduced the real part of the triplet of moment
maps defined similarly to the D3-D5 case:
\begin{align}
\mu_L=\xi Y_L^a \sigma^ a \xi^{\dagger} ,  
\end{align}
with
\begin{align}
\xi= \frac{1}{\sqrt{2}}\left( \begin{array}{c} A^{\dag}  \\ B \end{array} \right), \qquad \xi^{\dag} =  \frac{1}{\sqrt{2}}\left( \begin{array}{cc}  A & B^{\dag} \end{array} \right) \ .
\label{xidef}
\end{align}

We also have the $F$-term equation from the fields $A$ and $B$:
\beq
B: \mathcal{Y}_L A - A \mathcal{Y}_R &=&0  \ ,\\
A: B\mathcal{Y}_L - \mathcal{Y}_R B &=& 0  \ .
\eeq

Note that the $F$-term equation for the field $\mathcal{X}$ gives
complex part of
\beq
D_3 Y_L^a = Y_L^a \delta(y-y_0) \ ,
\eeq
which effectively states that $Y_L^a$ is gauge-covariantly-constant in
the bulk and takes an arbitrary value at the boundary.  This is just
what one expects for a D3-brane ending on an NS5-brane. Similar
conclusion applies to $Y_R^a$.

Crucially, because the fields are not required to be continuous across
an NS5-interface, we have independent gauge symmetries acting on each
side (unlike the D5-brane boundary condition.)  This is necessary so
that the D3-brane segments between any pair of NS-type branes give
rise to an independent 3-dimensional gauge symmetry.

\bigskip

To summarize, at NS5 oriented along 012789 with $N$ D3 ending from the
left and $M$ D3 ending from the right, we impose the junction
conditions
\beq
{\cal X}_L &=& AB \ ,\label{XAB}\\
{\cal X}_R &=& BA  \ ,\\
i X_{6,L} &=& A A^\dag -  B^\dag B  \ ,\\
i X_{6,R} & = & A^\dag A - B B^\dag \ ,\\
\mathcal{Y}_L A &=& A \mathcal{Y}_R  \ ,\\
B\mathcal{Y}_L &=& \mathcal{Y}_R B \ . \label{BYYB} 
\eeq
To gain some intuition for these equations, consider the case where
the gauge group is $U(N)$ on the left, and $U(1)$ on the right.  Then
$A$ and $B$ are constrained to be either a row or column eigenvector
of $\Y_L$.  If $A,B$ do not both vanish, then $\Y_R$ (which is a complex scalar) is equal to one
of the eigenvalues of $\Y_L$, and generic expectation values for $A$ or $B$ will
break all the gauge symmetry at the interface.  However, if $A$ and
$B$ do both vanish, there is no constraint and $\Y_R$ can take any
value.  In this latter case, $U(N)$ is only broken to $U(1)^N$ by
generic expectation values for $\Y_L$.  We see also that $\X$ is only
allowed to jump if some of the D3-branes are continuous across an NS5
interface.

We can also generalize these conditions for the case of
NS5$^\ensuremath{\prime}$ brane junction oriented along 012459 by
exchanging ${\cal X}$ and ${\cal Y}$.
\beq
{\cal Y}_L &=& AB \ , \label{YAB}\\
{\cal Y}_R &=& BA  \ , \\
i X_{6,L} &=& A A^\dag -  B^\dag B \ , \\
i X_{6,R} & = & A^\dag A - B B^\dag \ , \\
\mathcal{X}_L A &=& A \mathcal{X}_R \ , \\
B\mathcal{X}_L &=& \mathcal{X}_R B \ .  \label{BXXB}
\eeq


\subsubsection{Mass/FI Deformations}\label{sec323}

So far, we have assumed all the 5-branes to be located at the origin
in the transverse coordinates. It is however possible to consider
generalizations where we move the position of these D5 branes. These
deformations are interpretable as Fayet-Iliopolous terms and quark
masses of the low energy field theory and was mapped out in
\cite{deBoer:1997ka}.  These deformations also have the effect of
slightly modifying the junction condition.

For example, for D5-branes which are extended in the 456 directions,
so far we have assumed that they are located at $x^{7,8,9}=0$.
Displacing a D5-brane in the 78 directions can naturally be
implemented by modifying the boundary conditions
\beq
(\mathcal{Y} - m_{\mathbb{C}} \mathbb{I})Q = \tilde{Q}(\mathcal{Y} - m_{\mathbb{C}} \mathbb{I}) =0  \ ,
\eeq
which can be reproduced from extra defect superpotential
\begin{align}
\delta\mathcal{L}_{\rm defect}=m_{\mathbb{C}} \tilde{Q}Q \ .
\end{align}
We can also include real mass terms, giving expectation values to
$X_9$.  However since the real mass term is in the $D$-term, it is
expected that the precise form of the equation can be corrected
quantum mechanically.

For NS5-branes, the position of the NS5-brane gives the FI parameter,
which modifies the bulk Lagrangian \eqref{Lbulk} by the standard
$D$-term $ \mathcal{L}_{\rm FI}=\int d^4\theta\, \zeta V$.  This
naturally gives
\beq
iX_{6,L} &=& A A^{\dag} - B^{\dag} B+ \zeta_{\mathbb{R}} \mathbb{I}_L \ , \label{x6ns5bc1}
 \\
iX_{6,R} &=& A^{\dag} A - B B^{\dag}+\zeta_{\mathbb{R}} \mathbb{I}_R \ ,
\label{x6ns5bc2}
\eeq
for the real part, which is supplemented by the complex counterparts
\beq
\mathcal{X}_L &=& AB + \zeta_{\mathbb{C}} \mathbb{I}_L \ ,\\
\mathcal{X}_R &=& BA+\zeta_{\mathbb{C}} \mathbb{I}_R \ .
\eeq
In general, there can be quantum corrections to the real equations involving
$X_6$.  We have defined $\mathbb{I}_L$ and $\mathbb{I}_R$ as the
identity matrices on the left-hand and right-hand sides of the
NS5-brane.  Of course there are analogous expressions for the
NS5$^\ensuremath{\prime}$-brane boundary conditions.

In analyzing 2+1 field theories with ${\cal N}=2$ supersymmetries, it
will be instructive to explore how the moduli space depends on these
deformation parameters.

\subsection{Moduli Space and Complexified Gauge Symmetry}

A powerful technique which we will employ in analyzing the moduli space
of some of these boundary/defect systems is to complexify the gauge
symmetry.  The essential idea behind this technique is that because
the moduli space is a K\"{a}hler quotient, it can be computed by
promoting the gauge group $G$ to a complexified gauge group
$G_{\mathbb{C}}$ (modulo the issue of stability \cite{Mumford} which
will turn out to be important in the analysis of 3$d$ gauge theory
\cite{ToAppear}).  A discussion of this technique in the context of the
Nahm equations can be found in \cite{Kronheimer}.

To determine the moduli space, the mathematical problem of interest is
that of solving a set of differential equations (\ref{qnahm1}),
(\ref{qnahm2}), and (\ref{qnahm4}) subject to an algebraic constraint
(\ref{qnahm3}) and the boundary and junction conditions (which are
also algebraic), and up to gauge equivalence under the gauge group $G$.
This problem can be solved directly, but we can take advantage of the
fact that some of the equations, such as the differential equations
(\ref{qnahm1}), (\ref{qnahm2}), (\ref{qnahm3}), and some of the
boundary conditions, are manifestly in a complex form.  These complex
equations are actually invariant under a larger gauge symmetry group
than the full set of equations -- we may take $\X \rightarrow g^{-1} \X
g$, $\Y \rightarrow g^{-1} \Y g$, $\mathcal{A} \rightarrow g^{-1}
\mathcal{A} g + g^{-1}dg$, where $g$ is valued in the complexified
gauge group $G_{\mathbb{C}}$.  On the other hand, the real equation
(\ref{qnahm4}) is only invariant under the real gauge symmetry $G$ and
transforms nontrivially under $G_{\mathbb{C}}$.

It is a beautiful mathematical fact that solving the full system of
equations (\ref{qnahm1})--(\ref{qnahm4}), modded out by the gauge
symmetry $G$, is equivalent to solving the complex system
(\ref{qnahm1})--(\ref{qnahm3}), modded out by the complexified gauge
symmetry $G_{\mathbb{C}}$, and where we completely ignore the real
equation (\ref{qnahm4}).  The technical point is that for a particular
choice of $G_{\mathbb{C}}$ gauge, a typical solution of the complex
equations will not solve the real equation, but because (\ref{qnahm4})
transforms under $G_{\mathbb{C}}$, there will be a gauge-equivalent
point which does solve the real equation.  As a practical matter, this
procedure of using $G_{\mathbb{C}}$ and ignoring the real equation
proves to be a drastic technical simplification.

\subsection{Summary of Section \ref{sec:localized}}\label{sec34}

Let us pause to summarize the main results of section
\ref{sec:localized}. In this section, we have explicitly worked out
\begin{itemize}
\item Generalized Nahm equations  (\ref{qnahm1})--(\ref{qnahm4}), 
\item Junction condition for D5 (\ref{dX})--(\ref{QY}) and for D5$^{\ensuremath{\prime}}$ (\ref{dY})--(\ref{QX}),
\item Junction condition for NS5 (\ref{XAB})--(\ref{BYYB}) and for NS5$^{\ensuremath{\prime}}$ (\ref{YAB})--(\ref{BXXB}). 
\end{itemize}

The moduli space of field configurations subject to the boundary
conditions consists of solutions to these equations modulo gauge
equivalences. To make this notion completely precise, we need to
specify the boundary/junction condition also for the gauge parameters.
On D5 junctions with the same number of D3's on each sides, we will
require the gauge parameter to be continuous
\cite{Weinberg:2006rq}. On NS5-branes, where bifundamental degrees of
freedom live, we allow the gauge parameter to be discontinuous and
take arbitrary values on either side.

We have also described how the K\"{a}hler structure comes about naturally
from the complexified gauge formalism. These relations will be the
main ingredient for our subsequent analysis in the remainder of this
paper as well as in the companion paper \cite{ToAppear}.

\section{Boundary Conditions on a Half-Space}

A particularly simple class of boundary conditions are those which are
defined on a half-space; that is, we take the ${\cal N}=4$ theory
defined in $\mathbb{R}^{3,1}$ but restricted to $x^3 = y>0$.  At
$y=0$, we impose some conditions on the bulk fields consistent with
the amount of supersymmetry we wish to preserve.  This proves to be a
simple context in which we can understand the issues which arise from
reducing to supersymmetry to 1/4 BPS.

We will describe two important classes of boundary conditions -- those
which can be constructed from a sequence of D5 and
D5$^{\ensuremath{\prime}}$ branes, which we can think of as a
generalization of Dirichlet boundary conditions, and their S-duals,
which we can construct from a sequence of NS5 and
NS5$^{\ensuremath{\prime}}$-branes.  We will see that the various
boundary conditions impose different constraints on the bulk fields.

The spaces of bulk field configurations allowed by a given boundary
condition are similar to moduli spaces, but to give a truly well-posed
problem for the field configurations on a half-space we need to also
specify a boundary condition at infinity.  In GW \cite{Gaiotto:2008sa}
a canonical choice of boundary condition at infinity was used, namely
that the $\vec X$ fields should be valued in a maximal torus of the
gauge group, with the $\vec Y$ fields all set to zero.  As it turns
out, this boundary condition is equivalent to coupling the theory on a
finite interval to a quiver gauge theory $T[SU(N)]$.  This quiver
gauge theory admits a 1/4 BPS generalization which we will describe.

Throughout this section, we will place D5-branes at the origin of $X_{7,8,9}$ and D5$^\ensuremath{\prime}$-branes at the origin of $X_{4,5,9}$ for simplicity; generic positions can be easily restored.

In this section we start with a 1/2 BPS boundary conditions in section \ref{sec41}.
We then discuss 1/4 BPS boundary conditions in sections \ref{sec42} and \ref{sec43}, starting with 
simple examples and then moving on to more complicated examples.

\subsection{1/2 BPS Boundary Conditions}\label{sec41}

We review briefly some of the boundary conditions discussed in
\cite{Gaiotto:2008sa}.

\subsubsection{Ordinary Dirichlet Boundary Conditions}

\begin{figure}
\centering
\includegraphics[scale=0.8]{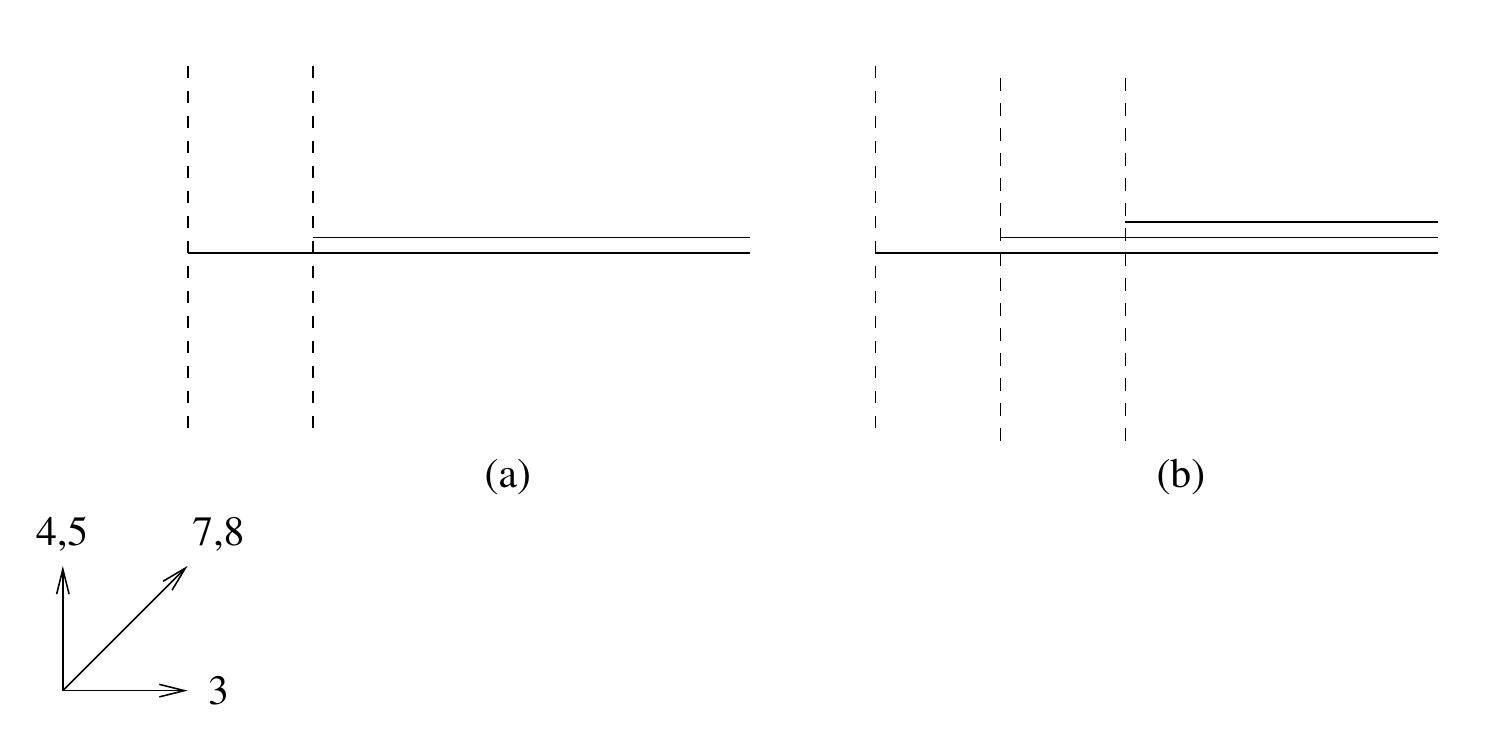}
\caption{Ordinary Dirichlet boundary conditions for $U(2)$ and $U(3)$ gauge theory in the bulk.}
\label{dirichlet}
\end{figure}

First, let us recall what GW called the ``ordinary'' Dirichlet boundary condition for $U(N)$ $\mathcal{N}=4$ SYM.  For the theory realized on $N$ D3-branes, it corresponds to having a stack of $N$ D5-branes, with one D3-brane ending on each D5-brane.  For $N=2$ and $N=3$, the corresponding configuration is shown in figure \ref{dirichlet}.
We assume that the fields are non-singular throughout the half-space
$y>0$, so that we can choose the $G_{\mathbb C}$ gauge $\A=0$.  The
conditions imposed on the scalars $\X$ and $\Y$ are that $\X$ takes
any finite value while $\Y=0$.

This boundary condition plays a central role in the discussion of
S-duality in \cite{Gaiotto:2008ak}.  There it was called an
``ungauging,'' because it removes the gauge fields at low energies but
imposes no other constraint on the scalar field $\X$.

\subsubsection{1/2 BPS $U(2)$ Boundary with a Pole}

\begin{figure}
\centering
\includegraphics[scale=0.8]{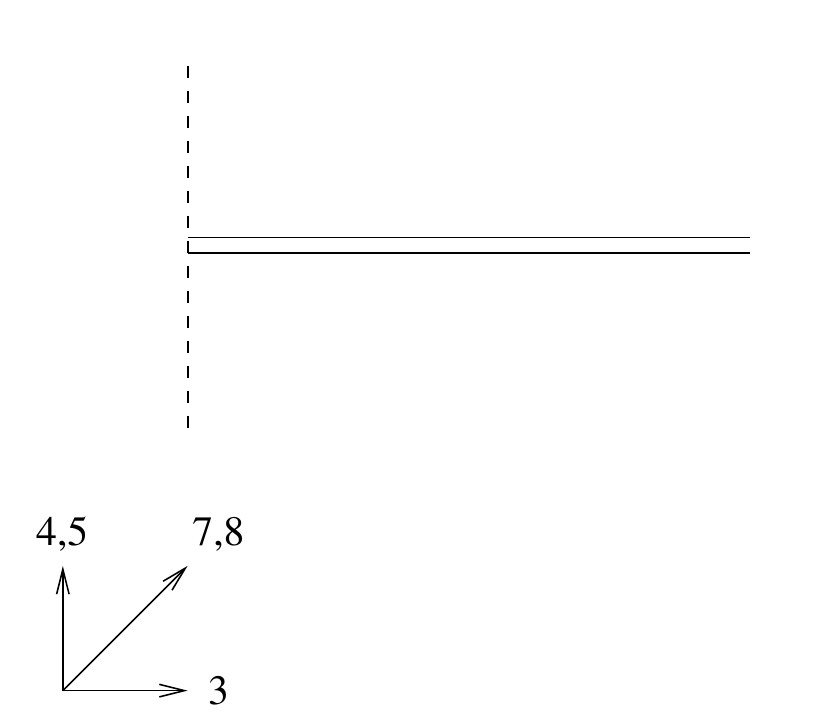}
\caption{$U(2)$ with D5-like boundary giving rise to an elementary Nahm pole.}
\label{dirichletU2}
\end{figure}

The simplest example where we allow a Nahm pole singularity arises for a bulk gauge group $U(2)$. This is the case illustrated in figure \ref{dirichletU2}.

In the analysis of Dirichlet boundary conditions, we chose the gauge
$\mathcal{A} =0$.  If the boundary has a pole we cannot choose this
gauge because we only allow non-singular gauge
transformations. However, we are allowed to gauge away all the
non-singular terms in $\mathcal{A}$ , leaving us with
\beq
\mathcal{A} =\frac{ it_3}{y} \ ,
\eeq
where $t_3$ is the Cartan generator of an $\mathfrak{sl}(2)$ subalgebra in the gauge algebra.  We satisfy Nahm's equations with\footnote{
The matrices $t_a=\frac{i \sigma_a}{2}$ are anti-Hermitian (recall  the $X_a$ are anti-Hermitian in our conventions) 
and satisfy $[t_a, t_b]=-\epsilon_{abc} t_c$. We define $t_{\pm}=t_1\pm i t_2=
\frac{i}{2}(\sigma_1\pm i \sigma_2)$, which satisfy
$\left[ t_+, t_-\right] = 2 i t_3$. }
\beq
\mathcal{X} = \frac{t_+}{y}  + ({\rm nonsingular}) \ .
\eeq
The normalization of the singular part is determined by the singular terms in the third Nahm equation.

We need to solve the BPS equation $\mathcal{D}\X=0$.
There are residual gauge transformations satisfying $g(0)=1$; after modding these out we can put $\X$ in the form
\beq
\mathcal{X} = \left( \begin{array}{cc} a & 1/y \\
by & a \end{array} \right) \ .
\eeq
We are left with two complex parameters, $a$ and $b$.  

From the action of S-duality in string theory, it is clear that the Nahm pole boundary condition is S-dual to a D3-brane ending on an NS5-brane, which gives ordinary Neumann boundary conditions (with no added degrees of freedom.)

\subsection{1/4 BPS Dirichlet Boundary Conditions}\label{sec42}

The simplest generalization of the 1/2 BPS Dirichlet boundary
conditions is obtained by rotating some of the D5-branes to
D5$^\ensuremath{\prime}$-branes.  Despite their simplicity, these
boundary conditions will already illustrate an important point --- unlike
the 1/2 BPS case, there is no canonical ordering of the 5-branes in
the mixed D5-D5$^\ensuremath{\prime}$ system.  This means, in
particular, that these boundary conditions do not have a simple
classification in terms of Young diagrams of the type illustrated in
figure \ref{figa}.

We consider some concrete examples for $U(2)$ and $U(3)$ gauge theory in the bulk.  We begin with boundary conditions without Nahm poles; for these we can consistently choose the $G_{\mathbb{C}}$ gauge $\mathcal{A}=0$.  Then the bulk equations are trivial except for the commutator $[\X,\Y]=0$.

\subsubsection{D5-D5$^\ensuremath{\prime}$ Boundary Conditions for $U(2)$ Gauge Theory}

\begin{figure}
\centering
\includegraphics[scale=0.8]{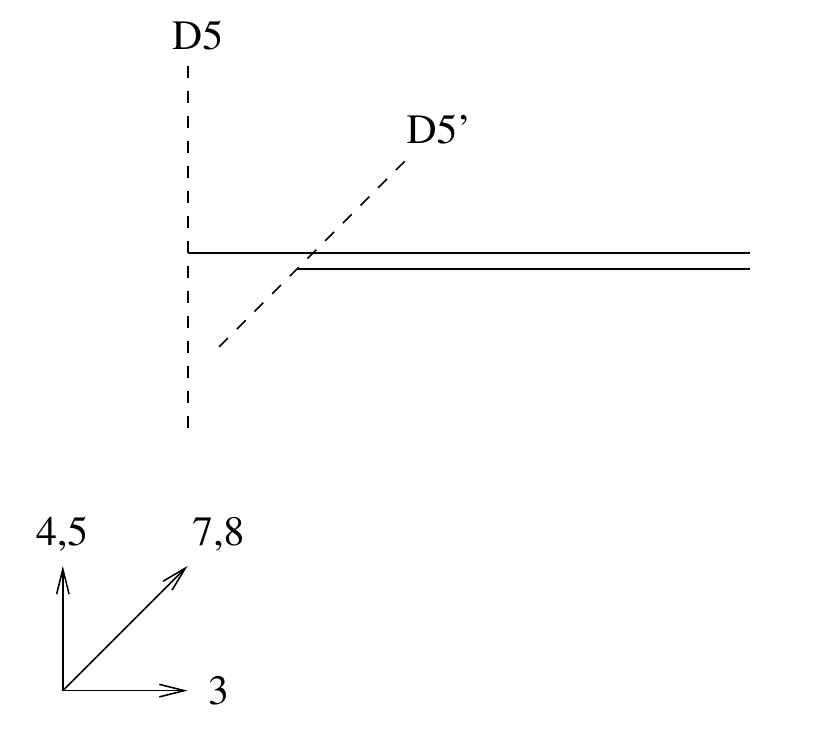}
\caption{Dirichlet boundary conditions for D5
and D5$^\ensuremath{\prime}$
branes in $U(2)$ gauge theory.}
\label{ddprimeu2}
\end{figure}

A natural class of 1/4 BPS generalizations of the ordinary Dirichlet
boundary condition can be obtained by rotating some of the D5-branes
to D5$^\ensuremath{\prime}$-branes; it is particularly interesting
because, like the 1/2 BPS ordinary Dirichlet condition, it acts on the
vectors as an ungauging.  The simplest such boundary condition arises
in $U(2)$ $\mathcal{N}=4$ gauge theory.  The brane configuration is
shown in figure \ref{ddprimeu2}.

We construct the boundary conditions from left to right.  Beginning
with the leftmost gauge theory segment, we have
\beq
\mathcal{X} = \left(a\right) \ , \qquad \qquad \mathcal{Y} = \left(0 \right) \ ,
\eeq
Proceeding to the $U(2)$ region, $\mathcal{X}$ grows an extra row and
column which vanish:
\beq
\mathcal{X} = \left(\begin{array}{cc} a & 0\\
0 & 0
\end{array} \right) \ ,
\eeq
but $\mathcal{Y}$ grows without constraint:
\beq
\mathcal{Y} = \left(\begin{array}{cc} 0 & b\\
c & d
\end{array} \right).
\eeq
Now we impose $[\mathcal{X},\mathcal{Y}] =0$ to find
\beq
b a = c a = 0 \ ,
\eeq
and there are two solution branches.  The first branch has $b=c=0$:
\beq
\mathcal{X} &=& \left(\begin{array}{cc} a & 0\\
0 & 0
\end{array} \right) \ , \label{firstbranch1}\\
\mathcal{Y} &=& \left(\begin{array}{cc} 0 & 0\\
0 & d
\end{array} \right) \ . \label{firstbranch2}
\eeq
For this branch, we have only diagonal elements. 

The other branch has  $a=0$ with no constraints on $b,c,d$:
\beq
\mathcal{X} &=& \left(\begin{array}{cc} 0 & 0\\
0 & 0
\end{array} \right) \ ,\label{secondbranch1} \\
\mathcal{Y} &=& \left(\begin{array}{cc} 0 & b\\
c& d
\end{array} \right) \ .\label{secondbranch2}
\eeq

\begin{figure}
\centerline{\includegraphics[scale=0.8]{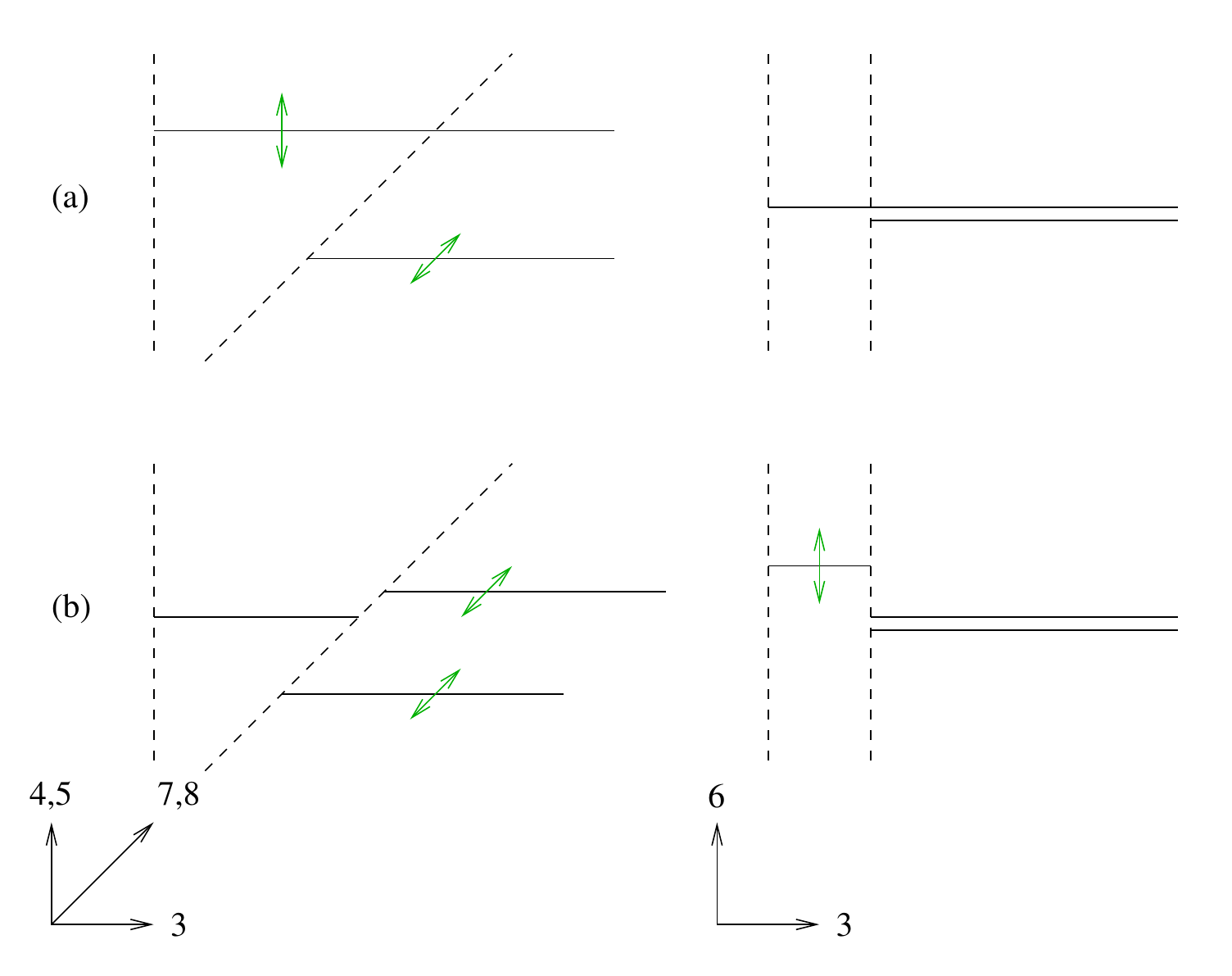}}
\caption{Brane configuration associated with the branch
  (\ref{firstbranch1})--(\ref{firstbranch2}) on top, and the branch
  (\ref{secondbranch1})--(\ref{secondbranch2}) on the bottom.  The
  positions of the semi-infinite D3 branes in the $X_6$ direction are obscured in the
  analysis in the main text.  They can be recovered by finding the $G_{\mathbb C}$ gauge transformation
  which solves the real part of the Nahm equations.
\label{figbb}}
\end{figure}

It is instructive to interpret the field configurations for $\X$ and
$\Y$ in terms of the allowed motions of branes, as shown in in figure
\ref{figbb}.  We have indicated each adjustable complex degree of
freedom with a green double arrow. We can see that how two parameters,
$a$ and $d$ affect the first branch and three parameters, $b$, $c$,
and $d$ affect the second branch. Some of these parameters affect the
configuration of the brane far from the boundary.  The data encoding
the position of the semi-infinite D3 branes in the $X_6$ direction are
not clearly reflected in this presentation because of our choice of
$G_{\mathbb C}$ gauge.  In particular, note that (\ref{secondbranch2})
is not diagonal.  The $X_6$ data are encoded in the off-diagonal
terms, and can be extracted by choosing a different gauge.  In general
these data are not totally geometric, because if the ``physical''
gauge has $\A\neq 0$, the scalars will not necessarily commute.

Of course, interchanging the D5 and D5$^\ensuremath{\prime}$
corresponds to interchanging $\X$ and $\Y$, and this illustrates an
important point -- after the exchange of the D5 and
D5$^{\ensuremath{\prime}}$, different boundary conditions are imposed
on the bulk fields.  In general, this will give rise to a phase
transition in the moduli space of vacua.

\subsubsection{$U(3)$ Examples without Poles}

\begin{figure}
\centering
\includegraphics[scale=0.8]{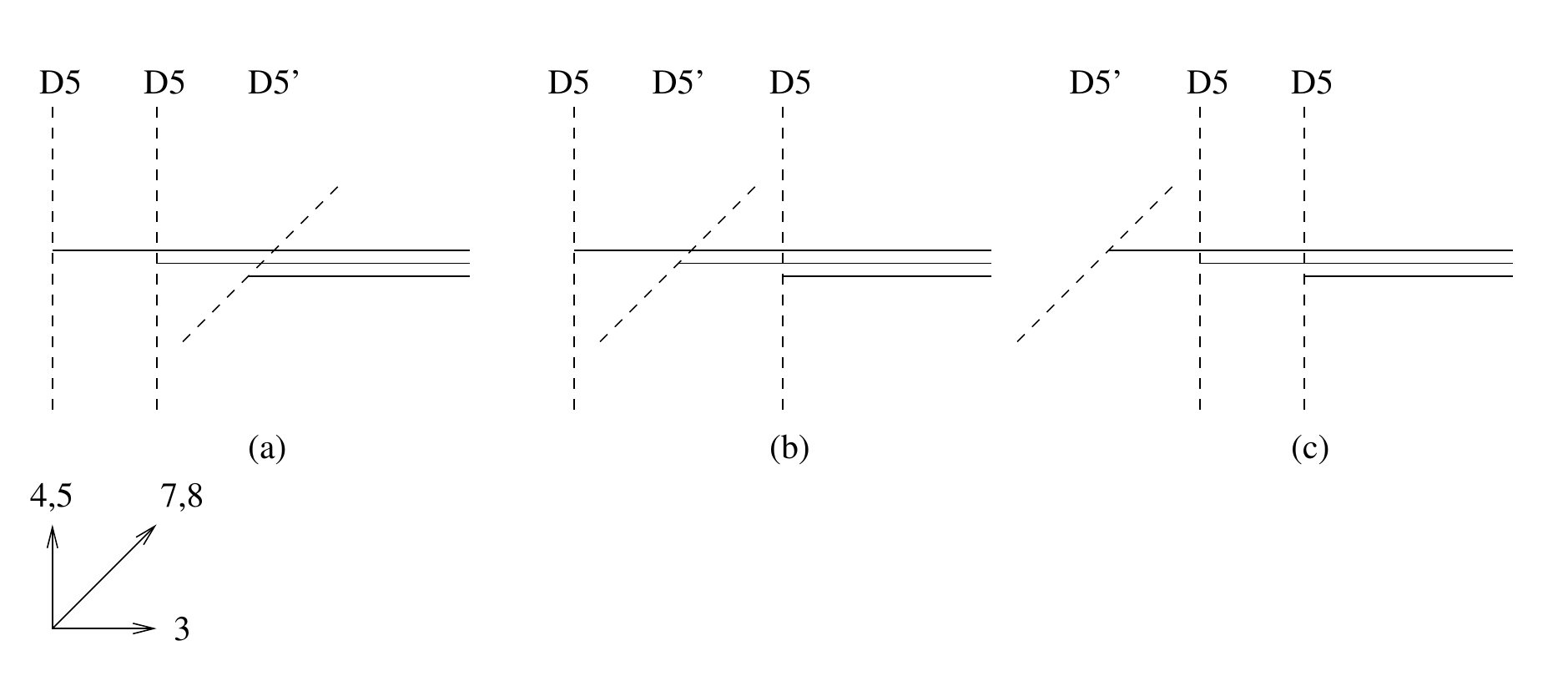}
\caption{Possible arrangements of D5 and D5$^\ensuremath{\prime}$
  branes boundary conditions for $U(3)$ gauge theory on a half-space.}
\label{ddprime}
\end{figure}

Now we turn to the examples in figure \ref{ddprime}. 
These analyses are included mainly to illustrate our methods to
broader set of examples.  We consider (a), (b), and (c) in turn.

\noindent {\bf D5-D5-D5$^\ensuremath{\prime}$}

\begin{figure}
\centerline{\includegraphics[scale=0.8]{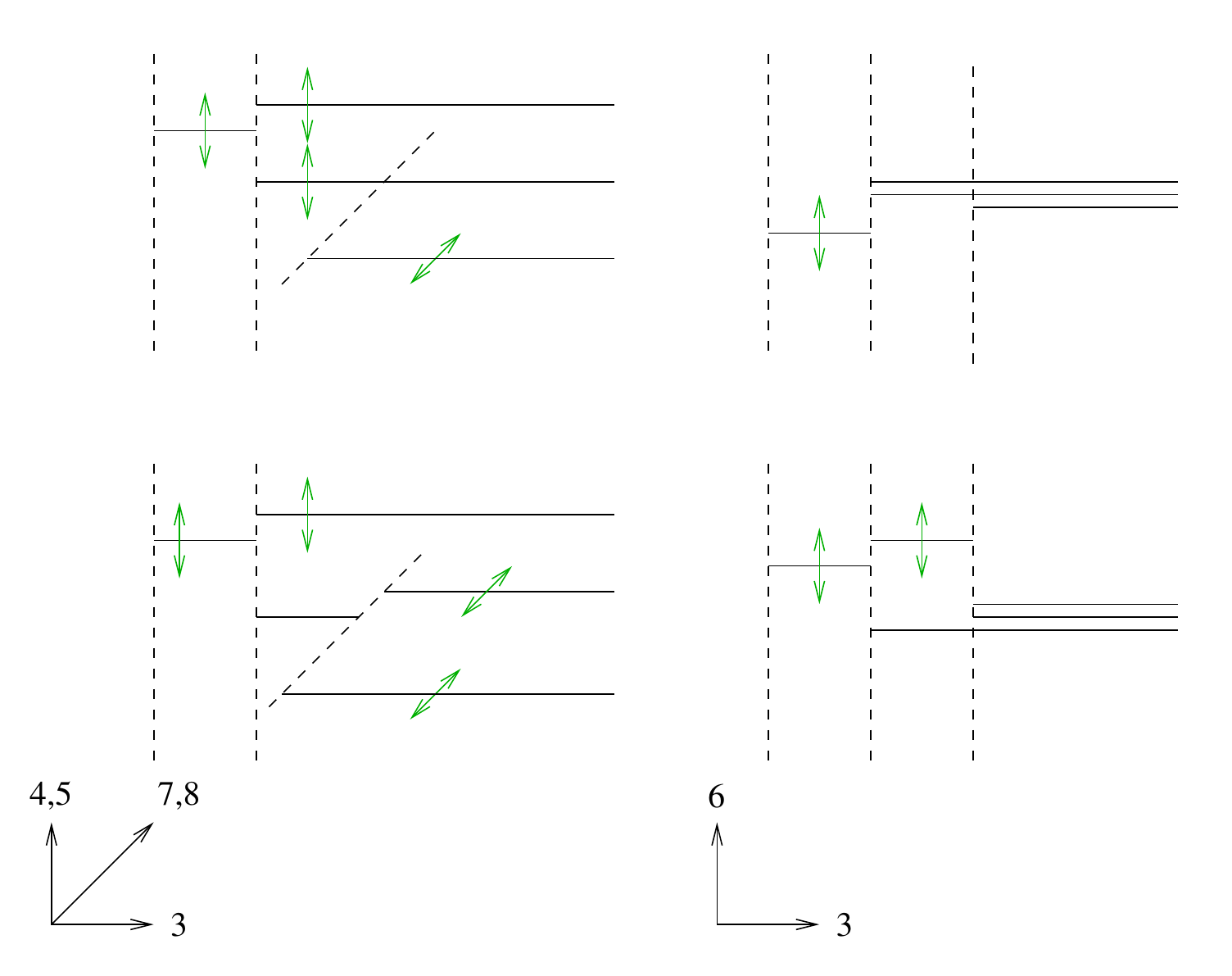}}
\caption{Allowed brane motions for the 2 solution branches of the D5-D5-D5$^{\ensuremath{\prime}}$ configuration. The top configuration has 5 adjustable parameters and the bottom has six, shown by green arrows.  \label{figbbb2}}
\end{figure}

Starting from the left, we have a $U(2)$ gauge theory with ordinary Dirichlet boundary conditions
\beq
\mathcal{X} &=& \left(\begin{array}{cc} a & b\\
c & d
\end{array} \right) \ , \\
\mathcal{Y} &=& \left(\begin{array}{cc} 0 & 0\\
0 & 0
\end{array} \right)  \ .
\eeq
Crossing the D5$^\ensuremath{\prime}$-brane, these become
\beq
\mathcal{X} &=& \left(\begin{array}{ccc} a & b & 0\\
c & d & 0 \\
0& 0& 0
\end{array} \right) \ , \\
\mathcal{Y} &=& \left(\begin{array}{ccc} 0 & 0& u\\
0 & 0 & v\\
r & s & t
\end{array} \right) \ .
\eeq
The commutation $[\mathcal{X},\mathcal{Y}] =0$ then gives the equations
\beq
\left(\begin{array}{cc} a & b\\
c & d
\end{array} \right) 
\left(\begin{array}{c} u \\ v \end{array}\right) &=& 0 \ ,\label{sec7.1eq1}\\
\left( \begin{array}{cc} r & s \end{array}\right) \left(\begin{array}{cc} a & b\\
c & d
\end{array} \right) &=& 0 \ .\label{sec7.1eq2}
\eeq
This gives rise to distinct branches of field configurations (see
figure \ref{figbbb2}). If $u=v=r=s=0$, there are 5 unconstrained
coordinates $a,b,c,d,t$.  On the other hand, if any of $u,v,r,s$ are
nonvanishing, then we have the constraint $ad-bc=0$, giving rise to
total of six parameters: one of $u$ or $v$, one of $r$ or $s$, $t$,
and three of $a$, $b$, $c$, and $d$.

\noindent {\bf D5-D5$^\ensuremath{\prime}$-D5}

In the $U(2)$ region we have
\beq
\mathcal{X} &=& \left(\begin{array}{cc} a & 0\\
0 & 0
\end{array} \right) \ , \\
\mathcal{Y} &=& \left(\begin{array}{cc} 0 & b\\
c & d
\end{array} \right) \ .
\eeq

\begin{figure}
\centerline{\includegraphics[scale=0.8]{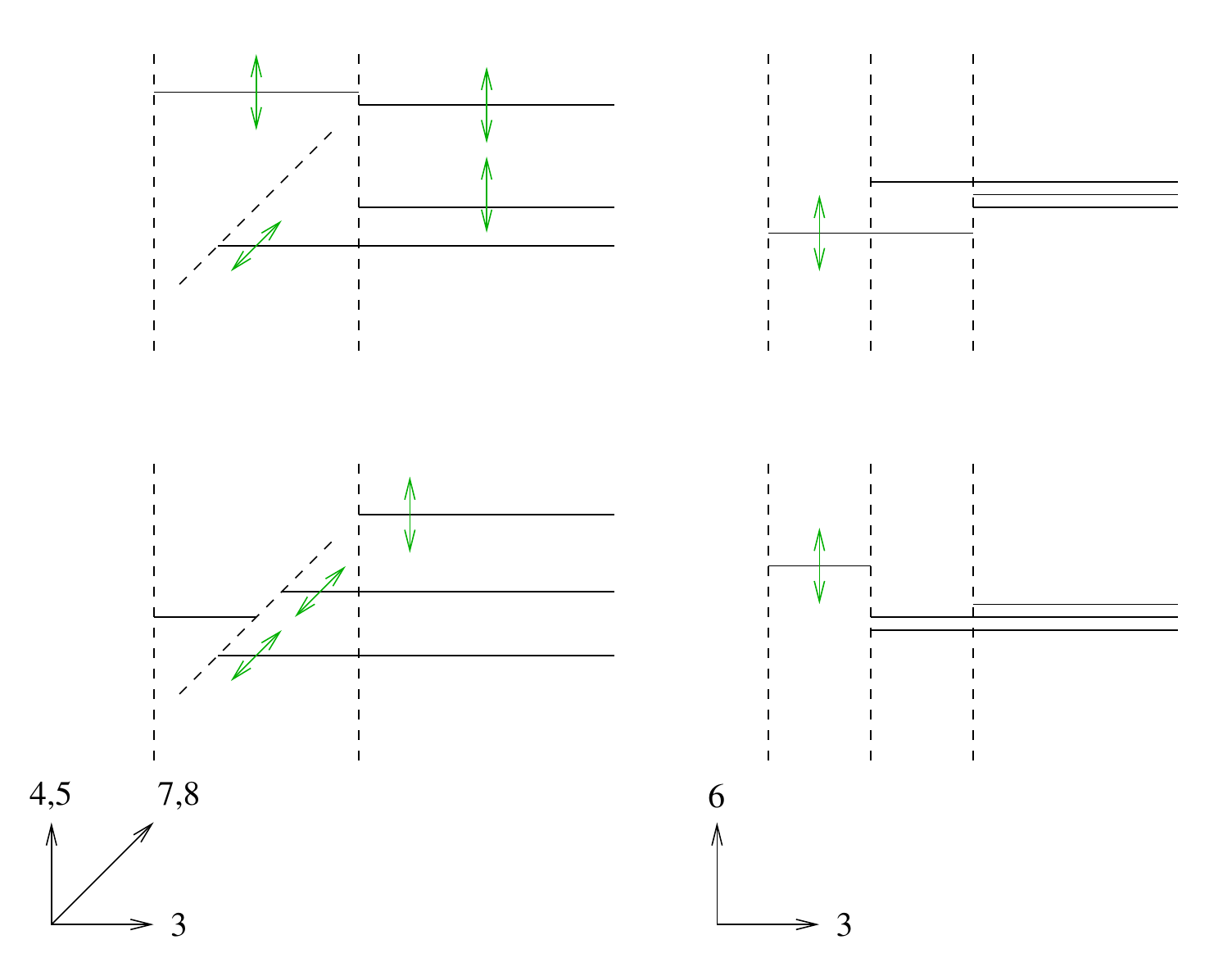}}
\caption{Brane configuration for the 2 solution branches of the D5-D5$^{\ensuremath{\prime}}$-D5 configuration. The top configuration has five adjustable parameters and the bottom has four, shown by green arrows.  \label{figbbb}}
\end{figure}

Crossing the D5 brane on the right, we have
\beq
\mathcal{X} &=& \left(\begin{array}{ccc} a & 0& r\\
0 & 0 & s\\
u& v& t
\end{array} \right) \ ,\\
\mathcal{Y} &=& \left(\begin{array}{ccc} 0 & b & 0\\
c & d & 0\\
0 & 0 & 0
\end{array} \right) \ .
\eeq
The commutation relation $[\mathcal{X},\mathcal{Y}] =0$ implies
\beq
\left(\begin{array}{ccc}  & b a& -b s\\
-ca & 0 & -c r- sd\\
cv & b u +vd &0
\end{array} \right)= 0  \ .
\eeq
There are various branches. One involves setting $b=c=v=s=0$ giving
rise to a five dimensional branch parameterized by $a$, $d$, $u$, $r$,
and $t$, illustrated in figure \ref{figbbb} on top.  Another involves
setting $a=r=s=u=v=0$ giving rise to a four dimensional branch
parameterized by $b$, $c$, $d$, and $t$, illustrated in figure
\ref{figbbb} on bottom.

\noindent {\bf D5$^\ensuremath{\prime}$-D5-D5}

\begin{figure}
\centerline{\includegraphics[scale=0.8]{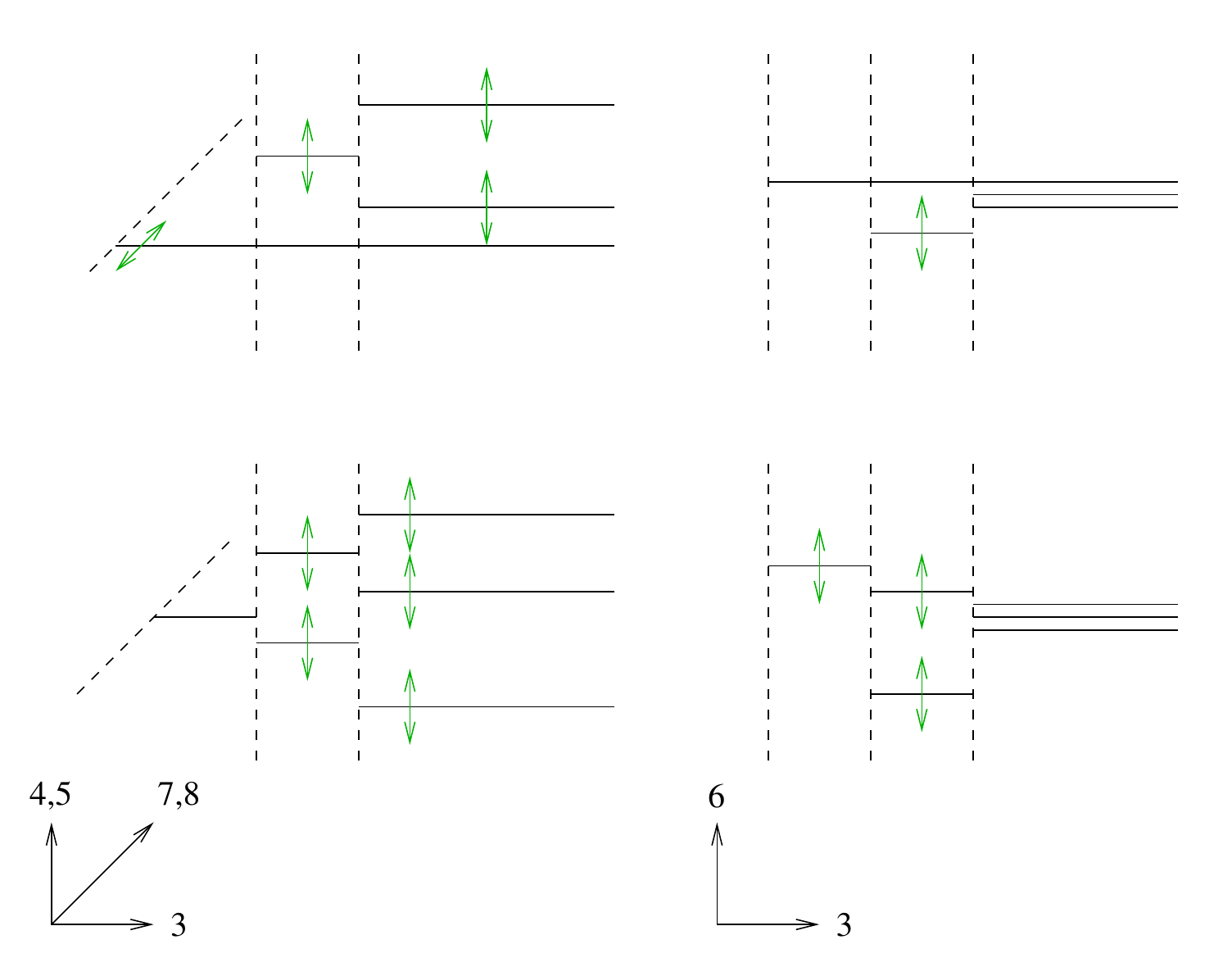}}
\caption{Brane configurations for the 2 branches of the D5$^{\ensuremath{\prime}}$-D5-D5 system. The top configuration with $a \ne 0$ has five adjustable parameters and the bottom with $a=0$ has eight, indicated by green arrows.  \label{figbbbb}}
\end{figure}

This is the easiest case for $U(3)$.  One component of $\mathcal{X}$ is fixed to a fiducial value, while for the rest of the components we have
\beq
\mathcal{X} &=& \left( \begin{array}{ccc}0 & b & c \\ d & e& f \\ g& h & j \end{array}\right) \ ,\\
\mathcal{Y} &=& \left( \begin{array}{ccc} a & & \\ & 0 & \\ & & 0\end{array}\right).
\eeq
If $a\neq 0$, we are forced to set $b=c=d=g=0$, and so we
  have a five dimensional space parameterized by $a$, $e$, $f$, $h$,
  and $j$.  If $a=0$, there are no additional constraints on
  $\mathcal{X}$ and so it is parameterized by eight variables $b$, $c$,
  $d$, $e$, $f$, $g$, $h$, and $j$. The brane configuration associated
  with these deformation is illustrated in figure \ref{figbbbb}.

\subsubsection{1/4 BPS Pole Boundary Conditions\label{poleBC}}

\begin{figure}
\centering
\includegraphics[scale=0.8]{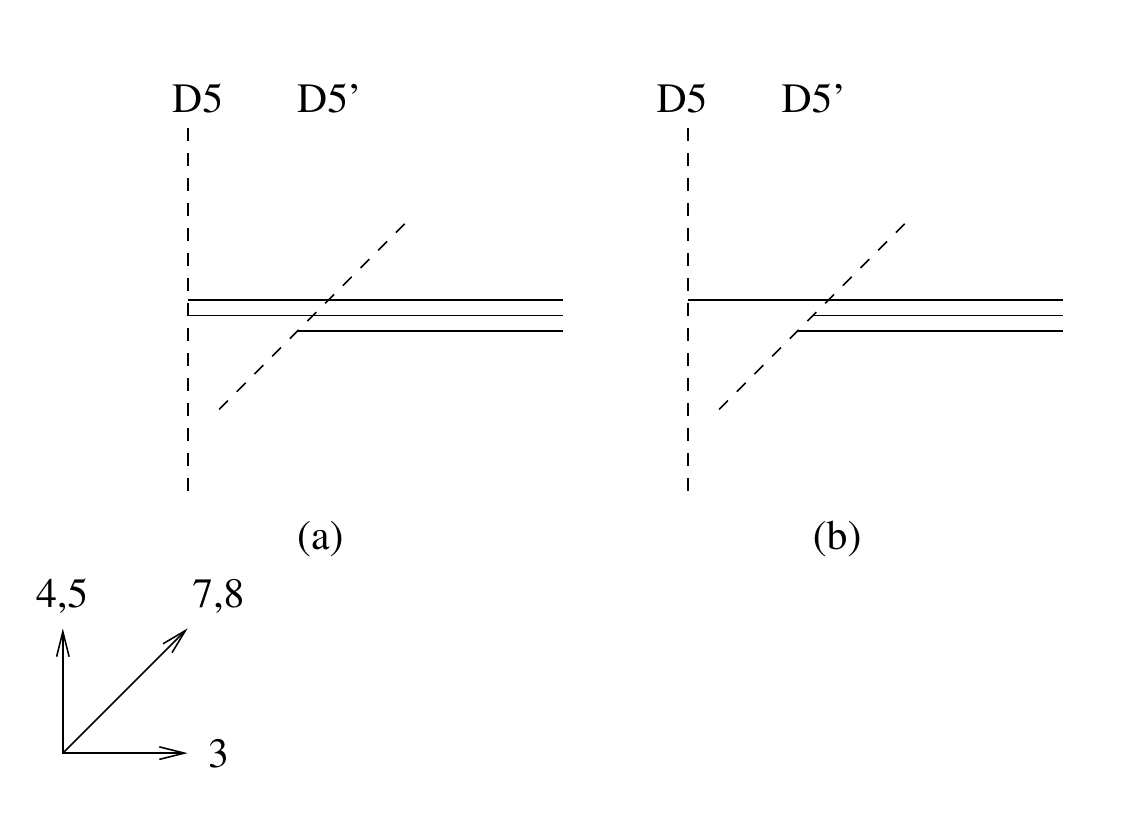}
\caption{1/4 BPS $U(3)$ boundary conditions with poles.}
\label{u3poles14bps}
\end{figure}

When we allow both D5 and D5$^\ensuremath{\prime}$ branes we can also have 1/4 BPS configurations with poles, as shown in figure \ref{u3poles14bps}. 
We want to generalize the procedure for handling poles to 1/4 BPS.  There are a few subtle points.  We will now have two matrices $\mathcal{X}$ and $\mathcal{Y}$.  Therefore we have to impose singularity constraints on both of them.  We will also have to be careful in modding out by residual gauge transformations, which will act on both $\mathcal{X}$ and $\mathcal{Y}$.

First we consider case (a).  We begin in the $U(2)$ region, $0< y< y_1$, where $y_1$ is the position of D5$^{\ensuremath{\prime}}$ in the $y$-direction.  According to our rules for treating D5 and and D5$^\ensuremath{\prime}$ boundary conditions, we should set $\mathcal{Y}  =0$.  We also have
\beq
\mathcal{A} = 
\frac{1}{2y}\left( \begin{array}{ccc} 1& 0 & 0 \\ 0& -1 & 0 \\0 & 0 & 0\end{array}\right) \ ,
\eeq
so that we will solve all the constraints in the $U(2)$ region with 
\beq
\mathcal{X} &=& \left( \begin{array}{ccc} a& 1/y & \\ by & a & \\ &  &  \end{array}\right) \ , \\
\mathcal{Y} &=& \left( \begin{array}{ccc} 0& 0 & \\ 0 & 0& \\ &  &  \end{array}\right)  \ .
\eeq
Proceeding to the $U(3)$ region, $y>y_1$, we should put zeros in $\mathcal{X}$ off the diagonal while $\mathcal{Y}$ grows in all of the new entries; at the boundary we may choose any values for these entries but they also have a $y$ dependence given by the complex equation $\mathcal{D} \mathcal{Y} = 0$:
\beq
\mathcal{X} &=& \left( \begin{array}{ccc} a& 1/y & 0\\ by & a &0 \\ 0& 0 & 0 \end{array}\right) \ , \\
\mathcal{Y} &=& \left( \begin{array}{ccc} 0& 0 & uy^{-1/2}\\ 0 & 0 & vy^{1/2} \\ ry^{1/2}& sy^{-1/2} & t \end{array}\right)  \ .
\eeq
Of course the $y^{-1/2}$ terms are nonsingular because they are defined only for $y\ge y_1>0$.

\begin{figure}
\centerline{\includegraphics[scale=0.8]{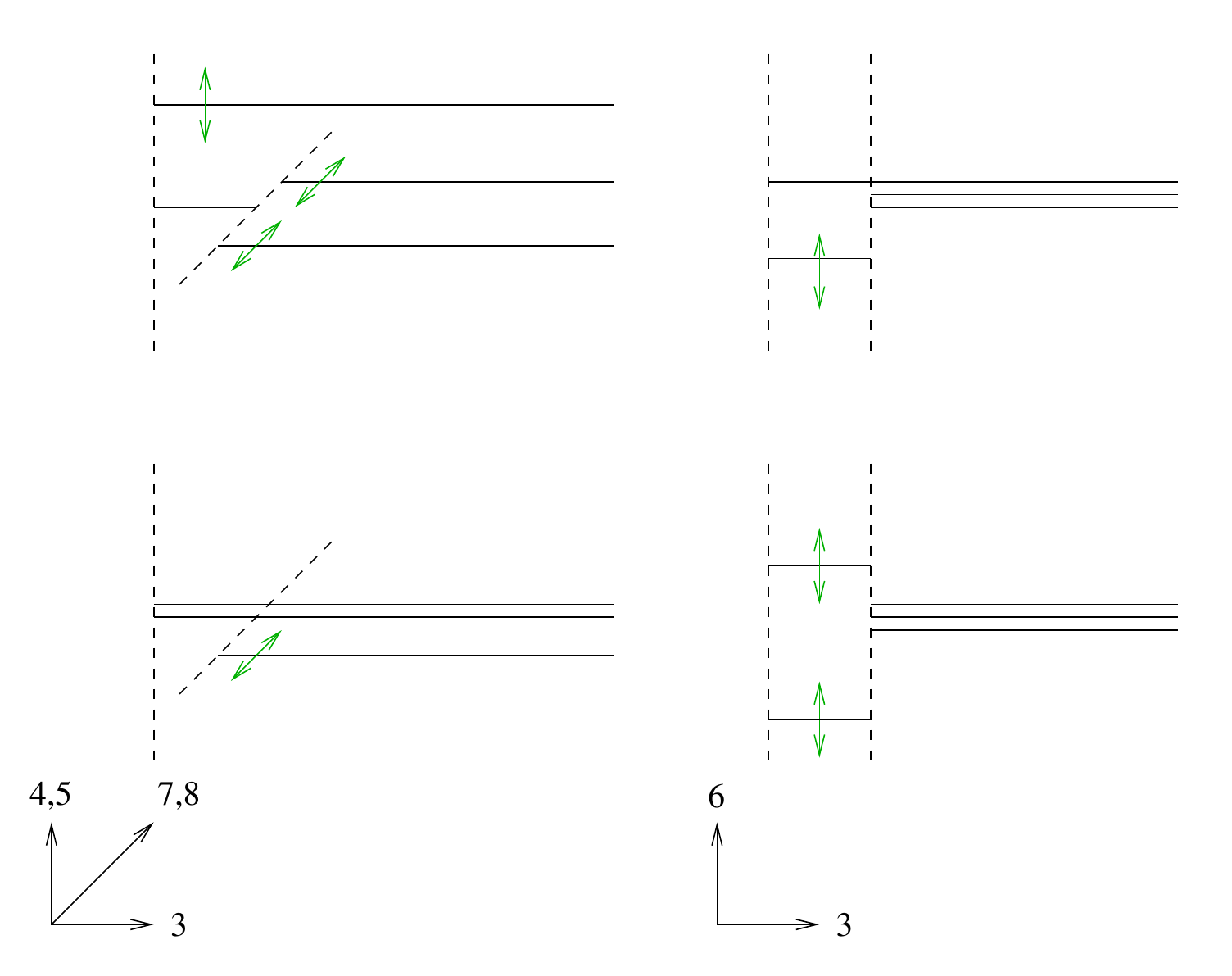}}
\caption{
Allowed brane motions
for the two branches associated with the  configuration of figure \ref{u3poles14bps}.a. The top configuration  has four adjustable parameters and the bottom with  has three, indicated by green arrows.  \label{figd52d3d5p3d3}}
\end{figure}

We are not done because we still have to impose the 1/4 BPS commutator equation $[\mathcal{X}, \mathcal{Y}] = 0$.  This gives the equations
\beq
au  + v  &=& 0 \ ,\\
av  + bu  &=& 0 \ ,\\
ar + bs  &=& 0 \ ,\\
as  + r  &=& 0 \ .
\eeq
This system of equations has solutions when
\beq
b=a^2 \ ,
\eeq
with $u$ determined by $v$ and $r$ determined by $s$. In
  this case the configurations are parameterized by four variables,
  $a$, $v$, $s$, and $t$. The other possiblity is for $a=b=0$ which
  also forces $r=v=0$ giving rise to a three parameter branch of
  solutions parameterized by $u$, $s$, and $t$, which we illustrate
  in figure \ref{figd52d3d5p3d3}.

Note that on this branch, if we try to take the limit $y_1 \rightarrow
0$ we encounter singular terms proportional to $y^{-1/2}$. (For finite
brane separations, this is not singular because it is evaluated at
$y_1$.)  In the 1/2 BPS analysis of \cite{Gaiotto:2008sa}, a similar
singularity was found for the analogous brane configuration with only
D5-branes. They pointed out that this configuration can be related to
the 1/2 BPS version of figure \ref{u3poles14bps} accompanied by some
decoupled free sectors by changing the ordering of the branes, and
argued that they are therefore redundant in their classification
scheme. Since branes can not be reordered with similar control in the
case with 1/4 BPS, and since the way in which decoupled sectors arise
is generally more intricate \cite{Kapustin:2010xq,Safdi:2012re}, it
seems less convincing to attribute the singularity here as a signature
of some free decoupled sector.

Another noteworthy feature of the three dimensional branch illustrated
on the bottom of figure \ref{figd52d3d5p3d3} is the fact that the
positions of three semi-infinite D3's along the ${\cal Y}=X_7+i X_8$
coordinates are not the most general ones allowed from the
consideration of the brane configurations. In fact, only one of the
three semi-infinite D3-branes are allowed to move in the ${\cal Y}$
direction.  This is one of the manifestations of the non-abelian
dynamics which we will explore in greater detail in the follow up
paper \cite{ToAppear}.

Next we consider case (b).  Here the pole is at $y=y_1$ and we set
\beq
\mathcal{A} = 
\frac{1}{2(y-y_1)}\left( \begin{array}{ccc} 1& 0 & 0 \\ 0& -1 & 0 \\0 & 0 & 0\end{array}\right)
 \ .
\eeq
We must also have (using our rules for crossing D5 and D5$^\ensuremath{\prime}$)
\beq
\mathcal{X} = \left( \begin{array}{ccc} 0& 0 & 0 \\ 0& 0 & 0 \\0 & 0 & a\end{array}\right) \ ,
\eeq
(we have reversed the ordering of rows and columns) while
$\mathcal{Y}$ has one element fixed at the boundary and one element
fixed by the singularity structure:
\beq
\mathcal{Y} = \left( \begin{array}{ccc} 0 & (y-y_1)^{-1} & \\ & 0 &  \\ &  & 0\end{array}\right) \ ,
\eeq
and the empty elements are still to be filled in. Solving the complex
Nahm equation and modding out by $G_C$ requires
\beq
\mathcal{Y} = \left( \begin{array}{ccc}d & (y-y_1)^{-1} &u(y-y_1)^{-1/2} \\e (y-y_1)  & d & v(y-y_1)^{1/2} \\ r(y-y_1)^{1/2}& s(y-y_1)^{-1/2} & 0\end{array}\right) \ .
\eeq
We see that there are unwanted singular coefficients at $y=y_1$ off
the diagonal.  These have to be set to zero:
\beq
\mathcal{Y} = \left( \begin{array}{ccc}d & (y-y_1)^{-1} &0 \\e (y-y_1)  & d &v(y-y_1)^{1/2} \\r(y-y_1)^{1/2}& 0 & 0\end{array}\right) \ ,
\eeq
and we should impose $[\mathcal{X}, \mathcal{Y}] = 0$.  There is a
branch of configurations with $a=0$ and $d,e,r,v$ unconstrained, and
another branch with $v=r=0$, and $a, d,e$ unconstrained.

\begin{figure}
\centerline{\includegraphics[scale=0.8]{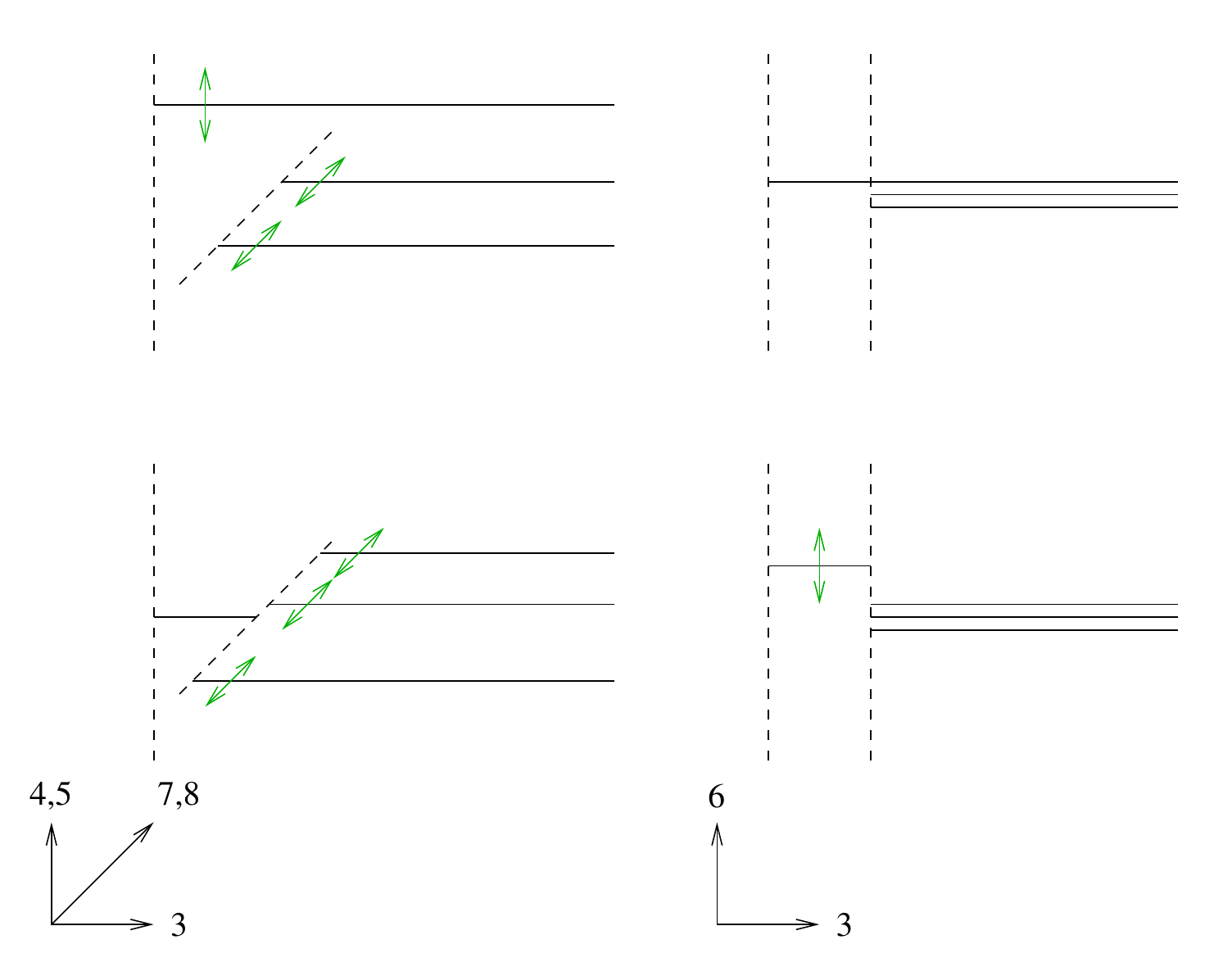}}
\caption{Allowed brane motions
for the two branches associated with the brane configuration 
of figure \ref{u3poles14bps}.b. The top configuration  has three adjustable parameters and the bottom with  has four, indicated by green arrows.  \label{figd5d3d5p3d3}}
\end{figure}

This configuration does have a simple $y_1 \rightarrow 0$ limit,
because the boundary conditions forced us to drop the potentially
singular terms.

\subsection{$T[SU(N)]$ and Its 1/4 BPS Generalizations}\label{sec43}

\begin{figure}
\centerline{\includegraphics[scale=0.8]{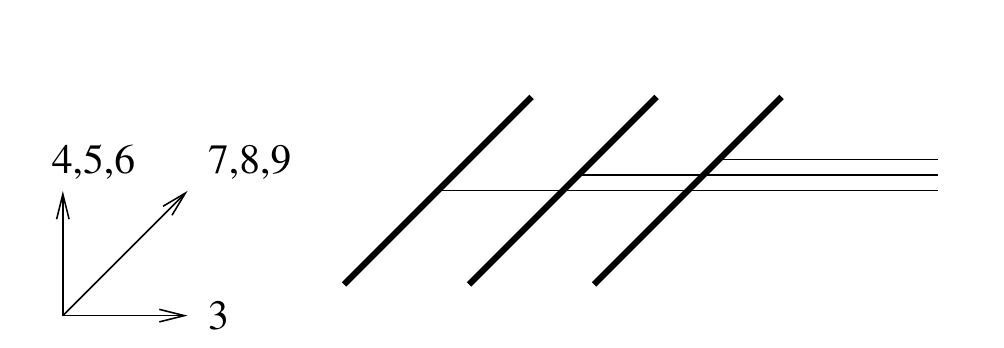}}
\caption{Basic configuration  of the boundary theory generally referred to as $T(SU(N)$. Here, the case for $N=3$ is illustrated. \label{figTsuN}}
\end{figure}

In this subsection, we consider boundaries made out of NS5-branes in
place of the D5-branes which appeared in the previous subsection.
Typical boundaries which arise in this way are illustrated in figure
\ref{figTsuN}.  They are related to the analysis of the previous
section by S-duality. We will first review the case where 1/2 of the
supersymmetry is preserved, and then proceed to generalize to the 1/4
BPS case.

\subsubsection{$T[SU(N)]$}

We can understand the NS5/NS5$^\ensuremath{\prime}$ boundary condition in terms fo
a coupling  of  the bulk $\mathcal{N}=4$ SYM theory
to a boundary $\mathcal{N}=2$ theory via bifundamentals.
Note that the bifundamental couplings can look different depending on
whether the right-most boundary is NS5 or NS5$^\ensuremath{\prime}$
(unlike 3$d$ $\mathcal{N}=4$, the bifundamental coupling is not
universal when only four supersymmetries are preserved.)

In the 1/2 BPS case, the Nahm pole boundary conditions ($N$ D3-branes
ending on a D5) are S-dual to ordinary Neumann boundary conditions
($N$ D3-branes ending on NS5.)  The ordinary Dirichlet boundary
conditions, on the other hand, are S-dual to a coupling to a quiver
gauge theory which GW called $T[SU(N)]$, corresponding to $N$
D3-branes ending on $N$ NS5-branes. It is not hard to see that
coupling to $T[SU(N)]$ fixes the characteristic polynomial of
$\mathcal{X}$.

\noindent {\bf NS5-NS5}

\begin{figure}
\centerline{\includegraphics[scale=0.8]{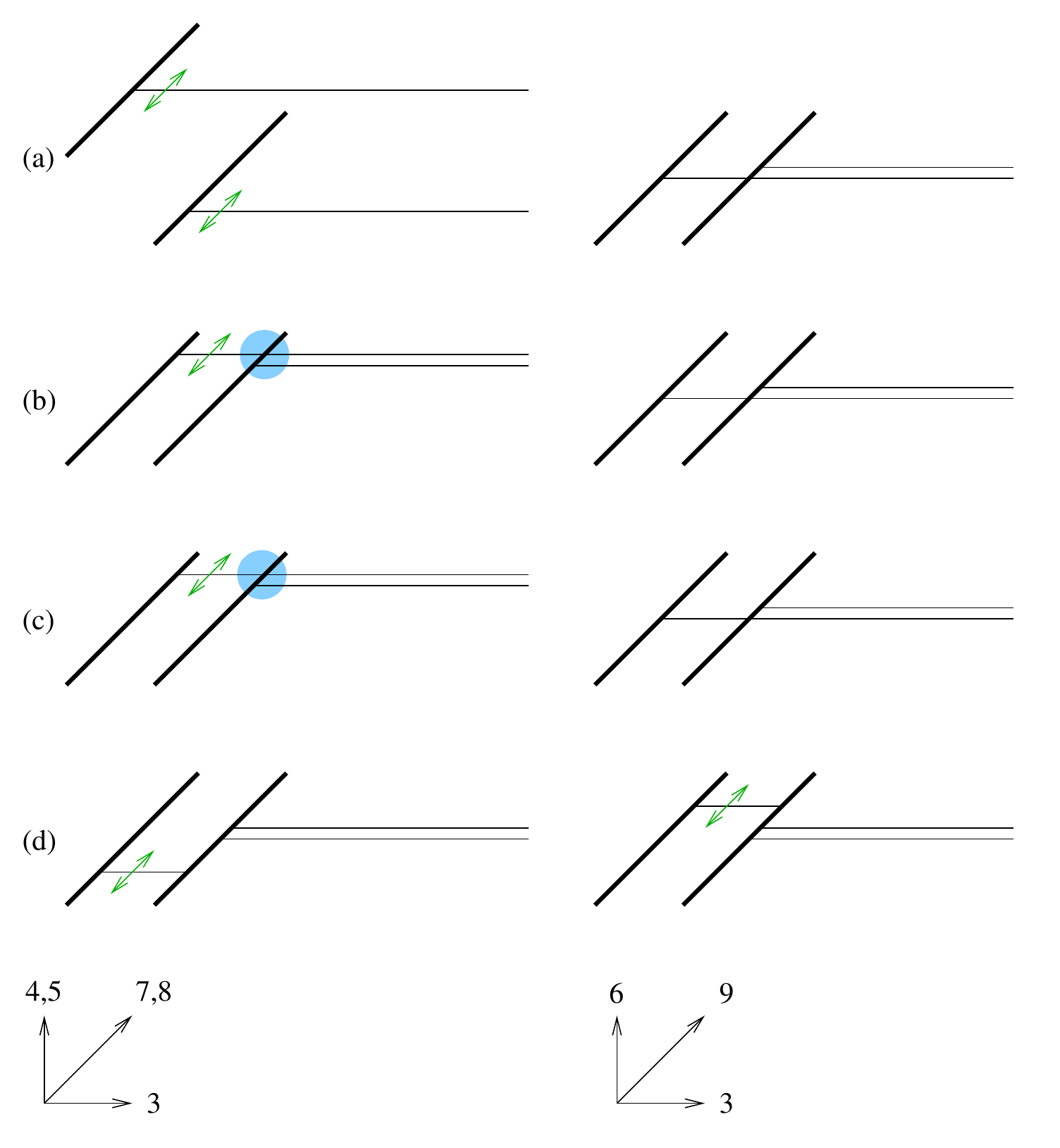}}
\caption{Brane configuration for $T[SU(2)]$ with distinct FI terms,
  and three case with coincident FI terms. The blue disk indicates one
  complex degree of freedom associated with a deformation localized at
  the defect which does not deform the branes geometrically in ${\cal
    X}$ and ${\cal Y}$ directions, but in some cases may affect the
  $X_6$ embedding (not drawn in figure). The blue disk also has the
  effect of forcing the D3 interval and two semi infinite D3 branes to
  bind and move collectively in the ${\cal Y}$ direction as indicated
  by the green arrows in (b) and (c).  For the case illustrated in
  (d), there will be an unbroken $U(1)$ gauge symmetry giving rise to
  a two complex dimensional moduli from the ${\cal N}=4$ vector
  multiplet illustrated using the green arrows.
\label{figns5d3ns52d3}}
\end{figure}

Let us revisit coupling to $T[SU(2)]$.  The brane configuration is
NS5---1D3(1)---NS5---2D3(2)---(semi-infinite). See also figure
\ref{figns5d3ns52d3}.  We will assume that generic complex FI terms
are turned on.  Note that when we couple this boundary condition to
some configuration on the right there can be a $U(2)$ gauge
transformation relating the two conditions.

From the first NS5, we have
\beq
\X_1 = -\zeta_1 \ , \qquad \Y_1 = a \ ,
\label{bc8eq1}
\eeq
and at the 1-2 interface we find
\beq
\X_1 &=& A_1 B_1  -\zeta_2  \ , \label{bc8eq2}\\
\X_2 &=& B_1 A_1 - \zeta_2 \left(\begin{array}{cc} 1 & 0\\0& 1\end{array} \right) \ .\label{bc8eq3}
\eeq
$A_1$ is a row vector and $B_1$ is a column vector in this example.  This sets
\beq
\X_2 &=& \left(\begin{array}{cc}- \zeta_1 & 0\\0& -\zeta_2\end{array} \right) \ ,\\
\Y_2 &=& \left(\begin{array}{cc} a & 0 \\0 & b \end{array} \right)  \ ,\\
A_1 &=& (\zeta_2-\zeta_1,0) \ ,\\
B_1 &=& \left( \begin{array}{c} 1\\ 0\end{array} \right)
 \ ,
\eeq
up to a gauge transformation, provided that $\zeta_1 \neq
\zeta_2$. There are two parameters, $a$ and $b$. This configuration
and the geometric interpretation of the two parameters is illustrated
in figure \ref{figns5d3ns52d3}.a.

If, instead, the two  FI parameters take equal values, then we can have
\beq
\X_2 &=& \left(\begin{array}{cc}- \zeta_1 & 1\\0& -\zeta_1\end{array} \right)\label{X2equalzetaJordan} \ ,\\
\Y_2 &=& \left(\begin{array}{cc} a& b\\0& a\end{array} \right) \ ,\\
A_1 &=& (0,1) \ ,\\
B_1 &=& \left( \begin{array}{c} 1\\ 0\end{array} \right) \ .
\eeq
Here, the form of the matter fields introduces an off-diagonal element
in (\ref{X2equalzetaJordan}); this possibility is consistent with the
fact that when two or more eigenvalues of a matrix coincide one cannot
always diagonalize the matrix, but it can always be put in the upper
triangular Jordan normal form.  Once again, we find two parameters,
$a$, and $b$, but their physical interpretation, illustrated in figure
\ref{figns5d3ns52d3}.b, is different. The parameter $b$ does not
affect the eigenvalues of $\cal X$ or $\cal Y$ and as such does not
deform the branes geometrically in these directions. Instead, it
encodes aspects of the embedding in the $X_6$ coordinate which
requires additional care. In addition, having this parameter
non-vanishing binds the finite D3 segment and the two semi-infinite D3
branes so the entire collection of D3-branes move together. This is
one of the novel features of the $T[SU(2)]$ not previously seen in the
S-dual configuration built using the D5-branes. It should also be
viewed as a subtle consequence of S-duality being non-local. We will
elaborate further on this point below.

Another possible branch arises from the case
\beq
\X_2 &=& \left(\begin{array}{cc}- \zeta_1 & 0\\0& -\zeta_1\end{array} \right) \ ,\\
\Y_2 &=& \left(\begin{array}{cc} a& 0\\0& a\end{array} \right) \ ,\\
A_1 &=& (0,b) \ ,\\
B_1 &=& \left( \begin{array}{c} 0\\ 0\end{array} \right) \ ,
\eeq
whose geometric interpretation also involves the deformation of the
semi-infinite D3 in the $X_6$ direction (there is a potential issue
with stability of the complex gauge quotient for this branch, which
will be described in detail in \cite{ToAppear}.) For now we will
illustrate it as figure \ref{figns5d3ns52d3}.c.  There will be another
closely related branch with non-vanishing $B_1$ and $A_1=0$.

Finally, if we set $\zeta_1 =\zeta_2$ without including the Jordan
form terms, there will be an unbroken gauge symmetry which we
illustrate in figure \ref{figns5d3ns52d3}.d.  It is the branches (a)
and (d) which has immediate counterparts in the S-dual. However, since
(d) leaves some gauge symmetry unbroken, and involves turning on
$X_9$, we expect the classical picture to receive corrections. The
fact that we have access to enumerating boundary deformation
classically in one description and the S-dual will eventually enable
us to infer the self-consistent quantum corrected description of the
moduli space when these boundaries are used to define a system on a
finite interval, which we will explore in detail in \cite{ToAppear}.

This structure generalizes to $T[SU(N)]$.  We will have that $\X_N$ is
a matrix in Jordan normal form.  $\Y_N$ will then be forced to be a
commuting matrix, also in Jordan normal form, except that we do not
have enough gauge freedom to set the off-diagonal elements to 1.

\subsubsection{1/4 BPS Generalization of $T[SU(N)]$}

\noindent {\bf NS-NS$^{\ensuremath{\prime}}$}

\begin{figure}
\centerline{\includegraphics[scale=0.8]{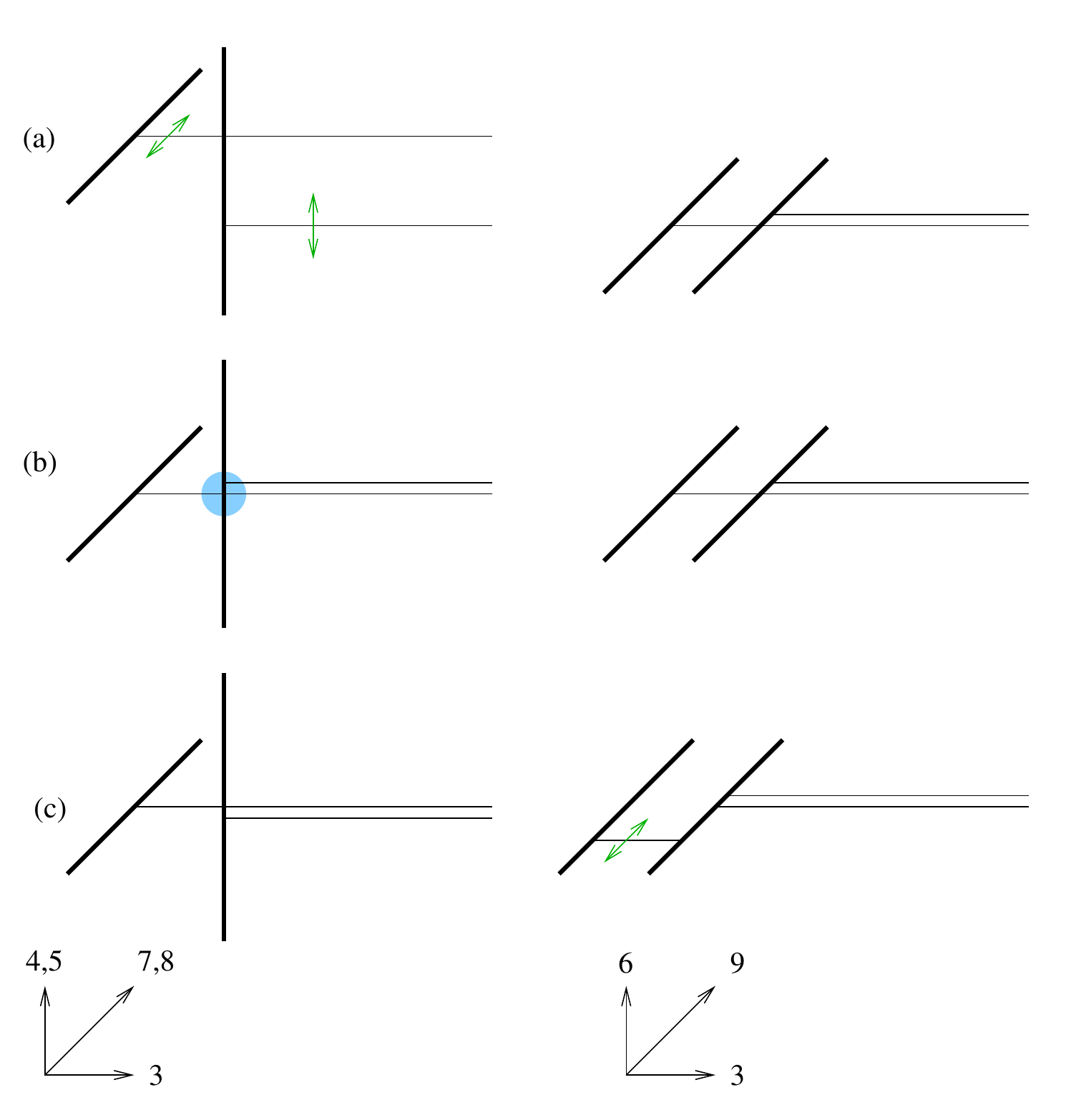}}
\caption{Brane configuration for NS5--D3--NS5$^{\ensuremath{\prime}}$--2D3--. This is the S-dual of configuration illustrated in figure \ref{figbb}. Unlike in the case of figure \ref{figbb}, some data pertaining to the embedding in the $X_6$ direction such as the real component of the FI parameter is contained in the blue disk in (b). 
 \label{figns5d3ns5p2d3}}
\end{figure}

Now suppose we rotate the second NS5 to NS5$^{\ensuremath{\prime}}$.
Then we have
\beq
\Y_1 &=& A_1 B_1  -\zeta_2 = a \ ,\\
\Y_2 &=& B_1 A_1 - \zeta_2 \left(\begin{array}{cc} 1 & 0\\0& 1\end{array} \right) \ .
\eeq
This implies that up to a gauge transformation we must be able to have
\beq
\Y_2 &=& \left(\begin{array}{cc}  a & 0\\0& -\zeta_2\end{array} \right) \ ,\\
\X_2 &=&  \left(\begin{array}{cc}- \zeta_1 & 0\\0& b\end{array} \right) \ ,
\eeq
and gives rise to a branch with two degrees of freedom, $a$, and $b$,
illustrated in figure \ref{figns5d3ns5p2d3}.a. There is a special case
when $a=-\zeta_2$ and $b= -\zeta_1$.  Then we can have a one complex
dimensional branch parameterized by $c$
\beq
\Y_2 &=& \left(\begin{array}{cc}  -\zeta_2 & 1\\0& -\zeta_2\end{array} \right) \ ,\\
\X_2 &=&  \left(\begin{array}{cc} -\zeta_1 & c\\0& - \zeta_1\end{array} \right) \ ,
\eeq
illustrated in figure \ref{figns5d3ns5p2d3}.b. There is also a branch
where ${\cal X}$ will not be in the Jordan normal form, for which the
gauge symmetry is not broken. The brane configuration for that branch
is illustrated in figure \ref{figns5d3ns5p2d3}.c up to gauge
transformations. Note also that the freedom to one of the
semi-infinite D3 freely in the ${\cal X}$ direction is not captured in
this branch.

How does this generalize to $N$ 5-branes?  It appears that in this
case, for each NS5 or NS5$^{\ensuremath{\prime}}$, one eigenvalue of
$\X_N$ or $\Y_N$, respectively, is fixed by the FI parameter of that
5-brane, while correspondingly one gets a meson in $\Y_N$ or $\X_N$.
When some of the eigenvalues coincide one can only put the matrix in
Jordan normal form, and then there is only enough freedom to set the
off-diagonal elements to 1 in either $\X$ or $\Y$ but not both.

\noindent {\bf NS5-NS5-NS5}

Let us write it out explicitly for $T[SU(3)]$.  We will find
\beq
\X_3 &=& \left(\begin{array}{ccc} -\zeta_1 & & \\ & -\zeta_2 & \\ & & -\zeta_3 \end{array} \right) \ ,\label{eq464} \\
\Y_3 &=& \left(\begin{array}{ccc} \;\;a & & \\ & \;\;b & \\ & & \;\;c \end{array} \right) \ , \label{eq465}
\eeq
unless some of the $\zeta_i$ are equal.  If two are equal, then we also have to consider
\beq
\X_3 &=& \left(\begin{array}{ccc} -\zeta_1 & & \\ & -\zeta_2 & 1\\ & & -\zeta_2 \end{array} \right) \ , \label{eq466} \\
\Y_3 &=& \left(\begin{array}{ccc} \;\; a & & \\ & \;\; b & \;\;c \\ & & \;\;b \end{array} \right) \ . \label{eq467}
\eeq

\begin{figure}
\centerline{\includegraphics[scale=0.8]{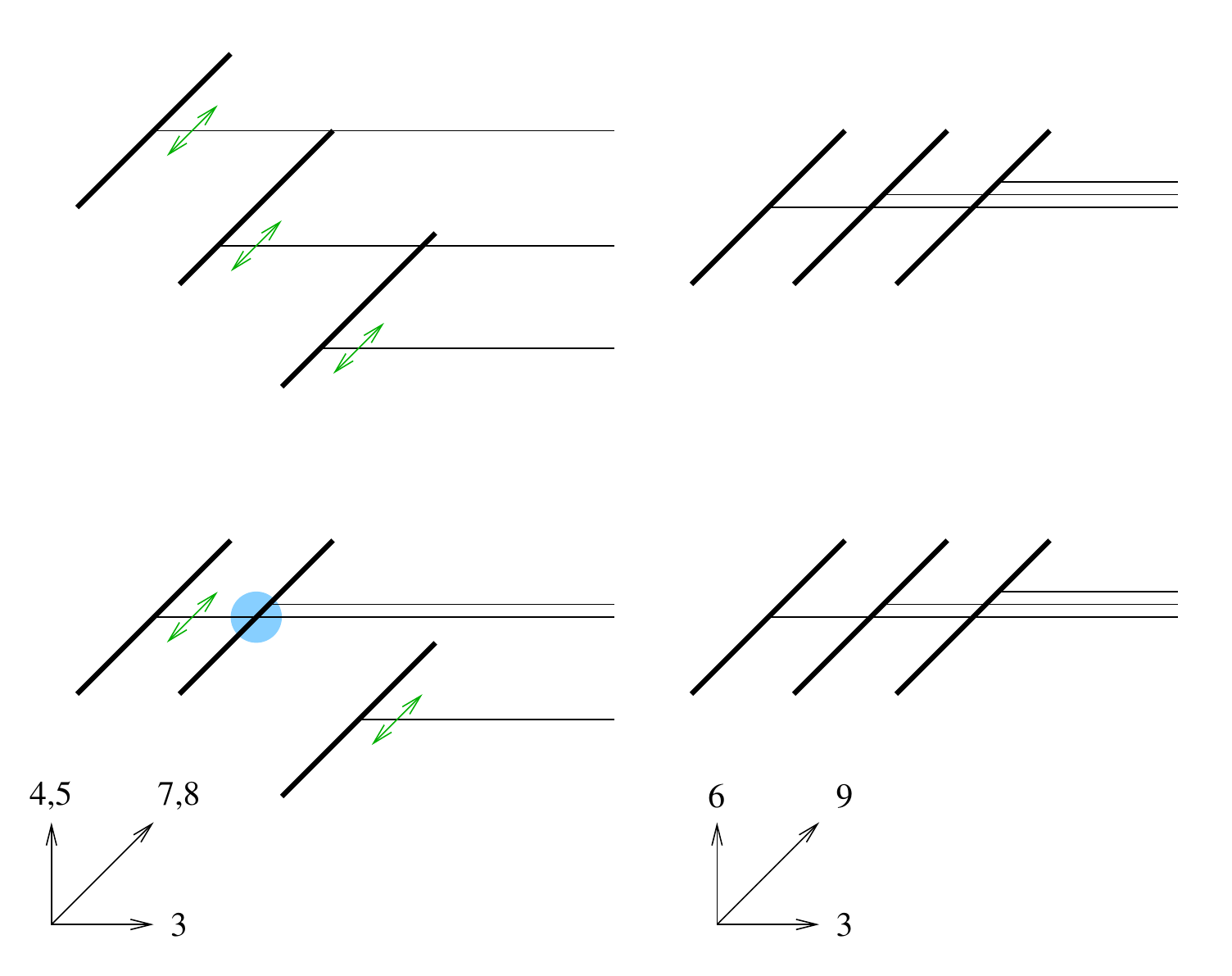}}
\caption{Brane configuration for NS5--D3--NS5--2D3--NS5--3D3--. \label{figns5d3ns52d3ns53d3}}
\end{figure}

The three parameters $a$, $b$, and $c$ in (\ref{eq464})--(\ref{eq465})
and (\ref{eq466})--(\ref{eq467}) are represented by green arrows and
blue disks in figure \ref{figns5d3ns52d3ns53d3}.

\noindent {\bf NS5-NS5-NS5$^{\ensuremath{\prime}}$}

\begin{figure}
\centerline{\includegraphics[scale=0.8]{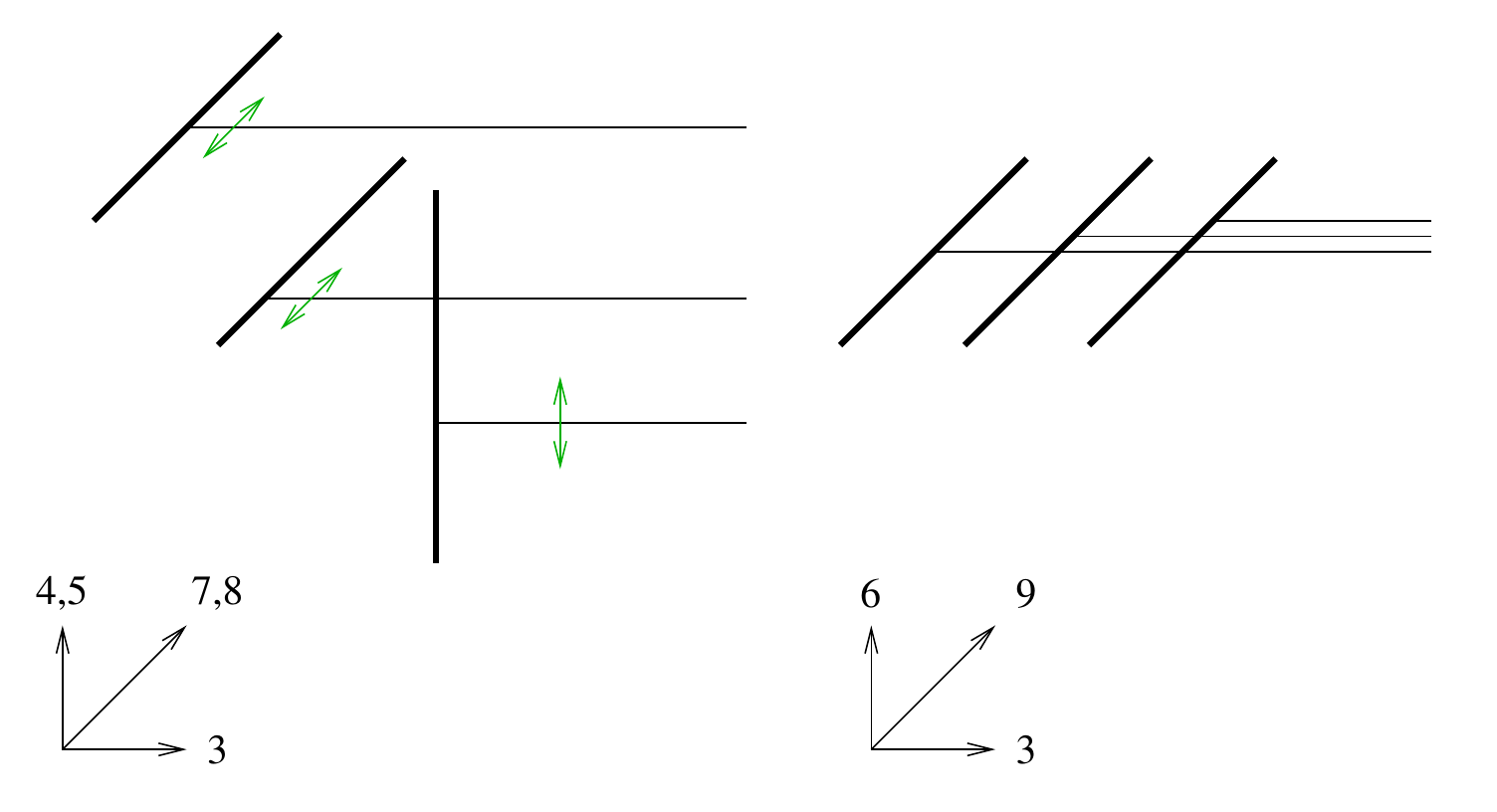}}
\caption{Brane configuration for NS5--D3--NS5--2D3--NS5$^{\ensuremath{\prime}}$--3D3--. \label{figns5d3ns52d3ns5p3d3}}
\end{figure}

If we have two NS5 and one NS5$^{\ensuremath{\prime}}$, we should write
\beq
\X_3 &=& \left(\begin{array}{ccc} -\zeta_1 & & \\ & -\zeta_2 & \\ & & c \end{array} \right) \ ,\label{eq468}\\
\Y_3 &=& \left(\begin{array}{ccc} \;\;a & & \\ & \;\;b & \\ & & -\zeta_3 \end{array} \right) \ , \label{eq469}
\eeq
assuming the eigenvalues are all distinct.  The three parameteres $a$,
$b$, and $c$ in (\ref{eq468})--(\ref{eq469}) corresponds to the green
arrows and blue disks in figure \ref{figns5d3ns52d3ns5p3d3}.

It should be clear now that we can prove our statement about the
structure of the boundary conditions for general $N$ by the method of
induction.  We assume the form of the boundary condition for
$T[SU(N-1)]$ (that is, $N-1$ NS5-branes with one D3-brane ending on
each NS5), so $\X_{N-1}$ and $\Y_{N-1}$ are some $(N-1)\times (N-1)$
diagonal matrices (if the eigenvalues are distinct.)  Then we couple
it to an NS5 brane (or NS5$^{\ensuremath{\prime}}$) with the $N-1$
D3-branes on one side and $N$ D3-branes on the other.  We then have
(for NS5)
\beq
\X_{N-1} &=& AB - \zeta I_{N-1} \ ,\\
\X_N &=& BA - \zeta I_{N} \ .
\eeq
We need to determine the eigenvalues of $\X_N$ in terms of those for
$\X_{N-1}$.  This can be done easily as follows.  Suppose $v$ is an
eigenvector of $\X_{N-1}$, with
\beq
\X_{N-1} v = \lambda v \ .
\eeq
As long as $Bv\neq 0$, one then checks that $Bv$ is an eigenvector of
$\X_{N-1}$ with eigenvalue $\lambda$.  If the gauge symmetry is
unbroken, one can always find $N-1$ such eigenvectors, either as
column eigenvectors $v$ in which case we multiply to form $Bv$ or row
eigenvectors $u$ and we construct $uA$.  So $N-1$ of the eigenvalues
of $\X_N$ are equal to the eigenvalues of $\X_{N-1}$.  To fix the last
eigenvalue, we compute
\beq
\tr(\X_N) - \tr(\X_{N-1})  =\tr( BA - \zeta I_{N})- \tr(AB - \zeta I_{N-1}) = -\zeta \ ,
\eeq
so the $N$-th eigenvalue of $\X_N$ is $-\zeta$.  

Now we need to fix the eigenvalues of $\Y_N$ in terms of the
eigenvalues of $\Y_{N-1}$.  We use
\beq
B\Y_{N-1}  =  \Y_{N}B \ ,
\eeq
and once again, we see that if $v$ is an eigenvector of $\Y_{N-1}$
with eigenvalue $\lambda$, $Bv$ is an eigenvector of $\Y_N$ with
eigenvalue $\lambda$.  This allows us to fix $N-1$ eigenvalues of
$\Y_N$, provided that the gauge symmetry is fully broken.  The $N$-th
eigenvalue of $\Y_N$ is unfixed and free to vary.  Note that each
eigenvalue of $\X$ which is fixed by an FI term is paired with a free
eigenvalue of $\Y$, and vice versa.

Because $[\X,\Y]=0$, we can simultaneously diagonalize $\X_N$ and
$\Y_N$.  Note that this assumes that we have a $U(N)$ gauge symmetry
to the right of the last NS5-brane.  If the gauge symmetry is broken
by the boundary conditions at infinity, then we can only impose the
$T[SU(N)]$ boundary conditions up to a $GL(N)$ rotation.

If some of the eigenvalues are equal then we have to consider the
corresponding Jordan normal forms as well.  In fact, one can easily
see that if we assume that $\X$ and $\Y$ are purely diagonal with
mutually degenerate eigenvalues, the interface matter fields preserve
some amount of unbroken gauge symmetry.  Demanding that this gauge
symmetry is broken forces us to {\it only} consider Jordan normal
form.

\section{Summary and Discussion}

Let us pause and summarize the status of this paper. Our primary goal
was to explore the boundary conditions which one can impose on ${\cal
  N}=4$ SYM in 3+1 dimensions such that the boundary preserves 1/4 of
the supersymmetries, generalizing the earlier analysis for the case
where 1/2 of the supersymmetries were preserved
\cite{Gaiotto:2008sa,Gaiotto:2008ak}.

Just as in the earlier work, we found that
\begin{enumerate}
\item A large class of boundaries can be constructed starting with the
  D3-brane intersecting with and terminating on NS5 and D5-branes
  arranged to preserve the correct number of supersymmetries.
\item The boundaries are equipped with parameters which can be viewed
  as deformation parameters, and unfixed degrees of freedom which can
  give rise to moduli of the theory at low energies.
\item For a given boundary, there exists a notion of the S-dual
  inherited from the S-duality of type IIB string theory.
\end{enumerate}
In this article, we generalized the results of
\cite{Gaiotto:2008sa,Gaiotto:2008ak} by including a broader class of
orientations of the NS5 and D5-branes consistent with the reduced
number of supersymmetries.

The main technical results which form the basis of our subsequent analysis are 
\begin{itemize}
\item Generalized Nahm equations  (\ref{qnahm1})--(\ref{qnahm4}),
\item Junction condition for D5 (\ref{dX})--(\ref{QY}) and for D5$^{\ensuremath{\prime}}$ (\ref{dY})--(\ref{QX}),
\item Junction condition for NS5 (\ref{XAB})--(\ref{BYYB}) and for NS5$^{\ensuremath{\prime}}$ (\ref{YAB})--(\ref{BXXB}). 
\end{itemize}
The analysis of the boundary conditions amounts to solving these sets
of equations which are a combination of algebraic and first order
differential relations.

As a practical tool to analyze the supersymmetry equations, it is
convenient to employ the complex gauge formalism where the Nahm
equation and the junction conditions are separated into complex
equations and the real equations. This typically breaks the analysis
into to two parts, where the complex part can be treated
separately. Most of the interesting features are already contained in
the analysis of the complex equations, without reference to the real
equations.  However there are situations (primarily related to real
mass or FI deformations) where some important additional information
relies on the analysis of the real part.

What we find from these analyses is that in many cases, the parameters
characterizing the boundary correspond to moving brane segments around
in the Hanany-Witten like picture (corresponding to the case where the
fields are in some sense nearly Abelian and the notion of positions is
well-defined.)  However, when the fields have some intrinsically
non-Abelian structure, there are some departures from the naive
expectations based on brane diagrams alone. This is to be expected
since one expects more quantum corrections with few preserved
supersymmetries. It is, after all, well known that there are more
quantum corrections to ${\cal N}=2$ supersymmetric gauge theories in
2+1 dimensions compared to ${\cal N}=4$ which can dynamically generate
superpotentials and lift various branches of moduli space
\cite{deBoer:1997ka,deBoer:1997kr,Aharony:1997bx}.

Some of the key departures and subtleties identified from our analysis
are as follows.

\begin{enumerate}
\item Embedding of the semi-infinite bulk in the $X_6$ coordinate is
  hidden as is illustrated in figure \ref{figbb}. This feature is
  also shared by the 1/2 BPS constructions of
  \cite{Gaiotto:2008sa,Gaiotto:2008ak} but was not emphasized in their
  treatment.   
\item Some configurations which are naively allowed from the
  consideration of naive Hanany-Witten brane constructions are not
  realized in the Nahm analysis.  An example of this is illustrated in
  figure \ref{figd52d3d5p3d3} in section \ref{poleBC} where only one
  of the three deformations of the semi-infinite D3 segments are
  allowed to move in the ${\cal Y}=X_7+iX_8$ directions. This is an
  example of lifting of branches naively expected from brane
  considerations.
\item There are 
degrees of freedom, illustrated in figures
  \ref{figns5d3ns52d3}, \ref{figns5d3ns5p2d3}, and
  \ref{figns5d3ns52d3ns53d3} which do not have any geometric
  interpretation as movement in the ${\cal X}= X_4+iX_5$, ${\cal Y} =
  X_7 + i X_8$, or $X_9$ directions. We will
  see in a followup paper \cite{ToAppear} that these deformations have
  a role in adjusting the $X_6$ coordinates. This further illustrates
  the subtleties involving the embedding in the $X_6$ coordinate.
\item Another feature, noted in section \ref{poleBC}, is the fact that
  the limit of coincident D5 and D5$^{\ensuremath{\prime}}$ branes is
  singular depending on the direction in which this limit is
  approached. This is in contrast to the half BPS cases where the
  positions of the 5-branes are irrelevant in the infrared. One
  important consequence of this observation is the fact that the
  classification of 1/4 BPS boundaries along the lines of GW's
  treatment \cite{Gaiotto:2008sa,Gaiotto:2008ak} based on the
  decoupling of 5-brane positions does not immediately generalize.
\end{enumerate}

It appears though that generally, these boundaries have parameters
characterizing deformations consistent with the supersymmetry
constraints (\ref{qnahm1})--(\ref{qnahm4}), (\ref{dX})--(\ref{QY}),
(\ref{dY})--(\ref{QX}), (\ref{XAB})--(\ref{BYYB}), and
(\ref{YAB})--(\ref{BXXB}).

An important question one must address is whether these
constraints derived at the classical level receive quantum
corrections. Generally, one expects a variety of perturbative and
non-perturbative corrections for dynamics of 2+1 dimensional systems
with ${\cal N}=2$ supersymmetry
\cite{deBoer:1997ka,deBoer:1997kr,Aharony:1997bx,Affleck:1982as}. These
boundary systems have the same amount of isometries and
supersymmetries, so we should expect similar corrections.

A powerful tool to assess the role of quantum corrections is
S-duality. Through S-duality, a model in the weak coupling limit
probes the strong coupling limit of the S-dual theory.  Using
S-duality, one can therefore access the theory in two opposing limits.
By identifying features such as the dimension of space of variables
parameterizing the boundaries, one can assess if any of these features
are likely to be constant with respect to the change in coupling and
therefore protected against quantum corrections.

Furthermore, one can argue that if the gauge symmetry is completely
broken in one of the duality frames, the supersymmetry constraints
(\ref{qnahm1})--(\ref{qnahm4}), (\ref{dX})--(\ref{QY}),
(\ref{dY})--(\ref{QX}), (\ref{XAB})--(\ref{BYYB}), and
(\ref{YAB})--(\ref{BXXB}) are protected against certain corrections
(such as instanton corrections.)  Using these ingredients, one can
attempt to patch together a self-consistent picture of the complex
structure of the parameter space which takes quantum corrections into
account.

We have considered the simplest 1/4 BPS boundary consisting of D5 and
D5$^{\ensuremath{\prime}}$ branes illustrated in figure \ref{figbb}
and its S-dual illustrated in figure \ref{figns5d3ns5p2d3}. We in fact
see that under S-duality, branch (a) of figure \ref{figbb} is
naturally mapped to branch (a) of figure \ref{figns5d3ns5p2d3} in the
S-dual. Similarly, branch (b) of figure \ref{figbb} appears to
correspond to branch (c) of figure \ref{figns5d3ns5p2d3}. For this
branch, we see that there are unbroken gauge symmetries in the latter
description, so we take the former to be the more reliable
description.

The status of branch (b) in figure\ref{figns5d3ns5p2d3} is somewhat
curious. There is no counterpart of this branch in figure
\ref{figbb}. Nonetheless, gauge symmetry is completely broken in the
duality frame depicted by figure \ref{figns5d3ns5p2d3} and as such one
expects this branch to be reliable. So why does it not show up in the
figure \ref{figbb} frame?

The answer to this question has to do with the subtlety of the
treatment of the $X_6$ coordinate as well as the accounting of the
boundary condition infinitely far away in the $X_3$ direction. The
point is that the bulk scalar fields need to be specified at infinity
in order to define the 
moduli space
rigorously.  In addition, in comparing the moduli spaces of two
different S-dual descriptions of the same system, it is necessary to
properly S-dualize the boundary condition at infinity consistently. In
particular, because S-duality is strictly not a local transformation
in space-time, some features specifying the boundary condition at
infinity on one end in one duality frame might map to some features
localized near the other boundary.

One way to systematically resolve these issues is to carefully
construct the system on an interval terminated by boundary conditions
on both ends, and to compare the moduli space on both sides. Such a
construction is precisely the prescription to engineer ${\cal N}=2$
field theories in 2+1 dimensions along the line illustrated in figure
\ref{figb}. Taking advantage of S-duality and protection against
potential quantum correction in cases where gauge symmetry is
completely broken, we should be able to piece together the complicated
branching structure of moduli spaces of field theories in 2+1
dimensions with 4 supercharges. Some of these issues have been studied
previously using tools such as brane construction, mirror symmetry,
and some inspired guesswork
\cite{deBoer:1997ka,deBoer:1997kr,Aharony:1997bx}. The 1/4 BPS
conditions (\ref{qnahm1})--(\ref{qnahm4}), (\ref{dX})--(\ref{QY}),
(\ref{dY})--(\ref{QX}), (\ref{XAB})--(\ref{BYYB}), and
(\ref{YAB})--(\ref{BXXB}) turns out to offer a new systematic approach
to analyze these issues and offer some new insights. We will report on
our findings on these matters in part II of this paper
\cite{ToAppear}.

\section*{Acknowledgments}

We would like to thank P.~Argyres, T.~Clark, D.~Gaiotto, P.~Goddard,
K.~Intriligator, S.~Khlebnikov, M.~Kruczenski, G.~Michalogiorgakis,
N.~Seiberg, E.~Weinberg, B.~Willett, E.~Witten, and D.~Xie for discussion.

This work is supported in part by the DOE grant DE-FG02-95ER40896
(A.~H.), by DOE grant DE-FG02-91ER40681 (P.~O.) and partly by
Princeton Center for Theoretical Science, by Institute for Advanced
Study and by WPI program, MEXT, Japan (M.~Y.).

This work was initiated at the SPOCK meeting (University of
Cincinnati, 2011) and we would like to thank the organizers for the
meeting.  M.~Y. would like to thank Aspen Center for Physics, Newton
Institute (Cambridge), Simons Center (Stony Brook), YITP (Kyoto), and
KITP (UCSB) for hospitality where part of this work has been
performed.  The content of this paper was presented at presentations
by A.~H. (Michigan, Dec.\ 2012, KITP Apr.\ 2014), P.~O. (Purdue,
Nov.\ 2013, Madison Apr.\ 2014) and M.~Y. (IAS, Apr.\ 2013,
Mar.\ 2014; Kavli IPMU, Aug.\ 2013; Perimeter, Jan.\ 2014; KITP,
Feb.\ 2014), and we thank the audiences of these talks for feedback.


\appendix

\section{Analysis of BPS Boundary Conditions}\label{appendixA}

\subsection{Spinor Algebra}
\label{appendixA1}

To perform explicit computations with the fermions, it is often useful
to have an explicit basis for the $\Gamma$-matrices which realizes the
global symmetries of the system of interest in a natural way.  For
boundaries with $\mathcal{N}=4$ supersymmetry, we have the global
symmetry ${SO(2,1)} \times {SO(3)} \times {SO(3)}$ which acts
naturally on the {012}, {456}, {789} directions.  For our purposes, a
convenient basis may be defined as follows:
\beq
{
\Gamma_0} &=& \; {i\sigma_2} \otimes 1\otimes 1\otimes 1 \otimes \sigma_2 \ ,\\
{\Gamma_1} &=& \; {\sigma_1} \otimes 1 \otimes 1 \otimes 1 \otimes \sigma_2  \ ,\\ 
{\Gamma_2} &=& \; {\sigma_3} \otimes 1 \otimes 1 \otimes 1 \otimes \sigma_2   \ ,\\ 
\Gamma_3 &=& \;\; 1 \otimes 1 \otimes 1 \otimes \sigma_2 \otimes \sigma_1   \ ,\\ 
{\Gamma_4} &=& -1  \otimes {\sigma_1} \otimes 1 \otimes \sigma_3 \otimes \sigma_1 \ , \\ 
{\Gamma_5} &=& -1  \otimes {\sigma_2} \otimes 1 \otimes \sigma_3 \otimes \sigma_1 \ ,  \\ 
{\Gamma_6} &=& -1 \otimes {\sigma_3} \otimes 1 \otimes \sigma_3\otimes \sigma_1 \ , \\ 
{\Gamma_7} &=& \;\;1 \otimes 1  \otimes {\sigma_1}  \otimes \sigma_1 \otimes \sigma_1  \ , \\
{\Gamma_8} &=& \;\;1 \otimes 1  \otimes {\sigma_2}  \otimes \sigma_1 \otimes \sigma_1 \ , \\ 
{\Gamma_9} &=& \;\; 1 \otimes 1  \otimes {\sigma_3}  \otimes \sigma_1 \otimes \sigma_1    \ ,
\eeq
from which we compute
\beq
{B_0} & = & \Gamma_{456789}=1\otimes 1\otimes 1\otimes {i\sigma_2}\otimes 1 \ , \\
{B_1} & = & \Gamma_{3456}=1\otimes 1\otimes 1\otimes {\sigma_1}\otimes 1 \ , \\
{B_2} & = & \Gamma_{3789}=1\otimes 1\otimes 1\otimes {\sigma_3}\otimes 1 \ ,\\
{\Gamma_{11}} &=&\bar{\Gamma}=1\otimes 1\otimes 1\otimes 1\otimes {\sigma_3} \ .
\eeq
The convenience of this basis lies in the fact that the global
symmetry ${SO(2,1)} \times {SO(3)} \times {SO(3)}$ acts simply on the
first three entries of this tensor basis, the fourth entry transforms
under an $SL(2,\bR)$ algebra (with generators $B_0, B_1, B_2$), and
the fifth entry labels the 10-d chirality.  The 16 component spinor
$\epsilon_L$ of 4$d$ $\mathcal{N}=4$ is in representation $({\bf 2} ,
{\bf2} ,{\bf 2} , {\bf 2})$.

There is also a Majorana condition which we should impose on the 10-d
spinor, but it only fixes the phase multiplying the components of the
spinors.  In our analysis, this phase plays no role, but to be
explicit, a possible Majorana condition is
\beq
\zeta^* = \Gamma_{4679} \zeta \ .
\eeq

\subsection{Half BPS Boundary Conditions}

Let us first discuss the half BPS case, and analyze the boundary
supersymmetry condition \eqref{boundary_half}.  Since we are
interested in boundary conditions which preserve the $SO(2,1)\times
SO(3)\times SO(3)$ global symmetry, we should not fix the spinors
$\Psi$ and $\varepsilon$ in the the first three tensor indices of the
basis in appendix \ref{appendixA1}; also the Weyl condition fixes the
fifth index.  Therefore we fix half the components of $\Psi$ at the
boundary by fixing the fourth spinor index.  Specifically, in $\Psi$
and $\varepsilon$ we label the fourth spinor as ${\varepsilon_0}$ and
${\vartheta}$ (with arbitrary $v$ and $\psi$ labeling the first three
spinor indices)
\be\varepsilon = v \otimes {\varepsilon_0} \ ,\qquad \Psi = \psi \otimes {\vartheta} \ .\ee
Now, we expand $({\bf 2},{\bf 2},{\bf 2})$ components of $F\!\!\!\!/$
\beq
 {1 \over 2} \Gamma\cdot  F &=& 
({\sigma_\mu},{1},{1}) {R^\mu}+
({\sigma_\mu},{\sigma_a},{1}) {U_{\mu a}}+
({\sigma_\mu},{1},{\sigma_m}) {V_{\mu m}}\cr
&& +  ({1},{\sigma_a},{1}) {S_a} + ({1},{1},{\sigma_m}) {T_m} + ({1},{\sigma_a}, {\sigma_m}) {W_{am}}
 \ .
 \eeq

The condition that half the supersymmetry is preserved is then simply
that all the components in this tensor decomposition of $F\!\!\!\!/$
vanish: ${R^\mu} = {U_{\mu a}} = {V_{\mu m}} = {S_a} = {T_m} =
{W_{am}} = 0$ with
\begin{align}
\begin{split}
R^\mu & =  \epsilon^{\mu \nu \lambda} F_{\nu \lambda} ({\bar \varepsilon_0 \vartheta}) + 2 F^{3 \mu} ({\bar \varepsilon_0 B_0 \vartheta})  \ ,\cr
U_{\mu a} & =  2 i D_\mu X_a ( {\bar \varepsilon_0 B_2 \vartheta}) \ , \cr
V_{\mu m} & =  -2 i D_\mu Y_m ({\bar \varepsilon_0 B_1 \vartheta})  \ ,\cr
W_{am} & =  2 [X_a,Y_m] ({\bar \varepsilon_0 B_0 \vartheta})  \ , \cr
S_a & =  2i [X_b,X_c] \epsilon_{abc} ({\bar \varepsilon_0 \vartheta}) - 2 i D_3 X_a ({ \bar \varepsilon_0 B_1 \vartheta} ) \ ,\cr
T_m & =  2i [Y_n,Y_p] \epsilon_{mnp} ({\bar \varepsilon_0 \vartheta}) - 2 i D_3 Y_m ({ \bar \varepsilon_0 B_2 \vartheta} ) \ ,\cr
{B_0} & =  i \sigma_2  \ , \cr
{B_1} & =   \sigma_1 \ , \cr
{B_2} & =   \sigma_3  \ .
\end{split}
\end{align}
To proceed further, we look for ${\varepsilon_0}$ and ${\vartheta}$
for which we can consistently solve this set of equations.  First, in
order to preserve the $SO(3)\times SO(3)$ R-symmetry, we have to
impose the same set of boundary conditions for $X_a$ for
$a=1,2,3$. The same applies to $Y_a$.

Up to an overall normalization we can write
$\bar{\varepsilon}_0=(1,a), \vartheta=(\alpha,1)^T$ (we allow $a$ and
$\alpha$ to be infinity.)  We then have
\beq
{\bar \varepsilon_0  \vartheta}=a+\alpha \ , \quad {\bar \varepsilon_0 B_0 \vartheta}=- a\alpha+1 \ , \quad {\bar \varepsilon_0 B_1 \vartheta}=a\alpha+1\ , \quad {\bar \varepsilon_0 B_2 \vartheta}=\alpha-a \ .
\label{vi}
\eeq

Suppose first that both $X_a$ and $Y_a$ both obey Dirichlet boundary
conditions.  Then from the second and the third equations we have
${\bar \varepsilon_0 B_1 \vartheta}={\bar \varepsilon_0 B_2
  \vartheta}=0$.  This cannot be satisfied for real $\epsilon_0$ and
$\vartheta$.

Suppose now that neither $X_a$ nor $Y_a$ obey Dirichlet boundary
conditions.  Then the fourth and the fifth equations tell us ${\bar
  \varepsilon_0 B_1 \vartheta}={\bar \varepsilon_0 B_2 \vartheta}=0$,
which is again a contradiction.

If $X$ is Neumann and $Y$ is Dirichlet, we obtain
\be D_3 X_a + {u \over 2} \epsilon_{abc} [X_b,X_c] = 0  \ , \ee
with
\be {0 = \bar \varepsilon_0 B_2 \vartheta = \bar \varepsilon_0 (1 + u B_1) \vartheta} \ .
\ee
We also have
\be \epsilon_{\lambda \mu \nu} F^{3 \lambda} = \gamma F_{\mu \nu} \ ,\ee
which implies
\be  {\bar \varepsilon_0 (1 + \gamma B_0) \vartheta = 0} \ .
\ee
This is solved by
\be \bar \varepsilon_0 = (1,a)  \ ,\qquad \vartheta = \left(\begin{array}{c} a  \\ 1 \end{array}\right) \ , \qquad \gamma = - {2a \over 1 - a^2} \ ,  \qquad u = -{2a \over 1 + a^2} \ .\ee

Different choices of $a$ correspond to different types of boundary conditions,
as discussed in the main text:

\bl $a=0$, $u=0$, $\gamma=0$: D3 ending on NS5 on ${012}{456}$

\bl $a=1$, $u=-1$, $\gamma=\infty$: D3 ending on D5 on ${012}{456}$

\bl  $\gamma=-{g_{\rm YM4}^2 \over 4 \pi}{q \over p}$: D3 ending on $(p,q)5$ on ${012}{456}$

We can rotate these boundary conditions with the $SO(6)$ R-symmetry of
the bulk, while preserving 1/2 BPS supersymmetry.  For example, an NS5
on ${012}{456}$ and a D5 on ${012}{789}$ preserve the same 
eight supersymmetries.  In the main text we mostly use these boundary
conditions.

\subsection{Rotated Branes}\label{appendixA3}

Our 4$d$ $\scN=4$ $U(N)$ gauge theory is the low-energy effective
theory on $N$ D3-branes extended along the 0123 directions. This
preserves the supersymmetry
\beq
\epsilon_L=\Gamma_{0123} \, \epsilon_R \ .
\label{D3epsilon}
\eeq
If we add a D5-brane in $012345$, we have another constraint
\begin{equation*}
\epsilon_L= \Gamma_{012456} \, \epsilon_R \ .
\end{equation*}
From \eqref{D3epsilon}, this is the same as 
\beq
\epsilon_L= \, \Gamma_{3456}\, \epsilon_L \ ,
\quad
\epsilon_R=- \, \Gamma_{3456}\, \epsilon_R \ .
\label{D5epsilon}
\eeq
The total system consisting of D3 and D5 branes preserves 8
supercharges.  Similarly, an NS5-brane in $0123456$-directions
preserves
\begin{equation*}
\epsilon_L=\Gamma_{012456} \,\epsilon_L\ , \quad
\epsilon_R=-\Gamma_{012456} \, \epsilon_R \ .
\end{equation*}
From the chirality constraint this is the same as 
\beq
\epsilon_L=\Gamma_{3789} \,\epsilon_L \ , \quad
\epsilon_R=-\Gamma_{3789} \, \epsilon_R \ .
\label{NS5epsilon}
\eeq
More generally, a $(p,q)$ 5-brane in
$012[47]_{\psi}[58]_{\varphi}[69]_{\rho}$ preserves (under
\eqref{D3epsilon}) $\epsilon_L, \epsilon_R$ satisfying
\beq
\epsilon_L=R_{\theta, \psi, \varphi, \rho} \Gamma_{3456} R_{\theta, \psi, \varphi, \rho}^{-1} \epsilon_L\ , \quad
\epsilon_R=-R_{\theta, \psi, \varphi, \rho} \Gamma_{3456} R_{\theta,
\psi, \varphi, \rho}^{-1} \epsilon_R \ ,
\label{RGR}
\eeq
or equivalently
\beq
(R_{\theta, \psi, \varphi, \rho}^2 -1)\, \epsilon_L=\epsilon_L, \quad
(R_{\theta, \psi, \varphi, \rho}^2 -1)\, \epsilon_R=\epsilon_R \ .
\label{RGR2}
\eeq
Here $\theta\equiv \pi \arctan \left(\frac{p}{q}\right)$, and 
the rotation matrix $R_{\theta, \psi, \varphi, \rho}$ given by
\beq
R_{\theta, \psi, \varphi, \rho}\equiv \exp\left(\frac{\theta}{2}\Gamma_{456789}
+\frac{\psi}{2}\Gamma_{47}+\frac{\varphi}{2}\Gamma_{58}+
\frac{\rho}{2}\Gamma_{69}\right) \ ,
\eeq
whose square is computed to be\footnote{We have \begin{align}
\label{SL2}
\begin{aligned}
\Gamma_{456789}&=1\otimes 1\otimes 1\otimes i\sigma_2\otimes 1\ , \\
\Gamma_{47}&=1\otimes \sigma_1\otimes \sigma_1\otimes \sigma_2\otimes  
 1\  , \\
\Gamma_{58}&=1\otimes \sigma_2\otimes \sigma_2\otimes \sigma_2\otimes  
 1 \ , \\
\Gamma_{69}&=1\otimes \sigma_3\otimes \sigma_3\otimes \sigma_2\otimes 
 1 \ ,
\end{aligned}
\end{align}
}
\beq
R_{\theta, \psi, \varphi, \rho}^2=\exp\left[ 1\otimes (\theta \, 1\otimes 1+ \psi\, \sigma_1\otimes \sigma_1+
\varphi\, \sigma_2\otimes \sigma_2 + \rho\, \sigma_3\otimes \sigma_3)
\otimes \sigma_2\otimes 1\right] \ .
\eeq
Therefore \eqref{RGR2} becomes the statement that 
the term inside the bracket, i.e.,
\beq
\theta\, 1\otimes 1+ 
\psi\, \sigma_1^{(2)}\otimes \sigma_1^{(3)}+
\varphi\, \sigma_2^{(2)}\otimes \sigma_2^{(3)} 
+ \rho\, \sigma_3^{(2)}\otimes \sigma_3^{(3)}
\ ,
\label{shouldbetrivial}
\eeq
acts trivially on the second and the third components.  The operator
\eqref{shouldbetrivial} has eigenvectors
\begin{align}
\begin{aligned}
|1\rangle&\equiv \frac{1}{\sqrt{2}}
\left[
|\! \!\uparrow \uparrow \rangle+
|\! \!\downarrow \downarrow \rangle
\right] \ , 
\\
|2\rangle& \equiv
\frac{1}{\sqrt{2}}
\left[
|\! \!\uparrow \uparrow \rangle-
|\! \!\downarrow \downarrow \rangle
\right]\ ,
\\
|3\rangle& \equiv\frac{1}{\sqrt{2}}
\left[
   |\! \!\uparrow \downarrow \rangle + 
   |\! \!\downarrow \uparrow \rangle
\right] \ ,
\\
|4\rangle&\equiv \frac{1}{\sqrt{2}}
\left[
  |\! \!\uparrow \downarrow \rangle-
  |\! \!\downarrow \uparrow \rangle 
\right] \ , 
\label{eigenstates}
\end{aligned}
\end{align}
with eigenvalues
\begin{align}
\begin{aligned}
\theta_1&\equiv \theta+\rho+ \psi-\varphi \ , \\
\theta_2&\equiv \theta+\rho- \psi+\varphi \ , \\
\theta_3&\equiv \theta-\rho +\psi+\varphi \ , \\
\theta_4&\equiv \theta-\rho-\psi-\varphi  \ .
\label{eigenvalues}
\end{aligned}
\end{align}
When the angles $\theta, \psi, \varphi, \rho$ are generic this means
all the supersymmetries are broken. However, in special situations
some of \eqref{eigenvalues} are zero, and corresponding supersymmetry
is preserved. This gives Table \ref{KOOtable2}.  Note that the number
of unbroken symmetry is preserved under permutations of $\theta, \psi,
\varphi, \rho$.

\begin{table}
{\small
\begin{center}
\begin{tabular}{|c|c|c|l|}
\hline
&  condition &  3$d$ SUSY & 5-brane \\
\hline
\hline
I &  $ \textrm{all angles zero} $ & $\scN=4$
 & D5$\left(012456\right)$ \\
\hline 
II-1 &  $\theta=\psi=\varphi=-\rho$ &   \multirow{2}{*}{$\scN=3$}
 & $(p,q)5\left(012\rot{4}{7}{\theta}\rot{5}{8}{\theta}
 \rot{6}{9}{-\theta}\right)$ \\
 \cline{1-2}\cline{4-4} 
II-2& $-\theta=\psi=\varphi=\rho$ & 
 & $(p,q)5\left(012\rot{4}{7}{-\theta}\rot{5}{8}{-\theta}
 \rot{6}{9}{-\theta}\right)$ \\
\hline
III-1 &  $\varphi=\psi, \rho=-\theta$& \multirow{3}{*}{$\scN=2$}
 & $(p,q)5 \left(012\rot{4}{7}{\psi}\rot{5}{8}{\psi}\rot{6}{9}{-\theta}\right)$ \\
 \cline{1-2}\cline{4-4}  
III-2 &  $\theta=\rho=0, \psi=\varphi$& 
 & D5$\left(012\rot{4}{7}{\psi} \rot{5}{8}{\psi}6 \right)$ \\
 \cline{1-2}\cline{4-4} 
III-3 &  $\psi=\varphi=0, \rho=\theta$ &   & $(p,q)5\left(01245\rot{6}{9}{\theta}\right)$ \\
\hline
IV&  $\rho=\theta+\psi+\varphi$ &$\scN=1$&  $(p,q)5
\left(012\rot{4}{7}{\psi}\rot{5}{8}{\varphi}
\rot{6}{9}{\theta+\psi+\varphi}\right)$ \\
\hline
\end{tabular}
\end{center}
}
\caption{The classification of supersymmetries preserved, when we have
D3-D5(012456) and an extra 5-brane,
where we defined $\theta=\pi \arctan \left(p/q\right)$.
The number of remaining supersymmetries is determined simply from the number of zeros in the four angles \eqref{eigenvalues}. 
 This is a reformat of the similar table from \cite{Kitao:1998mf}.}
\label{KOOtable2}
\end{table}

\subsection{1/4 BPS Case}

In this appendix we explore the 1/4 BPS boundary conditions in a
framework more general than that of the main text, at least those
corresponding to the 1/4 BPS case of \eqref{branetable}.  What is
special about the brane configuration there is that the system has
$SO(2)_{45}\times SO(2)_{78}$ symmetry, corresponding to the rotation
symmetry of $45$ and $78$-planes; we expect that the general 3$d$
$\mathcal{N}=2$ theory to have only one $SO(2)$ symmetry, namely the
R-symmetry of the theory.

As we learn from the analysis of the previous subsection, the
supersymmetries preserved by the D3-D5-D5$^{\ensuremath{\prime}}$
system is spanned by $|1\rangle, |2\rangle$ (case III-2 in table
\ref{KOOtable2}).  Motivated by this, we choose the Ansatz
\beq
\bar{\epsilon}^T = \{ |1\rangle \otimes \stup+  a |1\rangle \otimes \stdown \ , \quad |2\rangle \otimes \stup+  a |2\rangle \otimes \stdown  \} \ ,
\eeq
and
\begin{align}
\begin{split}
\Gamma_3 \Psi \in 
\textrm{span} \Big\{
& \beta  |1\rangle \otimes \stup +   |  1\rangle  \otimes \stdown \ ,  \qquad 
  \beta  |  2\rangle \otimes \stup +   |  2\rangle  \otimes \stdown \ ,  \\ 
& \alpha |  3\rangle \otimes \stup + |  3\rangle \otimes \stdown \ , \qquad 
\alpha |  4\rangle \otimes \stup + |  4\rangle \otimes \stdown 
\Big\} \ .
\end{split}
\end{align}
The one parameter $a$ (two parameters $\alpha, \beta$) determines the
choice of two (four) out of ${\bf 2}\otimes {\bf 2}\otimes {\bf 2}$.
Note that this Ansatz is consistent with the $SO(2)_{45}\times
SO(2)_{78}$ symmetry, which mixes two states $|1\rangle$ and
$|2\rangle$, and also states $|3\rangle$ and $|4\rangle$.  The
equation \eqref{boundary_half} simplify considerably under this
Ansatz, and we obtain
\begin{align}
c F_{\mu \nu}+ c_0\, \epsilon_{\mu \nu\lambda} F_{3 \lambda}&=0  \ , \label{manyeq1}\\
c_1 D_{\mu} X_9 -c_2 D_{\mu} X_6&=0 \ , \label{manyeq2}\\
d_2 D_{\mu} X_4=d_2 D_{\mu} X_5&=0 \ , \label{manyeq3}\\
d_1 D_{\mu} X_7=d_1 D_{\mu} X_8&=0 \ , \label{manyeq4}\\
c_0 [X_6, X_9] &=0\ , \label{manyeq5}\\
c_0( [X_5,X_7]+[X_4, X_8])&=0 \ , \label{manyeq6}\\
c_0([X_5,X_8]-[X_4, X_7])&=0 \ ,\label{manyeq7}\\
d_1 D_{3} X_{4}+d_0 [X_9, X_5]+  d [X_6, X_5]&=0 \ ,\label{manyeq8}\\
d_1 D_3 X_{5}+d_0 [X_9, X_4]+ d [X_6, X_4]&=0 \ , \label{manyeq9}\\
d_2 D_3 X_7+d_0 [X_6, X_8]+ d [X_9, X_8]&=0 \ ,\label{manyeq10}\\
d_2 D_3 X_8+d_0 [X_6, X_7]+ d [X_9, X_7]&=0 \ , \label{manyeq11}\\
c ([X_4, X_5]+[X_7, X_8])-  c_1 D_{3}X_6-c_2 D_{3}X_9&=0 \label{manyeq12}\ , 
\end{align}
where we defined
\begin{equation}
\begin{split}
c&\equiv a+\beta \ , \quad c_0\equiv -a\beta+1 \ , \quad
c_1\equiv a\beta+1 \ , \quad c_2\equiv \beta-a  \ , \\
d&\equiv a+\alpha \ , \quad d_0\equiv -a\alpha+1 \ , \quad
d_1\equiv a\alpha+1 \ , \quad d_2\equiv \alpha-a \ .
\end{split}
\end{equation}
These are the 1/4 BPS generalizations of \eqref{vi}.

Let us analyze the equations. Due to the existence of
$SO(2)_{45}\times SO(2)_{78}$ symmetry $X_{4,5}$, and also $X_{7,8}$
have to obey the same boundary condition.

First, assume that $X_{4,5}$ and $X_{7,8}$ both obey the Dirichlet
boundary condition.\footnote{Dirichlet here means that the field
  vanishes on the boundary; more general boundaries can be discussed
  when we discuss mass/FI deformations as in section \ref{sec323}.}
Then from \eqref{manyeq8}--\eqref{manyeq11} we have $d_1=d_2=0$.  This
is not possible for real $a, \alpha$.

Second, assume that neither $X_{4,5}$ nor $X_{7,8}$ obey the Dirichlet
boundary condition.  Then from the \eqref{manyeq3} and \eqref{manyeq4}
we have $d_1=d_2=0$.  Again, this is not possible for real $a,
\alpha$.

Therefore we can assume that only one of $X_{4,5}$ and $X_{7,8}$ obey
Dirichlet boundary conditions.  Without losing generality we can
assume that $X_{7,8}$ obey Dirichlet boundary conditions.  Then
\eqref{manyeq3} gives $d_2=0$ and hence $\alpha=a, d_1\ne 0$.  The
remaining equations are
\begin{align}
c F_{\mu \nu}+ v_0\, \epsilon_{\mu \nu\lambda} F_{3 \lambda}&=0 \ , \label{lesseq1} \\
c_1 D_{\mu} X_9 -c_2 D_{\mu} X_6&=0  \ ,  \label{lesseq2} \\
c_0 [X_6, X_9]&=0 \ ,  \label{lesseq3} \\
d_1 D_{3} X_{4}+d_0 [X_9, X_5]+  d [X_6, X_5]&=0 \ ,  \label{lesseq4} \\
d_1 D_3 X_{5}+d_0 [X_9, X_4]+d [X_6, X_4]&=0 \ ,  \label{lesseq5} \\
c ([X_4, X_5])-  c_1 D_{3}X_6-c_2 D_{3}X_9 &=0\ .  \label{lesseq6} 
\end{align}

Let us now consider boundary conditions for $X_6$ and $X_9$. Suppose
that both of them obey Dirichlet boundary conditions.  Then we have
from \eqref{lesseq4} and \eqref{lesseq5} that $d_1=0$, which
contradicts with the condition $d_2=0$ we obtained previously.

Another case is that only one of $X_6$ and $X_9$ obey Dirichlet.
Suppose $X_9$ obey Dirichlet, but not $X_6$.  Then we have $c_2=0$
from \eqref{lesseq2}, and together with the previous results we have
$d_1, c_1\ne 0, c_2=d_2=0$, and the remaining equations are the Nahm
equations for $X_{4,5,6}$.  Hence the equations reduce to the 1/2 BPS
case. Similarly, if $X_4$ obey Dirichlet, but not $X_7$, then we have
$d_1, c_2\ne 0, c_1=d_2=0$, and the remaining equations are the Nahm
equations for $X_{5,6,7}$.  Again this is 1/2 BPS.

The last case is when neither $X_6$ nor $X_9$ obey Dirichlet boundary
conditions.  We still have to satisfy \eqref{lesseq2} and
\eqref{lesseq3}.  In \eqref{lesseq2} if $X_6$ and $X_9$ are
independent we have $c_1=c_2=0$, which is a contradiction.  This means
we need
\beq
c_2 X_6=c_1 X_9\ ,
\label{X47}
\eeq
which automatically solve \eqref{lesseq3}.  Three of the remaining
equations \eqref{lesseq4}--\eqref{lesseq6} then simplify to
\beq
D_3 X_6-{c'}[X_4, X_5]=0\ , \!\quad
D_3 X_4-{c''}[X_5, X_6]=0\ , \!\quad 
D_3 X_5-{c''}[X_6, X_4]=0 \ ,
\label{Nahmtmp}
\eeq
with
\begin{align}
\begin{split}
c'\equiv \frac{c_1 c}{c_1^2+c_2^2}=\frac{(a\beta+1)(a+\beta)}{(a^2+1)(\beta^2+1)}\ , \quad
c''\equiv \frac{d c_1 +d_0 c_2}{d_1 c_1}=\frac{a+\beta}{1+a\beta} \ .
\end{split}
\end{align}
If we define the rescaled variables $\tilde{X}_{4,5,6}$ by
\beq
\tilde{X}_6\equiv \frac{X_6}{f \cos{\varphi}} \ , \quad
\tilde{X}_{4,5}\equiv \frac{X_{4,5}}{f} \ ,
\eeq
with 
\beq
f =\frac{2a}{a+\beta} \sqrt{\frac{1+\beta^2}{1+a^2}} \ , \quad \cos\varphi=\frac{1+a \beta}{\sqrt{(1+a^2)(1+\beta^2)}}=\frac{c_1}{\sqrt{c_1^2+c_2^2}}  \ , 
\eeq
then \eqref{Nahmtmp} becomes the Nahm equation \eqref{Nahm}
\beq
D_3 \tilde{X}_{a}-\frac{u}{2}\epsilon_{abc}[\tilde{X}_b, \tilde{X}_c]=0 \qquad (a=4,5,6) \ ,
\eeq
with the value of $u$ being the same as in the 1/2 BPS case:
\beq
u\equiv \frac{c' f}{\cos \varphi}=\frac{c'' \cos\varphi}{f}=\frac{2a}{1+a^2} \ .
\eeq
The final remaining equation \eqref{lesseq1} for the gauge field is familiar from the 1/2 BPS case:
\beq
\epsilon_{\lambda\mu\nu} F^{3\lambda}+\gamma F_{\mu\nu}=0 \ ,
\eeq
with $\gamma=c/c_0$.

\bigskip

To summarize, we have a 2-parameter family of boundary conditions, 
parametrized by $a$ and $\beta$, or equivalently $\gamma$ and $\varphi$:
\begin{align}
\begin{split}
\epsilon_{\lambda\mu\nu}F_{3 \lambda}+\gamma F_{\mu\nu}\big|=D_3 (\cos \varphi X_{6}+\sin\varphi X_9) \big|=D_3 X_{4,5} \big| \\
\quad =(\sin\varphi X_6-\cos \varphi X_{9})\big|
=X_{7,8}\big|
=0 \ .
\end{split}
\end{align}

One of the parameters, $\gamma$ represents the $\theta$-angle of the
bulk 4$d$ ${\cal N}=4$ theory, just as in the 1/2 BPS case.  Another
parameter, $\varphi$, represents the rotation in the $47$-planes. In
fact, we can see from \eqref{X47} that such a rotation by $\varphi$ in
the $69$-plane sets $X_9=0$ (notice that $\tan
\varphi=c_2/c_1=X_9/X_6$).  Alternatively, we could choose to do
simultaneous rotations by the same angle in $47$ and $58$-planes,
while keeping the $47$-plane.  This is consistent with the 1/4 BPS
case of Table \ref{KOOtable2}, which excludes pure D5-like boundary
condition preserving 3/8 BPS supersymmetry.

Let us choose to rotate by $\varphi$ in $47$ and $58$-planes instead
of the $69$-plane.  In the absence of the $\theta$-angle, we find
NS5-like boundary conditions
\begin{align}
\begin{split}
\textrm{NS5-like}: F_{3 \mu}\big|=D_3 X_{6}\big|=D_3 (\cos\varphi X_{4,5}+\sin\varphi X_{7,8}) \big| \\
\quad =X_{9}\big|
=(\cos\varphi X_{7,8}-\sin\varphi X_{4,5} )\big|
=0 \ ,
\end{split}
\end{align}
as well as the D5-like boundary condition.\begin{align}
\begin{split}
\textrm{D5-like}: F_{\mu \nu}\big|&= X_{6}\big|=(\cos\varphi X_{4,5}+\sin\varphi X_{7,8}) \big| \\
&=D_3 X_{9}\big|
=D_3(\cos\varphi X_{7,8}-\sin\varphi X_{4,5} )\big|=0\ .
\end{split}
\end{align}

Note each of these boundary conditions are 1/2 BPS, not 1/4 BPS, and is
simply the previous 1/2 BPS boundary condition rotated by an angle
$\varphi$. However we can construct more general 1/4 BPS boundary conditions
as the composite of 1/2 BPS boundary conditions
as in figure \ref{defect_collide} (see section \ref{sec23}).

\subsection{1/8 BPS Equation}

By following the same method as in the main text,
we can derive the $1/8$ 
bulk BPS equations.
The equation involves a component $A_3$
of the gauge field and all the six scalars $X_{4}, \ldots, X_{9}$
(c.f.\ \cite{Constable:2002yn}):
\begin{align}
\begin{split}
D_3 X_{4}&=[X_5,X_6]+[X_7,X_8]  \ , \\
D_3 X_{5}&=[X_6,X_4] +[X_7,X_9] \ , \\
D_3 X_{6}&=[X_4,X_5] +[X_9,X_8]\ , \\
D_3 X_{7}&=[X_8,X_4] +[X_9,X_5]\ , \\
D_3 X_{8}&=[X_4,X_7] +[X_9,X_6] \ , \\
D_3 X_{9}&=[X_5,X_7]+[X_8,X_6] \ , \\
[X_5, X_8]&+[X_6, X_7]+ [X_9, X_4]=0 \ . 
\label{1/8Nahm}
\end{split}
\end{align}
The equations represent the brane configuration
\begin{align}
\begin{tabular}{c||ccc|c|cccccc}
       & 0& 1  & 2& 3& 4& 5& 6& 7& 8& 9 \\
       \hline
D3 & $\circ$  &  $\circ$ & $\circ$ & $\circ$ &    &   &    &   &   &     \\
D5 & $\circ$  &  $\circ$ & $\circ$ &  &  $\circ$ & $\circ$ & $\circ$   &   &  &     \\
D5 & $\circ$  &  $\circ$ & $\circ$ &  &   & $\circ$ &    &  $\circ$ &   & $\circ$     \\
D5 & $\circ$  &  $\circ$ & $\circ$ &  &  &  &$\circ$   &  & $\circ$  & $\circ$    \\
$\overline{\rm D5}$ & $\circ$  &  $\circ$ & $\circ$ &  &   $\circ$ &   &    &  $\circ$ & $\circ$  &     
\label{manyD5}
\end{tabular}
\end{align}
where $\overline{\rm D5}$ means the D5 with orientation reversal.

The equation \eqref{1/8Nahm} can be obtained from the dimensional
reduction of the higher-dimensional generalization of the self-duality
equations.  For the 1/2 BPS case this states that the Nahm equation is
the dimensional reduction of 4$d$ self-duality equations. For the 1/8
BPS case the relevant equations are
\begin{align}
&F_{12}+F_{34}+F_{56}=0 \ , \\
&F_{35}+F_{17}-F_{64}=0 \ , \\
&F_{15}+F_{34}+F_{62}=0 \ , \\
&F_{13}+F_{57}-F_{42}=0 \ , \\
&F_{27}+F_{34}+F_{56}=0 \ , \\
&F_{47}+F_{17}-F_{64}=0 \ , \\
&F_{76}+F_{34}+F_{56}=0 \ , 
\label{1/8SD}
\end{align}
which is one of the the higher dimensional generalizations of
instanton equations discussed in \cite{Corrigan:1982th}, coming from a
reduction of an octonionic instanton.  The field strength obeys
\beq
F_{\mu \nu}=\frac{1}{2}c_{\mu\nu\rho\sigma}F_{\rho \sigma} \ ,
\eeq
where $c_{\mu\nu\rho \sigma}$ is Hodge dual of the 3-form determined
from the structure constant of the imaginary octonion.  (Amusingly,
the octonionic instanton has an interpretation in terms of a
seven-dimensional generalization of the Euler top
\cite{Fairlie:1997sj,Ueno:1998en}, much as the $SU(2)$ version of
Nahm's equations are the three-dimensional Euler top equations.)

Since we have only two supercharges (3$d$ $\mathcal{N}=1$) the moduli
space of the 1/8 BPS equation will not be K\"{a}hler, and it is not
clear if we can control the quantum corrections in any way.  However
we could still hope to extract useful data on 1/8 BPS boundary
conditions or the boundary 3$d$ $\mathcal{N}=1$ theories.  For example,
we can again trivially solve the equations by superimposing Nahm
poles. In the gauge $A_3=0$
\begin{equation}
\begin{split}
X_4&=\frac{1}{y}\, \textrm{diag}(\rho_1(t_1),0, 0, \rho_4(t_1))  \ ,  \\
X_5&=\frac{1}{y}\, \textrm{diag}(\rho_1(t_2),\rho_2(t_1),0,0)  \ , \\
X_6&=\frac{1}{y}\, \textrm{diag}(\rho_1(t_3) ,0,\rho_3(t_1),0)  \ , \\
X_7&=\frac{1}{y}\, \textrm{diag}(0, \rho_2(t_2),0,\rho_4(t_2)) \ , \\ 
X_8&=\frac{1}{y}\, \textrm{diag}(0, 0,\rho_3(t_3),\rho_4(t_3)) \ , \\
X_9&=\frac{1}{y}\, \textrm{diag}(0, \rho_2(t_3),\rho_3(t_2),0) \ .
\end{split}
\end{equation}
where $\rho_{i}: \mathfrak{sl}(2)\to \mathfrak{g}_{i}\, (i=1, \cdots,
4)$ represent mutually commuting Nahm poles and $t_a$ are again
generators of $\mathfrak{sl}(2)$.  Such a quadruple of Nahm poles,
generalizing a pair of Nahm poles for the 1/4 BPS equation, should be
the crucial ingredient in the specification of the 1/8 BPS boundary
condition and the associated 3$d$ $\mathcal{N}=1$ theory.


\section{Boundary Degrees of Freedom in $\mathcal{N}=1$ Superspace}\label{sec.boundaryN1}

In section \ref{sec:localized} we presented the junction conditions
for D5 and NS5 branes in a manifestly $\mathcal{N}=2$ supersymmetric
language.  The $\mathcal{N}=2$ superfields realize half of the full
supersymmetry of the junction and are the natural objects for
considering 1/4 BPS boundary conditions.

It can also be useful to write the boundary effective actions in terms
of 3$d$ $\mathcal{N}=1$ superfields.  Although this only realizes a
quarter of the supersymmetry of the interface, it is possible to
realize the $SO(3)$ global symmetry explicitly, which is not possible
in $\mathcal{N}=2$ language.  From this point of view it is easier to
understand how the supersymmetry enhances to $\mathcal{N}=4$.

For D5-branes, the interface action with $N$ D3-branes on each side 
\beq
W_{\rm interface} = \left( \omega^{\dag}\phi_1^a \sigma^a \omega \right) \delta(y-y_0) \ ,
\eeq
as can be read off from (4.23) and (4.28) of
\cite{DeWolfe:2001pq}. 
The bulk superfield $\Phi$ are defined such
that its lowest component
\be \Phi^a = \phi^a_1 + i \phi^a_2   \ee
as defined in (4.22) of \cite{DeWolfe:2001pq}. The real and imaginary
parts of $\Phi$ are:
\beq 
\phi^a_1 &=& Y^a + \Theta \psi^a + \Theta \bar \Theta (F^a - D_3 X^a)  \ ,\\
\phi^a_2 &=& X^a + \Theta \chi^a + \Theta \bar \Theta (\tilde F^a - D_3 Y^a) \ .
\eeq
The $F$-terms associated with varying $\phi$ are
\beq
D_3 X^a = F^a = \frac{1}{2}\epsilon^{abc}[ X_b, X_c]-\frac{1}{2}\epsilon^{abc}[ Y_b, Y_c] + \omega^{\dag} \sigma^a \omega \delta(y-y_0)  \ .
\eeq
Because we have only realized $\mathcal N =1$ supersymmetry, the
$F$-term equations contain both the commutators of $X_a$ and of $Y_a$,
with an $SO(3)$ symmetry manifest.  The enhancement to $\mathcal N =4$
supersymmetry, with an $SO(4)$ $R$-symmetry, requires that the
commutators of $Y_a$ vanish. We also have $F$-terms from varying the
$Q$ fields which imply \eqref{D5Ycondition1}.

For NS5-branes, the effective superpotential at the interface is
\be
W = \tr \left(\phi_{1,L}^a \phi_{2,L}^a - \phi_{1,L}^a \xi^{\dag}\sigma^a \xi  - \phi_{1,R}^a \phi_{2,R}^a +\phi_{1,R}^a \xi^T (\sigma^a)^T \xi^* \right) \delta(y-y_0).
\ee
where $\xi, \xi^{\dagger}$ are defined in \eqref{xidef}.  It is
straightforward to check that the $F$-term equations from this
superpotential reproduce the equations (\ref{XAB})-(\ref{BYYB}).

\bibliography{boundary}
\bibliographystyle{utphys}

\end{document}